\documentclass[aps,prc,onecolumn,notitlepage,nofootinbib]{revtex4-2}

\usepackage{ulem} 
\usepackage[utf8]{inputenc}
\usepackage{textgreek}
\usepackage{graphicx}
\usepackage[figuresright]{rotating}
\usepackage{makecell}
\usepackage{bm,amsmath,amssymb}
\usepackage[mathscr]{eucal}
\usepackage{multirow}
\usepackage{xcolor}
\usepackage{mathrsfs}
\usepackage{mathtools}
\usepackage{slashed}
\usepackage{graphics}
\usepackage{graphicx}
\usepackage{subfigure}
\usepackage{dsfont}
\usepackage{longtable}
\usepackage{bbm} 
\usepackage{float}
\usepackage{tcolorbox}
\usepackage{lipsum}
\usepackage{cancel}
\usepackage{enumerate}
\usepackage{diagbox}
\usepackage{booktabs}
\usepackage{tabularx}
\usepackage{array}
\usepackage[breaklinks,colorlinks,citecolor=citcolor,urlcolor=blue,linkcolor=lcolor]{hyperref}

\definecolor{lcolor}{rgb}{0.5,0,0}
\definecolor{citcolor}{rgb}{0,0.3,0.0}

\newcommand{\dd}{\mathrm{d}}
\newcommand{\pT}{p_{\rm T}}

\newcommand{\Th}{T_{\rm H}}
\newcommand{\Tpc}{T_{\rm pc}}

\newcommand{\ie}{\textit{i.e.}}

\newcommand{\raa}{R_{\rm AA}}
\newcommand{\QQb}{Q\bar{Q}}

\newcommand{\rmd}{{\rm{d}}}

\newcommand{\Ds}{\mathcal{D}_s}

\newcommand{\NN}{\mathrm{NN}}

\begin{document}


 \title{ Open Heavy Flavor in Nuclear Collisions: From Lattice QCD to Observables}

\author{Yu Fu}
\affiliation{Department of Physics, Duke University, Durham, NC  27708-0305, U.S.A}

\author{Tharun Krishna}
\affiliation{Department of Physics and Astronomy and Cyclotron Institute, Texas A\&M University, College Station, TX 77843-3366, U.S.A}

\author{Weiyao Ke}
\affiliation{Key Laboratory of Quark and Lepton Physics (MOE) \& Institute of Particle Physics, Central China Normal University, Wuhan, Hubei 430079, China}
\affiliation{Southern Center for Nuclear-Science Theory (SCNT), Institute of Modern Physics, Chinese Academy of Sciences, Huizhou, Guangdong 516000, China}

\author{Steffen A. Bass}
\affiliation{Department of Physics, Duke University, Durham, NC  27708-0305, U.S.A}

\author{Ralf Rapp}
\affiliation{Department of Physics and Astronomy and Cyclotron Institute, Texas A\&M University, College Station, TX 77843-3366, U.S.A}

\date{\today}

\begin{abstract}
Heavy quarks serve as excellent probes of the Quark-Gluon-Plasma (QGP) produced in ultra-relativistic heavy-ion collisions. Here, we report on a comprehensive framework that transports charm quarks from their initial production through the QGP medium, hadronization, and the hadronic phase to final observables in experiments. For the transport through the QGP fireball, which is described by 2+1D relativistic viscous fluid dynamics, we employ an extended Langevin equation that combines elastic diffusion processes with medium-induced radiation, including interference effects. The charm-quark transport coefficients are derived from nonperturbative $T$-matrix interactions, which are constrained by lattice QCD 
and also produce an equation of state for the thermal medium which is consistent with the one employed in the hydrodynamic simulations. Pre-hadronic resonant states which emerge in the $T$-matrix near the QGP transition temperature naturally lead to a transition to hadronic degrees of freedom. This transition is realized in two different approaches: a coalescence model augmented with 4-momentum conservation and the resonance recombination model, both supplemented with fragmentation functions constrained by charm-hadron spectra in proton-proton collisions at the LHC and RHIC. The late-stage rescattering of charm hadrons is assessed with schematic heavy-light cross sections implemented into a hadronic transport model. We report results for the nuclear modification factors and elliptic flows of $D^0$ and $D_s^+$ mesons and $\Lambda_c^+$ baryons, together with the $D_s^+/D^0$ and $\Lambda_c^+/D^0$ yield ratios. The calculations are compared with measurements from the ALICE and CMS collaborations in Pb--Pb collisions at $\sqrt{s_{\NN}}=5$\,TeV and with STAR data in Au-Au collisions at $\sqrt{s_{\NN}}=200$\,GeV. 
\end{abstract}

\maketitle
\tableofcontents

\section{Introduction}
Quantum Chromodynamics (QCD), the theory of the strong nuclear force, predicts that at energy densities of around 0.5\,GeV/fm$^3$, hadronic matter undergoes a gradual transition to a new state where quarks and gluons are deconfined, i.e., the quark-gluon plasma (QGP)~\cite{HotQCD:2014kol,Borsanyi:2013bia}. Over the past three decades, ultra-relativistic heavy-ion collisions (URHICs) at the Super Proton Synchrotron (SPS), Relativistic Heavy-Ion Collider (RHIC) and Large Hadron Collider (LHC) have provided fairly robust evidence that this phase change is realized in experiments. Due to its transient existence in a relatively small space-time volume, the QGP produced in the laboratory is not directly observable, but a variety of final-state observables support its formation. These observables include collective flow patterns, jet quenching phenomena, and electromagnetic radiation, all of which suggest the formation of a hot and dense medium exhibiting very large collision rates and transport parameters suggestive of a near-perfect fluidity. Heavy-flavor (HF) observables are playing a central role in these diagnostics. The suppression and regeneration in quarkonium production~\cite{Andronic:2024oxz,Andronic:2025jbp} and the large elliptic flow of open HF hadrons~\cite{Rapp:2018qla,He:2022ywp} have corroborated that the large interaction rates in the medium provide more access to the underlying interactions than bulk observables. Ongoing efforts, in both experiment and theory, are improving the determination of transport parameters but also offer unique opportunities to study their microscopic origin while establishing direct connections to first-principle results from lattice QCD~\cite{HotQCD:2025fbd}.   
The large masses of the heavy charm and bottom quarks ($Q=c,b$) play a key role in this endeavor. Firstly, the heavy quark (HQ) mass is much larger than the nonperturbative QCD scale, $m_Q \gg \Lambda_{\text{QCD}}$. This means that HQ production requires large momentum transfer processes, allowing for perturbative calculations of their production cross-section in the initial hard nucleon-nucleon (NN) collisions. Secondly, since the HQ mass is significantly greater than the typical temperature of the QGP medium, $m_Q \gg T_{\text{QGP}}$, HQ pair production is essentially limited to the initial stages of the collision, with negligible thermal production from the QGP itself~\cite{Zhang:2007yoa}. Thirdly, the thermal relaxation time of HQs is expected to be comparable to or longer than that of light quarks, parametrically by a factor of $m_Q/T_{\text{QGP}}$.
This implies that, after the bulk matter reaches local thermal equilibrium, the HQ interactions with the medium, governed by the strong force and thus conserving quark flavor, impart noticeable modifications to the initial momentum spectra but retain a memory of the interaction history, including the hadronization process and subsequent rescattering in hadronic matter.
Finally, the large HQ mass facilitates the construction of reliable interactions, such as HQ effective theories and their implementation into quantum many-body theory, where the reduction of the Bethe-Salpeter equation renders the problem much more tractable, with controlled approximations.

Due to the nonperturbative and dynamic nature of the QCD medium in heavy-ion collisions, a fully first-principles description of the space-time evolution of the strongly coupled QGP and the transport of HQs embedded within remains elusive. Effective models have been developed to provide tractable and phenomenologically meaningful descriptions of the system. On the one hand, the evolution of QGP has been found to be adequately described by viscous relativistic hydrodynamics, incorporating proper initial conditions for the energy and entropy deposition during the initial impact of the incoming nuclei. On the other hand, the large HQ mass enables the use of a Brownian motion treatment where the typical momentum kicks from the medium are small. The stochastic nature of HQs' Brownian motion motivates the use of diffusion-based approaches, especially at low and intermediate momenta where elastic interactions dominate~\cite{Svetitsky:1987gq}; it is typically implemented via a Langevin process~\cite{Moore:2004tg,vanHees:2004gq}. At higher HQ momenta, the momentum transfer distribution increasingly deviates from the Gaussian form inherent in a stochastic framework, which can be better captured by the Boltzmann transport equation, although with the restriction of having to assume the medium is made of quasiparticles, typically coupled to perturbative two-body scatterings. This is not required for the transport coefficients in the Langevin approach~\cite{Liu:2018syc}, where off-shell properties of the medium partons and the pertinent HQ scattering amplitude can be accounted for.

In addition, HQs traversing the QGP may undergo inelastic scattering resulting from the emission (or absorption) of gluons. This process, known as medium-induced radiation, contributes significantly to the energy loss of HQs at high momentum. Unlike elastic processes, radiative processes can take out a substantial fraction of the HQ's energy in a single scattering event and are sensitive to the color structure and coherence effects in the medium.
{Theoretical treatments of medium-induced gluon radiation at weak coupling incorporate coherence effects associated with the Landau--Pomeranchuk--Migdal (LPM) effect~\cite{Migdal:1956tc}. Several theoretical approaches have been developed to describe medium-induced gluon radiation from HQs, incorporating finite-mass effects and multiple-scattering interference~\cite{Armesto:2003jh,Zhang:2003wk,Djordjevic:2003zk,Kang:2016ofv}. Beyond these weak-coupling treatments, complementary strong-coupling approaches based on gauge--gravity duality have also been employed to study HQ energy loss and drag~\cite{Herzog:2006gh,Gubser:2006bz}. These approaches differ in their assumptions concerning the microscopic structure of the medium, the treatment of multiple scattering and interference, and the kinematic regimes in which they are applicable.}

Most simulations, including a simultaneous treatment of elastic and radiative processes, are carried out with Monte Carlo techniques. However, depending on the approximations applied, the relative importance of collisional versus radiative energy loss, depending on the energy of the HQ, the path length traversed in the medium, and the transport coefficients characterizing the QGP, remains an open question.

Over the past two decades, substantial efforts have been devoted to developing quantitative transport descriptions of HF particles in QCD matter. A broad range of approaches has been pursued, including Langevin and Fokker--Planck dynamics, linearized and full Boltzmann transport, nonperturbative interactions, quasiparticle and off-shell descriptions, and different combinations of collisional and radiative processes, coupled to realistic bulk-medium evolution~\cite{He:2014cla,Cao:2013ita,Beraudo:2014boa,Scardina:2017ipo,Song:2015ykw,Uphoff:2014hza,Cao:2017hhk,Ke:2018tsh}. More recent developments have extended these approaches to quasiparticle and nonperturbative descriptions of the medium, multiscale and multistage evolution, event-by-event dynamics, and hybrid strong/weak-coupling implementations~\cite{Liu:2021dpm,Xing:2021xwc,JETSCAPE:2022hcb,Beraudo:2025nvq}. 
Recent studies have additionally begun to explore HQ dynamics in a medium that is beyond the near-equilibrium picture, including HQ evolution in the pre-equilibrium stage of the collision, non-Gaussian transport, as well as HQ interactions with collective excitations of the medium using effective field theory~\cite{Avramescu:2024xts,Boguslavski:2023fdm,Rajagopal:2025rxr,Kirchner:2025xdb}. For reviews of the study of HQ dynamics in QCD matter, see e.g. Refs.~\cite{Dong:2019byy,Dong:2019unq,Zhao:2020jqu,He:2022ywp,Apolinario:2022vzg}.

Within phenomenological transport approaches aimed at quantitative comparisons with heavy-ion data, systematic model comparisons have demonstrated that the extraction of HQ transport properties depends not only on the microscopic HQ--medium interaction, but also on the initial HQ distributions, the space-time evolution of the medium, the treatment of collisional and radiative processes, and the hadronization model~\cite{Rapp:2018qla,Cao:2018ews,Xu:2018gux,Zhao:2023nrz}. 

In this work, we develop an HF transport framework that combines nonperturbative HQ diffusion coefficients from a lattice-QCD-informed thermodynamic $T$-matrix approach with medium-induced radiation implemented through an effective treatment of interference effects. We investigate two $T$-matrix interaction scenarios, the Vector-Confining-Potential (VCP) and the Wilson-Line-Correlator (WLC) approaches, as well as two distinct hadronization prescriptions, improved sudden coalescence and resonance recombination, supplemented with fragmentation functions constrained by proton-proton ($pp$) data. The transport calculation is embedded in a viscous hydrodynamic background calibrated to light-hadron observables. We then systematically investigate the effects of the microscopic interaction, radiative energy loss, initial-state nuclear modifications, hadronization, and hadronic rescattering on open-charm observables. By comparing nuclear modification factors, elliptic flow, and charm-hadron yield ratios, we examine how these different ingredients of the HF transport in QCD matter affect the interpretation of experimental data.

The paper is organized as follows. In Sec.~\ref{sec:trans}, we introduce the HQ evolution model inside the QGP medium by recalling the transport equations (Sec.~\ref{ssec:trans-eq}), discussing pertinent coefficients derived from the nonperturbative $T$-matrix formalism for elastic interactions (Sec.~\ref{ssec:QMBT}), the implementation of radiative processes including interference effects (Sec.~\ref{ssec:rad}), hadronic transport (Sec.~\ref{ssec:hadron}), and the underlying bulk evolution (Sec.~\ref{ssec:bulk}).
In Sec.~\ref{sec:hadro} we describe the hadronization of HQs from the QGP, for which we will employ two different recombination models: an improved sudden-coalescence model (iSCM) in Sec.~\ref{ssec:coal} and the resonance recombination model in Sec.~\ref{ssec:rrm}, both supplemented with suitable fragmentation models. 
In Sec.~\ref{sec:obs} we deploy our framework to compute observables:  we constrain the initial HQ spectra, obtained from perturbative QCD, for each hadronization model via fits to experimental $pp$ spectra (Sec.~\ref{ssec:initial}), and  elaborate on our simulation results for key open HF observables, \ie,  the $D^0$, $D_s^+$, $\Lambda_c^+$ nuclear modification factors and elliptic flows (Secs.~\ref{ssec:raa} and \ref{ssec:v2}, respectively), as well as for $D_s^+/D^0$ and $\Lambda_c^+/D^0$ ratios (Sec.~\ref{ssec:chem}), thereby confronting our results to experimental data from the ALICE and CMS collaborations in Pb--Pb collisions at $\sqrt{s_{\NN}}=$ 5.02\,TeV, and from STAR data in Au--Au collisions at $\sqrt{s_{\NN}}=$ 200\,GeV; we also examine separately the effects of several ingredients entering the calculation, including the effects from charm quark hadronization, initial-state shadowing, rescattering in the hadronic phase, as well as the comparison of contributions from diffusion and radiation during the HQ in-medium transport state (Sec.~\ref{ssec:model-effects}). Finally, we summarize our main findings in Sec.~\ref{sec:sum}.

\section{Heavy Quark Transport in QGP Medium}
\label{sec:trans}
A comprehensive description of HQ transport through the QCD medium formed in URHICs is generally separated into several key components. We will start by discussing the transport equations in Sec.~\ref{ssec:trans-eq} for which we employ a Fokker-Planck approach for elastic interactions (where the nonperturbative microphysics is encoded in friction and diffusion coefficients), matched to a Boltzmann treatment for perturbative gluon radiation including interference effects. This will be followed by a more detailed account of the nonperturbative features in the interactions, specifically the self-consistent $T$-matrix approach for elastic HQ scattering off thermal partons with nontrivial spectral functions based on recent lattice-QCD constraints in Sec.~\ref{ssec:QMBT}, and the gluon radiation processes with an improved implementation of formation time effects in Sec.~\ref{ssec:rad}. 
Our model for HF hadron transport is based on rescattering processes in a hadronic transport approach and will be presented in Sec.~\ref{ssec:hadron}, and the main elements of the bulk evolution simulation, consisting of a viscous hydro model matched to UrQMD hadronic transport, are recapitulated in Sec.~\ref{ssec:bulk}.   
Finally, Sec.~\ref{sec:hadro} will be dedicated to the treatment of HQ hadronization.

\subsection{Transport Equations and Microscopic Inputs}
\label{ssec:trans-eq}
The transport of HQs in a QGP is typically studied in a linearized Boltzmann equation with few-body scatterings or with Fokker-Planck equations using a set of specific transport coefficients. 

In the linearized Boltzmann equation, the real-time dynamics of the HQ are obtained by solving the time evolution of its phase-space density $f_Q(t,x,p)$. Both the HQ and the medium constituents are treated as quasiparticles. The scattering between HQs and medium quasiparticles is assumed to be dominated by screened few-body scatterings computed via perturbation theory. These conditions rely on weak-coupling conditions. The term linearization is threefold: 1) scattering between heavy (anti-)quarks is neglected, 2) the back reaction from the HQ on the medium is ignored, and 3) the phase-space density of medium constituents is a local thermal distribution with temperature and flow velocity fields given by the evolution model.

The Fokker-Planck equation is a low-momentum transfer effective description of HQ transport provided that the HQ-medium interaction is frequent and soft and that the HF particles can be treated as quasiparticles. It makes no assumption on the nature of the medium,  which can include nonperturbative effects of the medium and the HQ-medium interaction. 

Both the linearized Boltzmann equation and the Fokker-Planck equation can be solved by evolving an ensemble of (weighted) particles, and all observables depending on their phase-space distributions can be evaluated as a weighted average over the particle ensemble. The particle dynamics corresponding to the linearized Boltzmann equation are described by a stochastic collision process governed by transition rates, whereas those corresponding to the Fokker–Planck equation can be described by Langevin equations.

In the LIDO (linearized Boltzmann with diffusion) transport model~\cite{Ke:2018tsh,Ke:2018jem}, employed in this work, both schemes are combined by incorporating the diffusion dynamics for soft interactions matched to a linearized Boltzmann equation that includes large-momenta-transfer processes, with a heuristic implementation of the LPM effect for gluon radiation.
If one neglects the non-local interference effect, the time evolution of the $f_Q(t,x,p)$ in this model can be formally expressed as
\begin{align}
\frac{\rmd f_Q}{\rmd t} = \mathcal{D}[f_Q] + C_{1\leftrightarrow 2}[f_Q] +  C_{2\leftrightarrow 2}[f_Q] + C_{2\leftrightarrow 3}[f_Q] \ .
\end{align}
In the standard LIDO formulation, the interaction kernel is separated by a momentum-transfer scale $Q_{\rm cut}$. Hard scatterings with momentum transfers above $Q_{\rm cut}$ are treated explicitly through $2\leftrightarrow 2$ and $2\leftrightarrow 3$ collisions. The microscopic input is the corresponding collision matrix elements obtained in perturbative QCD with $Q_{\rm cut}$ as the infrared cutoff of the momentum transfer between the HQ and the medium. Softer interactions below $Q_{\rm cut}$ are described within a Fokker-Planck approach, with the diffusion operator given by
\begin{align}
\mathcal{D}=  \frac{\partial}{\partial p_i}\Big(A(p) p_i + \frac{\partial}{\partial p^i}B(p)\Big)   \ , 
\end{align}
where $A(p)$ is the friction and $B(p)$ the momentum diffusion coefficient. It is 
further supplemented with a diffusion-induced radiation term $\mathcal{C}_{1\leftrightarrow2}$.  The diffusion process in the Fokker-Planck equation is routinely realized via a Langevin equation, which describes the time evolution of the trajectory and momentum of the HQ. We here discretize the Langevin equations as
\begin{align}
\vec{x}(t+\Delta t) &= \vec{x}(t) + \frac{\vec{p}}{E} \Delta t, \\
\vec{p}(t+\Delta t) &=\vec{p}(t) -\eta_D \vec{p} \Delta t + \vec{\xi}(t) \Delta t \ . 
\end{align}
The microscopic HQ-medium interactions are encoded in the drag coefficient $\eta_D$ and the momentum diffusion coefficient, $\kappa$, that sets the correlation function of the random force
\begin{align}
\int_0^{\Delta t} dt \int_0^{\Delta t} dt' \langle \xi_i(t)\xi_j(t')\rangle = \Delta t \left[\kappa_L \frac{p_ip_j}{p^2} + \kappa_T \left( \delta_{ij} - \frac{p_ip_j}{p^2} \right) \right] \ .
\end{align}
In this study, we enforce isotropic diffusion by equating the momentum diffusion coefficient in the longitudinal direction to that in the transverse direction, $\kappa_T =\kappa_L\equiv \kappa$. To ensure that the diffusion process in the Fokker-Planck equation reaches the correct thermal-equilibrium limit further requires an Einstein relation between diffusion and friction coefficients. We follow the standard procedure and set
\begin{align}
   \kappa=2TE \eta_D \ ,
\end{align}
which is supported by case studies using the Boltzmann equation~\cite{Rapp:2018qla}. We recall, however, that the physics encoded in the transport coefficients can go beyond the quasiparticle approximation implied by the Boltzmann equation. This, in particular, applies to situations where the medium is strongly coupled so that large collisional rates of its thermal constituents lead to nontrivial spectral functions. The latter may no longer be amenable to a quasiparticle approximation but can be accounted for in off-shell integrations over the spectral functions in the HQ transport coefficients.  

The spatial diffusion coefficient $\Ds$ and HQ jet transport coefficient $\hat{q}_{Q}$, which is due to soft interactions, are related to the momentum-space diffusion coefficient $\kappa$ as follows:
\begin{align}
\Ds=\frac{T}{M\eta_D(p=0)} = \frac{2T^2}{\kappa},\quad
\hat{q}_{Q}=2\kappa,
\end{align}
where $M$ is the HQ mass.

\subsection{Elastic Interactions: Quantum Many-Body Theory}
\label{ssec:QMBT}
The transport coefficients encode the information about the HQ interactions with the medium. In a strongly-coupled QGP, nonperturbative interactions are expected to be essential; in addition, it has been shown that the perturbative expansion of the HQ diffusion does not exhibit convergence at NLO, even for small coupling $g_s$~\cite{Caron-Huot:2007rwy}.  
Toward this end, the thermodynamic $T$-matrix approach has been developed~\cite{Mannarelli:2005pz,Riek:2010fk,Liu:2017qah} to provide a framework for HQ interactions in a strongly coupled QGP (sQGP).  
This framework incorporates several nonperturbative features that are instrumental in the present context: it resums the ladder series of a suitably defined interaction potential, solves the one- and two-body correlation functions self-consistently, accounts for their full off-shell properties by integrating over the relevant energy variables, treats scattering and bound states on equal footing, and can be implemented in a thermodynamically consistent framework of quantum many-body theory that describes both HQ physics and the QGP equation of state (EoS)~\cite{Liu:2016ysz}. Importantly, it is amenable to systematic improvements in both the $1/M$ expansion~\cite{Tang:2023lcn} and via increasingly advanced constraints from lattice QCD~\cite{Tang:2023tkm,Tang:2024dkz}.

Starting point is the four-dimensional (4D) Bethe–Salpeter equation~\cite{Salpeter:1951sz} which, taking advantage of the $1/M$ suppression in energy transfer, can be reduced to a 3D Lippmann-Schwinger equation~\cite{Blankenbecler:1965gx,Thompson:1970wt} and further partial-wave decomposed to arrive at the following 1D form~\cite{Riek:2010fk}:
\begin{align}
T_{ij}^{l,a}(E,p,p') &= V_{ij}^{l,a}(p,p') + \frac{2}{\pi} \int_{0}^{\infty} k^{2} \, \rmd k \, V_{ij}^{l,a}(p,k) 
G_{ij}^{0}(z,k) \, T_{ij}^{l,a}(E,k,p') \ , 
\label{eq:Tmat}
\end{align}
where $p$  and $p'$  denote the magnitudes of the incoming and outgoing parton momenta in the center-of-mass (CM) frame. The $V_{ij}^{l,a}$ are the interaction potentials between partons $i$  and  $j$  in color channels  $a$ with angular momentum $l$. The two-parton propagator, 
\begin{align}
G_{ij}^{0}(z,k) &= \int_{-\infty}^{\infty} \rmd\omega_{1} \, \rmd\omega_{2} \, \frac{[1 \pm n_{i}(\omega_{1}) \pm n_{j}(\omega_{2})]}{z - \omega_{1} - \omega_{2}} 
\rho_{i}(\omega_{1},k) \, \rho_{j}(\omega_{2},k) \ ,
\label{eq:Gij}
\end{align}
can be expressed as a convolution of two single-parton spectral functions,
\begin{align}
\rho_{i}(\omega,k) &= -\frac{1}{\pi} \, \text{Im} \, G_{i}(\omega, k) \  , 
\label{eq:rhoi}
\\
G_{i}(\omega,k) &= \frac{1}{\omega - \varepsilon_i(k) - \Sigma_i(k)} \ .
\label{eq:Gi}
\end{align} 
The $n_i$ denote a Bose or Fermi distribution function and with a pertinent $\pm$ sign for bosons (upper) and fermions (lower); $\varepsilon_i = \sqrt{M_i^2 + k^2}$  are the on-shell single-particle energies. The in-medium single-particle self-energies, $\Sigma_i(k)$,  in 
Eq.~\eqref{eq:Gi} are calculated by closing the $T$-matrix with an in-medium single-parton propagator~\cite{Liu:2017qah}.

The main input to this formalism is the interaction potential, $V$. The starting point is the Cornell potential, which is well-known and well-established by lattice QCD in vacuum~\cite{Bali:2000gf}, and leads to an effective field theory of QCD for heavy quarkonia known as potential non-relativistic QCD (pNRQCD). At finite temperature, however, the potential cannot be directly extracted from lattice QCD, but rather has to be inferred through suitable quantities that can be calculated in the $T$-matrix formalism. Toward this end, an ansatz for an in-medium screened potential is made~\cite{Riek:2010fk}, 
\begin{equation}
\widetilde{V}(r,T) = -\frac{4}{3} \alpha_{s} \left[\frac{e^{-m_{d} r}}{r} + m_{d}\right] -\frac{\sigma}{m_s} \left[e^{-m_{s} r-\left(c_{b} m_{s} r\right)^{2}}-1\right] \ ,
\label{Vstatic}
\end{equation}
where the strong coupling constant, $\alpha_s$, and string tension, $\sigma$, characterize the color-Coulomb and confining interactions, respectively, and are constrained by lattice QCD results for the vacuum HQ free energy. The $m_{d,s}$ are temperature-dependent screening masses, and $c_b$ is an effective string-breaking parameter. For a given choice
of the latter 3 parameters, specifying the in-medium potential ansatz, the $T$-matrices and self-energies are solved self-consistently in momentum space in quarkonium and heavy-light channels as well as the light-parton sector. The latter can be justified by the relatively large thermal parton masses that are required to fit the EoS from the lattice. In this way, the HQ interactions (and thus their transport in heavy-ion collisions) are embedded into the same EoS as is used in the hydrodynamic bulk medium evolution.

In the present paper, we focus on two versions of recent evaluations of $T$-matrices with constraints based on: 
\begin{enumerate}
    \item[(a)] HQ free energies, $F_{\QQb}(r; T)$, in connection with 1/$M$ corrections derived from spin-induced forces that improve the hyper-/fine splittings in the spectroscopy of charmonia and bottomonia in vacuum~\cite{Tang:2023lcn}; these studies, in particular, called for a vector component in the string force (which is usually assumed to have a scalar Lorentz structure). The pertinent relativistic corrections lead to a significant increase in HQ interaction strength compared to earlier calculations~\cite{Liu:2017qah} especially at high HQ momenta, while the loop integrals over the thermal-parton distributions, involving finite 3-momentum, also have an effect at low momenta. In the following, we will refer to this version as the ``vector confinement potential" ({\bf VCP}).
    \item[(b)] Wilson line correlators (WLCs), $W(r,\tau; T)$, which may be thought of as an extension of the HQ free energy into the Euclidean time direction (for $\tau=1/2T$ one recovers $F_{\QQb}$). It turns out that the WLCs exhibit, for the first time, an explicit sensitivity to the imaginary parts of the $\QQb$ self-energy (or potential)~\cite{Tang:2023tkm}; the main features of this scenario (which we refer to as {\bf WLC}) are a reduced screening at high temperature in connection with a preference for a smaller vector admixture in the string force (reduced by 50\% while still providing a fair description of vacuum quarkonium spectra) compared to the VCP scenario.
\end{enumerate} 

In both scenarios, the temperature-dependent HQ mass is determined from the asymptotic value of the corresponding color-singlet in-medium potential, \ie, 
\begin{equation}
M_Q(T)=M_Q^0+\frac{1}{2}\widetilde V(r\to\infty,T) \ ,
\end{equation}
where $M_Q^0=1.359$\,GeV is the bare mass fixed from the vacuum quarkonium spectrum~\cite{Tang:2023lcn}.

A key ingredient to the HQ friction coefficient, $A(p)$, is the heavy-light $T$-matrices. For an incoming HQ of momentum $\mathbf{p}$ and on-shell energy $\varepsilon_Q(p)=\sqrt{M_Q^2+\mathbf{p}^2}$ one has  
\begin{widetext}
\begin{equation}
\begin{aligned}
A(p)=& \sum_{i} \frac{1}{2 \varepsilon_{Q}(p)} \int 
d^4\tilde{q}_i \ d^4\tilde{p}'_Q \  d^4\tilde{q}'_i \ (2\pi)^4\delta^{(4)}(p_Q+q_i-p'_Q-q_i') 
\frac{1}{d_{Q}} \sum_{a,s}\Big|\sum_{l} \mathcal{M}^{a,l,s}_{Qi}\Big|^{2} \\
&\times \rho_{Q}\left(\omega^{\prime}, p^{\prime}\right) \rho_{i}(\nu, q) \rho_{i}\left(\nu^{\prime}, q^{\prime}\right) \hat{n}_{i}(\nu) \hat{n}_{Q}\left(\omega^{\prime}\right) \hat{n}_{i}\left(\nu^{\prime}\right)
  \left(1-\frac{\mathbf{p}\cdot\mathbf{p'}}{\mathbf{p}^2}\right) \ .
\end{aligned} 
\label{eq:Ap}
\end{equation}
\end{widetext}
This expression, derived from the Kadanoff-Baym equations~\cite{Liu:2018syc}, includes the effects of the parton nontrivial spectral functions, $\rho_{i,Q}$, for the thermal partons ($i$) and the outgoing HQ. The quantity $\mathcal{M}_{Qi}^{a,l,s}$ denotes the
heavy--light scattering matrix element which is directly related to the
corresponding in-medium $T$-matrix via kinematic factors. In Eq.~(\ref{eq:Ap}) we define the pertinent 4D phase-space differentials as
$d^4\tilde{p}'_Q= (d \omega^{\prime} d^{3} \mathbf{p}^{\prime})/(2 \pi)^{3} 2 \varepsilon_{Q}\left(p^{\prime}\right)$, and likewise for incoming and outgoing thermal partons with 4-momenta $(\nu,\mathbf{q})$  and $(\nu',\mathbf{q'})$, respectively. The off-shell integrations are essential to draw interaction strength from (broad) bound states that form in the attractive color channels (singlet and anti-triplet) below the nominal 2-particle threshold as the temperature approaches the pseudo-critical one from above. 
The first sum in Eq.~\eqref{eq:Ap} is over the thermal partons including three light-quark flavors and gluons, the second one runs over all available color ($a$, with appropriate color-Casimir coefficients for the pertinent potentials) and spin quantum numbers, and the third one over the relative angular momentum ($l\leq8$) in the heavy-light channels. Finally, $\hat{n}=1\pm n$ denotes the Bose-enhancement (upper sign) and Pauli-blocking (lower sign) factors applicable for final-state gluons and quarks, respectively, and $d_Q$ denotes the spin-color degeneracy of the incoming HQ.

\begin{figure}
    \centering
    \includegraphics[width=0.9\linewidth]{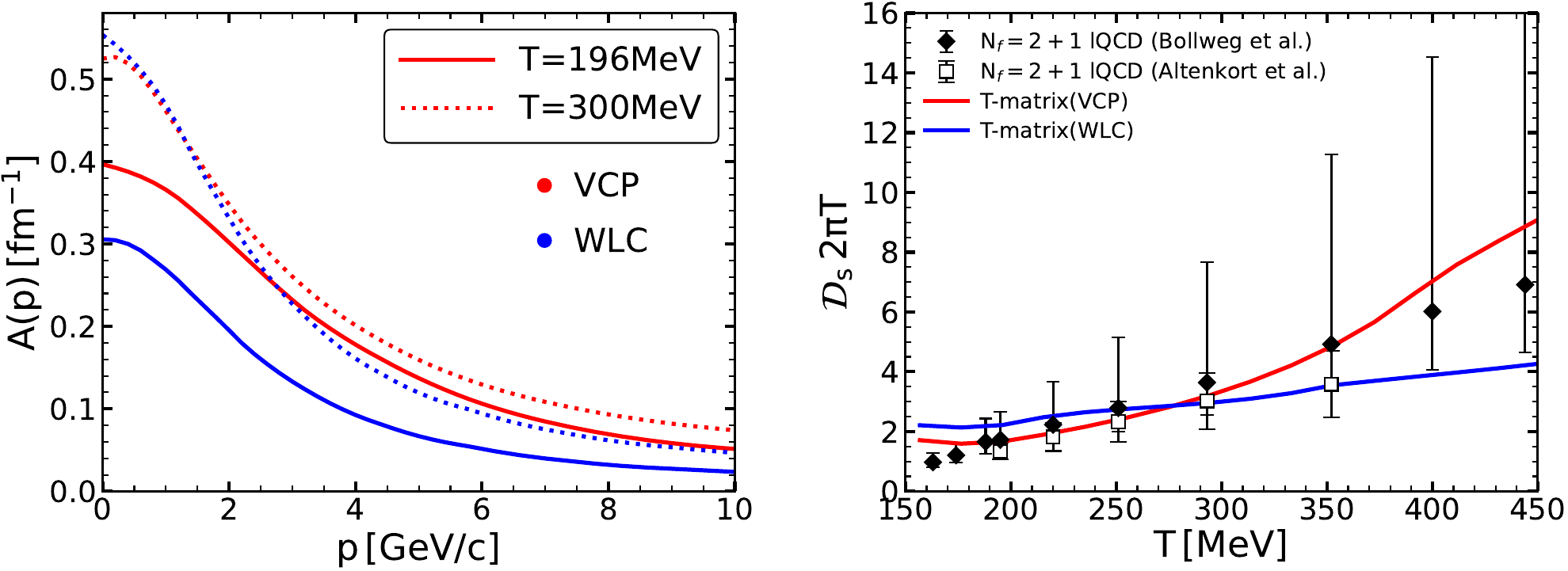}
    \caption{Left: Friction coefficient $A(p)$ of charm quarks obtained from the self-consistent $T$-matrix calculations with the VCP and WLC potentials. Red and blue curves denote the VCP and WLC scenarios, respectively, while solid and dotted curves correspond to $T=196$ and $300\,\mathrm{MeV}$. 
Right: The dimensionless spatial diffusion coefficient $2\pi T\mathcal{D}_s$, obtained from the zero-momentum limit of the friction coefficient. The red and blue curves show the VCP and WLC results, respectively, while the filled diamonds and open squares denote recent $(2+1)$-flavor lattice-QCD results from Bollweg et al.~\cite{HotQCD:2025fbd} and Altenkort et al.~\cite{Altenkort:2023oms,Altenkort:2023eav}, respectively. 
}
\label{fig:Ap_and_Ds2piT}
\end{figure}
The resulting charm-quark friction coefficients in the two potential scenarios are displayed in the left panel of Fig.~\ref{fig:Ap_and_Ds2piT} at two temperatures as functions of charm three-momentum. At the lower temperature, where both potentials remain relatively close to their vacuum forms, the larger vector component of the confining interaction in the VCP scenario produces a stronger interaction and, consequently, a larger thermalization rate. At the higher temperature, this difference is largely compensated by the weaker temperature dependence of the screening in the WLC scenario, resulting in rather similar friction coefficients for the two potentials.
The right panel of Fig.~\ref{fig:Ap_and_Ds2piT} shows the corresponding charm-quark spatial diffusion coefficient, obtained from the zero-momentum limit of the friction coefficient according to $\mathcal{D}_s=T/M(T) A(p=0;T)$.
The results are presented in terms of the dimensionless combination $2\pi T\mathcal{D}_s$ and are compared with recent lattice-QCD calculations~\cite{Altenkort:2023oms,Altenkort:2023eav,HotQCD:2025fbd}.
Within current uncertainties, the agreement with the lattice ``data" is fair, but the VCP result lies closer to the central values at low temperatures, whereas the WLC result exhibits a weaker temperature dependence that is better aligned with the trend suggested by the lattice data.

\subsection{Radiative Processes in the Bethe-Heitler and the deep LPM regimes}
\label{ssec:rad}
The elastic interactions between HQs and the medium as described within the thermodynamic $T$-matrix formalism are not affected by any restrictions on the momentum transfer in the scattering processes of Eq.~\eqref{eq:Tmat}. Consistent with this treatment, we retain only the Fokker-Planck component governed by $\mathcal{D}[f_Q]$ and the corresponding diffusion-induced radiation term $\mathcal{C}_{1\leftrightarrow 2}[f_Q]$ in LIDO, while disabling the matrix-element treatment of large-angle HQ--medium scatterings and the associated radiation processes. For the light-quark and gluon sectors, we continue to use the default LIDO framework, which incorporates both soft and hard interactions with the QGP.

For the HQ sector, the transport equation reduces to the schematic form
\begin{align}
\frac{\rmd f_Q}{\rmd t} = \mathcal{D}[f_Q] + C_{1\leftrightarrow 2}[f_Q] \ .
\end{align}
We treat the diffusion-induced radiation component using the collision-rate method. In the incoherent limit (Bethe-Heitler regime), the medium-induced $1\to 2$ splitting rate is obtained by taking the $q_\perp \ll k_\perp$ limit of the $2\to 3$ scattering formula in Ref.~\cite{Ke:2018jem}:
\begin{align}
\frac{dR_{1\to2}}{\dd x \dd k_{\perp}^2} = \hat q_S \times P_{q\to qg}(x) \times 
    \frac{\alpha_s(k_{\perp}^2)}{2\pi}\frac{1}
    {\left(k_{\perp}^2+m_\infty^2\right)^2} \times P_{\rm DC} \ ,
    \label{eq:R1to2}
\end{align}
where $P_{q\to qg}(x)=C_F[1+(1-x)^2]/x$ is the vacuum DGLAP splitting function, with $x$ denoting the light-cone momentum fraction carried by the emitted gluon. The asymptotic thermal-gluon mass squared, $m_\infty^2=m_D^2/2$, regulates the collinear divergence of the integral.
Here, $\hat q_S$ denotes the perturbative momentum-broadening coefficient entering the diffusion-induced radiation rate, as specified in Appendix~\ref{app:rad}. It is distinct from the HQ momentum-diffusion coefficient obtained from the nonperturbative $T$-matrix calculation, which governs the elastic transport of HQs.
The factor $P_{\rm DC}$ accounts for the dead-cone~\cite{Dokshitzer:2001zm,ALICE:2021aqk,LHCb:2025mcq} suppression of HQ radiation, as discussed below.

The diffusion-induced radiation rate introduced above applies to the Bethe-Heitler regime, where the splitting formation time $\tau_f$ is short compared to the gluon mean-free-path (mfp), $\lambda$. In contrast, in the deep-LPM regime, characterized by $\tau_f\gg\lambda$ for collinear splittings, coherence effects among multiple scatterings suppress the radiation rate. Solutions of the Arnold-Moore-Yaffe (AMY) equation~\cite{Arnold:2002ja,Arnold:2002zm} show that the branching rate in the deep-LPM regime is reduced by a factor of order $\lambda/\tau_f$ relative to the Bethe-Heitler limit. Motivated by this behavior, a numerical prescription for incorporating the LPM effect was developed in Ref.~\cite{Ke:2018jem}. Specifically, each candidate $1\to2$ splitting is initiated at a time $t_0$, but the daughter partons are not immediately treated as independent degrees of freedom. Instead, the splitting system, consisting of the parent ($p_0$) and daughter partons ($p_1$, $p_2$), continues to undergo elastic scatterings with the medium. These interactions modify the transverse momenta of the partons and, consequently, the light-cone formation time evaluated at time $t$:
\begin{align}
    \tau_f(t) &=\frac{E_0}{p_0^+} \times \frac{1}{p_1^-(t)+p_2^-(t)-p_0^-(t)} = \frac{2x_1 x_2 E}{Q_\perp^2(t) + x_1 m_2^2 + x_2 m_1^2 - x_1 x_2 m_0^2}
    \,, \\
    Q_\perp^2(t) &= x_1 p_{2, \perp}^2(t) + x_2 p_{1, \perp}^2(t)-x_1x_2 p_{0, \perp}^2(t) \ ,
\end{align}
where $x_1$ and $x_2$ are the light-cone momentum fractions carried by the daughter partons relative to the parent, with $x_1+x_2=1$. The masses $m_0$, $m_1$, and $m_2$ are taken to be the current mass for HQs and the thermal mass for light quarks and gluons. The light-cone energies evolve in time as the partons interact with the medium.

The evolution of the splitting system continues until a time $t=t_*$ at which the condition $\tau_f<t_*-t_0$ is first satisfied. The splitting is then accepted with a probability given by
\begin{align}
P_{\rm LPM} = \min\left\{1,\quad  \frac{\alpha_s(Q_\perp^2(t_*))}{\alpha_s(Q_\perp^2(t_0))}\times b\times \frac{\lambda(t_0)}{\tau_f(t_*)}\sqrt{\frac{\ln(1+\tau_f(t_*)/\lambda(t_0))}{\ln(1+6ET(t_0)/[m_D(t_0)]^2)}}\right\} \ .
\label{eq:Plpm}
\end{align}
This acceptance probability suppresses emissions with long formation times, accounting for the destructive interference among multiple soft scatterings in the deep-LPM regime. If the splitting is accepted, the daughter partons continue their evolution as independent quasiparticles. Otherwise, the candidate splitting is discarded without affecting the physical state. When the formation time is sufficiently short that the expression for the acceptance probability exceeds unity, the probability is capped at one, recovering the incoherent Bethe-Heitler limit.

This effective prescription provides a smooth interpolation between the Bethe-Heitler and deep-LPM regimes. It was shown that choosing the parameter $b=0.75$ in Eq.~(\ref{eq:Plpm}), the theoretical deep-LPM radiation rate is reproduced at next-to-leading-logarithmic accuracy in an infinite static medium, while also capturing finite-size effects~\cite{Ke:2018jem}. The ratio of running couplings in the suppression factor accounts for the reduction in the coupling strength at the radiative vertex due to the additional transverse-momentum broadening accumulated between $t_0$ and $t_*$. Importantly, the gluon transport sector also influences HQ energy loss, since interactions between the radiated gluon and the medium modify its effective formation time and, consequently, the HQ radiative energy loss predicted by LIDO.

The prescription for incorporating the LPM effect was originally developed for light quarks and gluons. Its extension to HQs requires not only the use of massive kinematics and the inclusion of mass-dependent terms in the formation time, but also a modification of the incoherent radiation rate relative to the massless case. This modification is implemented through a dead-cone suppression factor,
\begin{align}
P_{\rm DC} &= \left[\frac{Q_\perp^2(t)}{Q_\perp^2(t) + x_1 m_2^2 + x_2 m_1^2 - x_1 x_2 m_0^2}\right]^n \ .
\label{eq:Pdc}
\end{align}
In the original implementation of LIDO~\cite{Ke:2018jem}, the default choice is $n=2$, motivated by the squared gluon propagator in the Bethe-Heitler radiation formula in the diffusion-induced limit. However, as demonstrated by the comparison presented in the Appendix~\ref{app:rad}, the choice $n=1$ provides better agreement with the HQ energy loss calculated using soft-collinear effective theory with massive quarks (SCET$_M$) at finite path length when the LPM effect is taken into account. This can be understood from the dominance of interference effects in this regime: the mass dependence associated with one of the propagators in the interference contribution becomes subleading compared to the finite-size LPM scale, $x(1-x)E/L \gg x^2 M^2$, for the typically small values of $x$ that dominate radiative energy loss. We also neglect an additional contribution proportional to $x^3 M^2$ in the gluon radiation spectrum, which is found to have a relatively small effect on the average energy loss.
In principle, a more refined prescription should interpolate between the dead-cone suppression in the Bethe-Heitler and deep-LPM regimes. Nevertheless, as shown in  Appendix~\ref{app:rad}, for charm quarks the choices $n=1$ and $n=2$ lead to negligible phenomenological differences over the momentum range considered in this study. We therefore retain the default choice $n=2$ throughout this work.

Regarding the running coupling constant used in the LIDO model, we assume that the renormalization scale should not fall below the characteristic medium scale proportional to temperature and adopt the following prescription for the running coupling at finite temperature:
\begin{align}
\alpha_s(Q^2) = \frac{4\pi}{\beta_0} \frac{1}{\ln\!\left( \max\{\,Q^2,(\mu \pi T)^2\,\}/\Lambda_{\mathrm{QCD}}^2 \right)} \ ,
\label{eq:alphas}
\end{align}
where $\beta_0 = 11 - \frac{2}{3}n_f$ with $n_f=3$, and $\Lambda_{\mathrm{QCD}} = 200~\mathrm{MeV}$. The scale $Q^2$ is taken to be the squared transverse momentum entering the vertex, measured relative to the direction of propagation of the energetic parton.
The principal adjustable parameter in the perturbative sector of LIDO is the dimensionless medium-scale parameter $\mu$, which sets the minimum renormalization scale of the running coupling and thereby controls its infrared behavior. Increasing $\mu$ raises the minimum scale at which $\alpha_s$ is evaluated, reducing the coupling for momentum transfers $Q\lesssim\mu\pi T$. This weakens perturbative parton--medium interactions and suppresses medium-induced radiation. Conversely, decreasing $\mu$ enhances the infrared coupling and strengthens these effects. In the present setup, the elastic HQ transport coefficients are supplied by the thermodynamic $T$-matrix calculation and are therefore not controlled by $\mu$. Instead, $\mu$ primarily affects diffusion-induced radiation and the subsequent interactions of radiated gluons with the medium, thereby influencing the radiative energy loss of HQs. We adopt a moderate value of $\mu=1.5$ throughout this study.

\subsection{Hadronic HF Transport}
\label{ssec:hadron}
After hadronization, which will be elaborated in Sec.~\ref{sec:hadro}, HF hadrons undergo further rescattering in the hadronic phase. A recent review of existing evaluations of these processes in URHICs at RHIC and the LHC indicates that noticeable effects on the $\raa$ and $v_2$ of $D$ mesons may occur~\cite{Das:2024vac}, at the 10-20\% level at the LHC (and even larger at RHIC). This will, however, depend on the concrete model implementation, most notably the hadronic HF interactions and the switching temperature from parton to hadron degrees of freedom.   

Since in the present framework the bulk evolution of the hadronic phase is described by the UrQMD transport model, we also adopt a Boltzmann transport simulation for the HF hadrons, focusing on $D$ mesons for the present purpose. A realistic description of HF rescattering off light hadrons in principle requires a large set of cross sections, many of which are not well known and are not accessible experimentally, including multiple resonance channels for a given hadron~\cite{He:2011yi,Tolos:2013kva,Ozvenchuk:2014rpa}. Thus, instead of using a microscopic description, we employ a constant effective cross section in the UrQMD evolution, adopting a conservative baseline of 5~mb for $D$-meson rescattering on mesons and 7.5~mb for baryons in the bulk medium, following estimates in Ref.~\cite{He:2011yi}. {These cross sections are applied only to two-body collisions and conservatively limited to center-of-mass energies of $\sqrt{s}<3~\mathrm{GeV}$.}

\subsection{Bulk Evolution Model}
\label{ssec:bulk}
A realistic bulk medium evolution in heavy-ion collisions is a prerequisite for a reliable simulation of HQ transport within it. The bulk medium evolution provides the local temperature and flow fields for the transport coefficients that govern the HQ dynamics.
Here, we couple our HQ transport model to a state-of-the-art 2+1D  viscous hydrodynamical medium evolution, which consists of the following stages:
\begin{itemize}
\item[1.] The TRENTO initial condition model generates event-averaged initial conditions at time $\tau = 0^+$ when the Lorentz-contracted nuclei collide~\cite{Moreland:2014oya}. 
\item[2.] A collisionless Boltzmann equation (free streaming) models the pre-equilibrium stage prior to the onset of the hydrodynamic evolution at $\tau_{\rm fs}=1.16\, \text{fm/c}$~\cite{Liu:2015nwa}, \ie, the thermalization time.
\item[3.] 2+1D  relativistic viscous hydrodynamics evolves the QGP with an up-to-date lattice equation of state (EoS)~\cite{Shen:2014vra,HotQCD:2014kol}.
\item[4.] An ensemble of light hadrons is sampled from the corresponding hypersurface of the hydrodynamic evolution at a fixed temperature of $T$=152\, MeV utilizing the Cooper-Frye formula, \ie, the hydrodynamic evolution of the system extends below the pseudo-critical phase transition temperature $\Tpc$ into the hadronic phase.
\item[5.] Finally, the ensemble of light hadrons is combined with the generated set of HF hadrons, which are allowed to rescatter and decay utilizing the Ultra-Relativistic Quantum Molecular Dynamics (UrQMD) model~\cite{Bass:1998ca,Bleicher:1999xi}. For the rescattering of HF hadrons with light hadrons, we use a fixed elastic cross section as explained in Sec.~\ref{ssec:hadron}. In the future, this can be made more precise by utilizing more realistic (microscopically calculated) cross sections to produce a HQ transport coefficient that can match the one on the QGP side at $\Th$.
\end{itemize}
The parameters of the bulk medium evolution have previously been calibrated to reproduce a vast array of bulk observables at RHIC and  LHC energies \cite{Bernhard:2019bmu}, which provide a description of the bulk evolution of the QGP with high precision and will be adopted here as is.

\section{Heavy-Quark Hadronization}
\label{sec:hadro}
The hadronization of HQs in URHICs, subsequent to their transport through the QGP phase, is believed to have a significant impact on the HF hadron momentum distributions and thus affect (the interplay of) observables such as $\raa$ and elliptic flow~\cite{PHENIX:2006iih}. A key role in this is played by the recombination of HQs with light quarks, which enables the resulting HF hadron to inherit (part of) the $v_2$ encoded in the bulk medium around the hadronization transition, where it is believed to be near its maximal value. 
In the present work, we focus on two recombination models that satisfy 4-momentum conservation and also feature a substantial number of excited states. The former is particularly relevant at low momentum and a prerequisite to approach and/or maintain the correct kinetic-equilibrium limit in the hadronization process~\cite{He:2011qa}, while the latter likely plays an important role in correctly describing the hadro-chemistry of the HF hadrons produced in nuclear collision systems (from $pp$ via $p$A to AA).
Both hadronization models will be supplemented by fragmentation models and fit to experimental data for HF hadrons in $pp$ collisions in Sec.~\ref{ssec:initial}.

The remainder of this section is organized as follows. We first discuss the fragmentation setup, which is common to both models apart from the specific fragmentation function employed. We then review the main features of the improved sudden coalescence model (iSCM)~\cite{Cao:2013ita,Cao:2015hia,Cao:2019iqs} in Sec.~\ref{ssec:coal}, followed by the resonance recombination model (RRM)~\cite{Ravagli:2007xx,He:2019vgs} in Sec.~\ref{ssec:rrm}. 

\subsection{Fragmentation in pp}
\label{ssec:frag}
The iSCM setup employs the Peterson fragmentation function~\cite{Peterson:1982ak}, 
\begin{align}
\mathcal{D}_{Q\to H}(z)
\propto
\frac{1}{z\left(1-\frac{1}{z}-\frac{\epsilon}{1-z}\right)^2} \ ,
\end{align}
where we allow for separate fragmentation parameters, $\epsilon$, for mesons and baryons. Specifically, we constrain the values by fits to the $pp$ baseline spectra at RHIC and the LHC, yielding $\epsilon_M=0.01$ and $\epsilon_B=0.03$ for charm mesons and baryons, respectively.

In the RRM, fragmentation is implemented via a function based on HQ Effective Theory (HQET)~\cite{Braaten:1994bz}, 
\begin{equation}
\begin{aligned}
\mathcal{D}_{Q \rightarrow H}(z) &= N \frac{r z (1-z)^2}{[1 - (1 - r) z]^6} \Big[6 - 18(1 - 2r)z \\
&\quad + (21 - 74r + 68r^2)z^2   \\
&\quad - 2(1 - r)(6 - 19r + 18r^2)z^3 \\
&\quad + 3(1-r)^2(1-2r+2r^2)z^4
\Big] \ ,
\end{aligned}
\end{equation}
where the parameter $r$ scales with the hadron mass according to
$\frac{r_M}{r_{D^0}}=
\frac{(m_M-m_c)/m_M}{(m_{D^0}-m_c)/m_{D^0}}$ for mesons,
with $r_{D^0}$ denoting the parameter for the ground-state $D^0$ meson, tuned to fit $pp$ data. A similar mass scaling is applied to baryons. The best-fit values are found to be $r_{D^0}=0.1$ and $r_{\Lambda_c^+}=0.16$.

To model the flavor dependence of the hadronic final state, both models utilize the statistical hadronization framework for fragmentation as introduced in Ref.~\cite{He:2019tik}. This is based on the premise that $pp$ collisions at LHC energies at midrapidity produce sufficient light-parton phase space. An important ingredient is the inclusion of additional charm-baryon states as predicted by the relativistic quark model (RQM)~\cite{Ebert:2011kk}, relative to the experimentally established states listed by the Particle Data Group (PDG). All charm quarks are fragmented into hadrons according to the corresponding thermal weights, after which excited states decay into ground-state hadrons. The relative abundances of the various charm-hadron species are determined from statistical thermal weights derived from their respective thermal densities,
\begin{equation}
\label{eqn:Thermal density}
n_i = \frac{g_i}{2\pi^2}\gamma_s^{S_i} \gamma_{\mathrm{HB}} m_i^2 T_H K_2\left(\frac{m_i}{T_H}\right).
\end{equation}
where $m_i$ denotes the mass of the hadron, $g_i$ its spin-isospin degeneracy, $S_i$ the number of valence strange quarks, and $K_2$ is the modified Bessel function of the second kind. Following Ref.~\cite{He:2019tik}, the statistical hadronization temperature in $pp$ collisions is set to $170$\,MeV to best reproduce the observed abundances of the various charmed hadrons. Canonical suppression effects for strange hadrons and baryons have been observed as a function of the charged-particle multiplicity across collision systems and energies~\cite{ALICE:2016fzo}. To account for these observations, we introduce a strangeness-suppression factor, $\gamma_s$, together with an HF baryon suppression factor, $\gamma_{\mathrm{HB}}$. For LHC energies, we use $\gamma_s=0.7$, while for RHIC energies we adopt $\gamma_s=0.5$. In Ref.~\cite{Dai:2024vjy}, it has been suggested that the suppression of all baryon yields can be related to baryon-number conservation in a canonical-ensemble calculation. In the present work, we employ a simpler, phenomenological treatment by allowing for a baryon fugacity factor in Eq.~\eqref{eqn:Thermal density} that suppresses HF baryon abundances. From the LHC $pp$ data at mid-rapidity, it has been found that $\gamma_{\mathrm{HB}}=1$ is compatible with the observed $\Lambda_c^+/D^0$ ratio of approximately $0.5$. At RHIC energies, we expect the $\Lambda_c^+/D^0$ ratio to be smaller due to the reduced light-parton phase space density. This is also supported by a lower $\Lambda_c/D^0$ ratio at forward rapidity as measured by LHCb~\cite{LHCb:2018weo} and has been explicitly established via the measured multiplicity dependence for the $\Lambda_B/B^0$ in $pp$ collisions~\cite{LHCb:2023wbo}. Pending data on $\Lambda_c$ production in $pp$ collisions at RHIC we therefore assume $\Lambda_c^+/D^0\simeq0.25$, amounting to $\gamma_{\mathrm{HB}}=0.5$. Once all charm hadrons are produced, excited states decay to their ground states using the corresponding branching ratios. For experimentally observed states, we use PDG branching ratios (BRs) and assign equal weights for observed (but not quantified) decay channels. To estimate the maximal contribution from additional RQM states of excited $\Lambda_c$ and $\Sigma_c$ states not listed by the PDG, we assume their decays into $\Lambda_c^+$ plus one or two pions, motivated by the dominant $\Sigma_c(2455)\to\Lambda_c+\pi$ decay. For excited $\Xi_c$ baryons, PDG channels include $\Lambda_c+K$; for additional RQM states we assume a $50\%$ branching to $\Lambda_c$ and $50\%$ to ground-state $\Xi_c$.

\subsection{Improved Sudden Coalescence Model}
\label{ssec:coal}
For the ``sudden-coalescence" model in AA collisions, we utilize the improved version introduced in Ref.~\cite{Cao:2019iqs}. It includes an extended set of $S$- and $P$-wave charm-hadron states with all allowed spin and spin-orbit combinations, which covers most of the states listed by the 2018 particle data group~\cite{ParticleDataGroup:2018ovx}, while the possible excited states predicted from the RQM are not included. It is evaluated in a 3-dimensional harmonic-oscillator calculation and achieves energy-momentum conservation through the decay of virtual states formed in the coalescence process into on-shell pion and HF hadron states. This prescription has been shown to be consistent with the formation of equilibrium hadron spectra when using equilibrium quark input distributions in a static medium at $T$=160\,MeV. More explicitly, the momentum distributions $\frac{\rmd N_H}{\rmd^3\vec{p}_H }$ of produced mesons and baryons with momentum $\vec{p}_H$ are given by 
\begin{align}
\frac{\rmd N_H}{\rmd^3\vec{p}_H } =\left(\prod_{i}^{} \int \rmd^3\vec{p}_i \frac{\rmd N_i}{\rmd^3\vec{p}_i}\right) W(\{\vec{p}_i\}) \delta^{(3)}(\vec{p}_H - \sum_{i}^{}\vec{p}_i ) \ ,
\label{eq:dNdph}
\end{align}
where $\frac{\rmd N_i}{\rmd^3\vec{p}_i }$ is the momentum distribution of each constituent quark $i=$1-2 for meson and $i$=1-3 for baryon production. The momentum distributions of the HQs are obtained from their transport through the QGP, while the light quarks are thermalized as part of the hydrodynamic evolution and follow the Fermi-Dirac distribution in the local-cell rest frame. The Wigner function, $W(\{\vec{p}_i\})$, averaged over the spatial volume $V$, represents the probability for partons to combine into a hadron and is determined from the bound-state wavefunction. For the case of meson formation, the meson wavefunction is of a simple quantum-mechanical harmonic oscillator form, yielding the Wigner functions for $S$- and $P$-wave states as 
\begin{align}
    W_s({\vec{p}_i} )&=g_H\frac{8\pi^{3/2}\sigma^3}{V} e^{-\sigma^2\vec{k}^2}, \nonumber\\ 
    W_p({\vec{p}_i })&= g_H\frac{8\pi^{3/2}\sigma^3}{V}\frac{2}{3}\sigma^2\vec{k}^2 e^{-\sigma^2\vec{k}^2}
    \label{eq:Wsp}
\end{align}
where $\vec{k} =\frac{m_2\vec{p}_1 - m_1\vec{p}_2 }{m_1+m_2}$ denotes the relative momentum of quarks in their center-of-mass frame, $g_H$ accounts for the spin-color degrees of freedom, and the parameter $\sigma \equiv 1/\sqrt{\mu\omega}$ is related to the oscillator frequency $\omega$ in the meson wave function with the reduced mass, $\mu$, of the quark and anti-quark. The light-quark masses for flavors $u,d,s$ are chosen as $m_u=m_d=0.3\,\text{GeV}$ and $m_s=0.4\,\text{GeV}$, and the charm-quark mass as  $m_c=1.8\,\text{GeV}$ for consistency with its value in the $T$-matrix formalism, which is used for the in-medium interactions in the diffusion process. 

To investigate the sensitivity of charm-hadron chemistry to the relative meson and baryon coalescence probabilities, we allow separate harmonic-oscillator frequencies for the two sectors. We choose the meson frequency to be $\omega_M=0.20\,\mathrm{GeV}$, corresponding to a modest increase in the effective $D$-meson wavefunction radius relative to the original common-frequency setup $\omega_M=\omega_B=0.24$ used in Ref.~\cite{Cao:2019iqs}. The baryon frequency is then determined by
requiring the total coalescence probability of a zero-momentum ($p_{c}^*=0$) charm quark in the fluid rest frame to remain approximately unity when all included $S$- and $P$-wave channels are summed, $P_{\mathrm{coal}}^{\mathrm{total}}(p_{c}^*=0;\omega_M,\omega_B)=1$.
This procedure gives $\omega_B=0.267\,\mathrm{GeV}$. The resulting
parameter set is a phenomenological variation of the hadron
wavefunction widths, not an independent
determination of in-medium hadron radii.

The resonance feeddown prescription is also improved compared to Ref.~\cite{Cao:2019iqs}. In particular, the decays of excited $D_s$ states are treated according to the experimentally established decay channels compiled by the Particle Data Group. While the original implementation effectively maps all recombination-produced excited $D_s$ states to the ground-state $D_s$, states such as $D_{s1}(2536)$ and $D_{s2}^{*}(2573)$ decay predominantly through $D^{*}K$ and/or $DK$ channels, ultimately producing non-strange $D$ mesons. We therefore allow these states to feed down to $D^0$ and $D^+$, rather than forcing them into the $D_s$ channel. This redistributes charm among the final hadron species without changing the underlying recombination probabilities. In particular, the original prescription can underestimate the recombination contribution to the $D^0$ yield, thereby affecting the intermediate-$\pT$ $D^0$ nuclear modification factor and the $\Lambda_c^+/D^0$ and $D_s^+/D^0$ ratios.

\subsection{Resonance Recombination Model}
\label{ssec:rrm}
The resonance recombination model~\cite{Ravagli:2007xx} is derived from the Boltzmann equation by evaluating recombination and dissociation rates in a linear response approximation, yielding the meson phase-space distribution (PSD). For HF meson formation in AA collisions, the recombination rate is computed using a resonant quark–antiquark cross section, $q + \bar{q} \to M$, via a relativistic Breit-Wigner form with width $\Gamma_M$, which can then be used to compute the meson PSD given by 
\begin{align}
\label{RRM-1}
f_M(x,\vec{p}_M) &= \frac{\gamma_M}{\Gamma_M} \int \frac{d^3\vec{p}_1}{(2\pi)^3} \frac{d^3\vec{p}_2}{(2\pi)^3} f_q(x,p_1) f_{\bar{q}}(x,p_2) \\ \nonumber
&\quad \times  \sigma(s) v_{\text{rel}} (2\pi)^3 \delta^3(\vec{p}_M - \vec{p}_1 - \vec{p}_2) \ ,
\end{align}
where $f_{q,\bar{q}}$ are the PSDs of quark and antiquark, $v_{\text{rel}}$ is their relative velocity, $\gamma_M = E_M / m_M$, and $\sigma(s)$ is the resonant cross section. For baryons, recombination is evaluated through a two-step process. First, quark-1 and quark-2 recombine into a diquark, $q_1 + q_2 \to dq$, similar to meson formation, and the diquark recombines with quark-3  yielding a baryon PSD,  
\begin{equation}
\begin{aligned}
f_B(\vec{x}, \vec{p})=&\frac{\gamma_B}{\Gamma_B} \int \frac{d^3 \vec{p}_1 d^3 \vec{p}_2 d^3 \vec{p}_3}{(2 \pi)^6} \frac{\gamma_{d q}}{\Gamma_{d q}} f_1\left(\vec{x}, \vec{p}_1\right) f_2\left(\vec{x}, \vec{p}_2\right) \\
& \times f_3\left(\vec{x}, \vec{p}_3\right) \sigma_{d q}\left(s_{12}\right) v_{\mathrm{rel}}^{12} \sigma_B(s) v_{\mathrm{rel}}^{d q 3} \delta^3\left(\vec{p}-\vec{p}_1-\vec{p}_2-\vec{p}_3\right)
\end{aligned}
\end{equation}
To effectively partition between recombination and fragmentation, we self-consistently determine a recombination probability from Eq.~(\ref{RRM-1}) for a charm quark with fluid rest frame momentum $p^*_c$ by
\begin{equation} 
\begin{aligned}\label{RRM-prob} P_{\text{M}}(p^*_c) &= P_0 \int \frac{d^3 \vec{p}_1^*}{(2\pi)^3} f_1(p_1) \frac{\gamma_M}{\Gamma_M} \sigma(s) v_{\text{rel}} \text{; }\text{ } \\P_B\left(p_c^*\right) &=P_0 \int \frac{d^3 \vec{p}_1^* d^3 \vec{p}_2^*}{(2 \pi)^6}f_{1}(p_1)f_{2}(p_2) \frac{\gamma_B}{\Gamma_B} \frac{\gamma_{d q}}{\Gamma_{d q}} \\
&\times\sigma\left(s_{12}\right) v_{\mathrm{rel}}^{12} \sigma\left(s_{d 3}\right) v_{\mathrm{rel}}^{d q 3}  \ . 
\end{aligned} 
\end{equation}
An overall parameter $P_0$ is introduced to adjust the total recombination probability to 1 for a charm quark at rest in the fluid rest frame~\cite{He:2019vgs} (as there is no energy available for fragmentation). The current version of RRM includes all charmed hadron states listed by the Particle Data Group plus charm-baryon states predicted by the relativistic quark model (RQM)~\cite{Ebert:2011kk}. The masses of the light, strange, and charm quarks are taken as $m_u = 0.3$\,GeV, $m_s = 0.4$\,GeV, and $m_c = 1.5$\,GeV, respectively. The meson width is set to $\Gamma_M = 0.1$ GeV. For baryons, the diquark masses are taken from RQM studies~\cite{Ebert:2011kk}, with  $m_{[u d]} = 0.71$ GeV for the scalar diquark within $\Lambda_c$’s; $m_{\{u d\}} = 0.909$\,GeV for the axialvector within $\Sigma_c$’s; $m_{[u s]} = 0.948$\,GeV and $m_{\{u s\}} = 1.069$\,GeV within $\Xi_c$’s; and $m_{\{s s\}} = 1.203$\,GeV within $\Omega_c$’s. The diquark width is taken as $\Gamma_{dq} = 0.1$\,GeV, while baryons with $m_B < 2.6$\,GeV are assigned $\Gamma_B = 0.1$\,GeV, and for heavier resonances $\Gamma_B = 0.3$\,GeV. The same decay and feeddown prescriptions as in the $pp$ system are employed here. In particular, the default RRM calculation assumes that excited $\Lambda_c$ states feed down to the ground-state $\Lambda_c$. However, for excited states with masses above the $DN$ threshold, $m_{\Lambda_c^*}>m_D+m_N$, strong decays into a ground-state $D$ meson and a nucleon become kinematically allowed. Since the branching fractions of these excited states above the $DN$ threshold are poorly constrained, we discuss the uncertainty associated
with possible $DN$ decay channels in Appendix~\ref{app:feeddown-uncertainty}.
In AA collisions, at the end of the diffusion through the QGP, each charm quark’s fluid rest frame momentum is used to compute the recombination probability $P_{\text{rec}}(p^*_c)$. This directly gives a ``fragmentation probability" as $1-P_{\text{rec}}(p^*_c)$, which effectively partitions the charm quarks going into recombination and fragmentation.

Finally, we compare the iSCM and RRM coalescence probabilities for a charm quark to form different charm-hadron species, including non-strange $D$-mesons ($D$), strange $D$-mesons ($D_s$) and $\Lambda_c$. The probabilities are evaluated at $\Th=160$\,MeV in the local fluid rest frame as a function of the charm-quark momentum, $p_c^*$, with feed-down contributions included. Overall, the RRM yields larger recombination probabilities than the iSCM, indicating that a larger fraction of charm quarks hadronizes through recombination with thermal partons. Consequently, compared with the iSCM, the RRM is expected to generate a larger recombination contribution to charm-hadron production in the intermediate-$\pT$ region and to transfer more of the collective flow of the medium to the final-state charm hadrons.
\begin{figure}[H]
    \centering
    \includegraphics[width=0.85\linewidth]{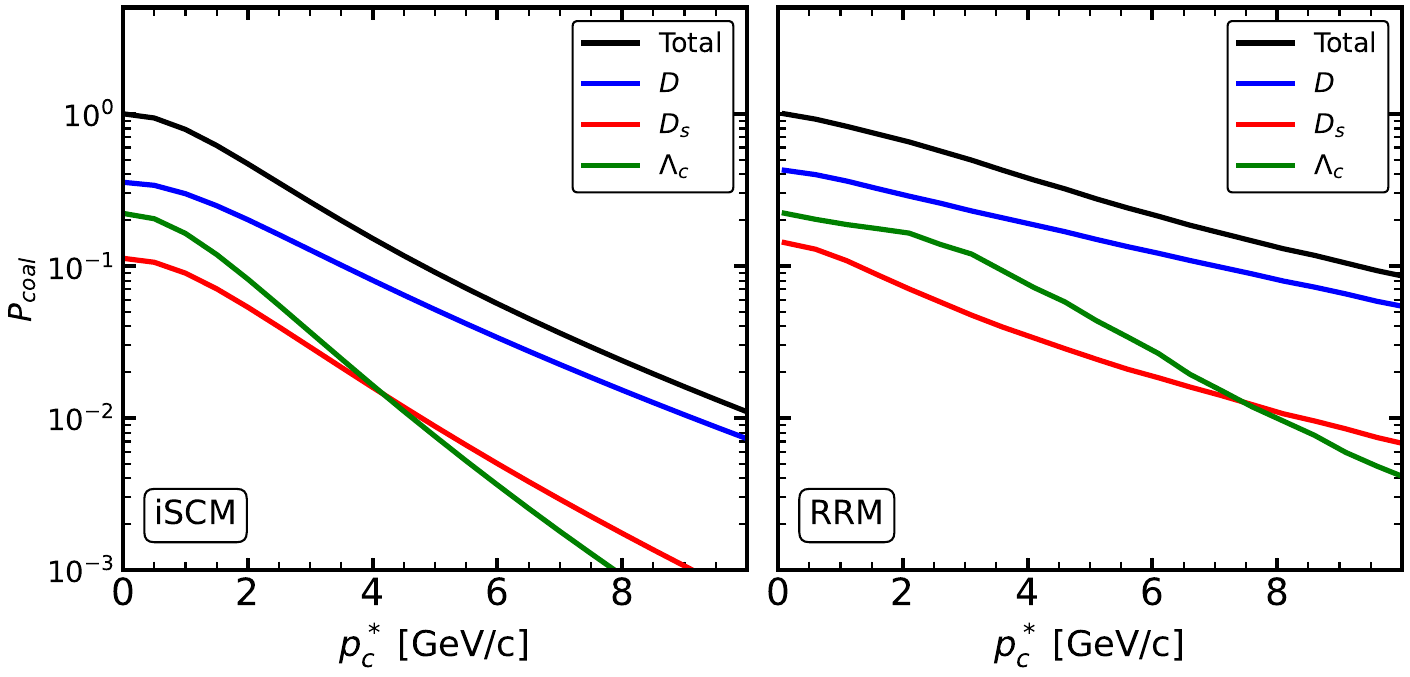}
    \caption{Comparison of the coalescence probabilities for a charm quark to form different charm-hadron species in the local fluid rest frame. The probabilities are calculated in the iSCM (left) and RRM (right) at $\Th=160$\,MeV as functions of the charm-quark momentum $p_c^*$, with feed-down contributions included.}
    \label{fig:Pcoal}
\end{figure}

\section{Comparisons to Observables}
\label{sec:obs}
In this section, we present and discuss our results for key open HF observables in both $pp$ and AA collisions at $\sqrt{s_{\NN}}=5.02~\mathrm{TeV}$ and $\sqrt{s_{\NN}}=200~\mathrm{GeV}$. We first establish the initial charm-quark momentum distributions and the fragmentation baselines by comparing the calculated $D^0$, $D^+$, $D_s^+$, and $\Lambda_c^+$ spectra in $pp$ collisions with experimental data in Sec.~\ref{ssec:initial}. The resulting initial conditions are subsequently propagated through our transport framework and coupled to the two hadronization prescriptions introduced in Sec.~\ref{sec:hadro}. In Secs.~\ref{ssec:raa} and \ref{ssec:v2}, we discuss the nuclear modification factor, $\raa$, and elliptic flow, $v_2$, respectively, for $D^0$, $D_s^+$, and $\Lambda_c^+$. Selected HF hadro-chemistry observables, in particular $D_s^+/D^0$ and $\Lambda_c^+/D^0$, are presented in Sec.~\ref{ssec:chem}.  We note that the model results in Secs.~\ref{ssec:raa}, \ref{ssec:v2} and \ref{ssec:chem} do not yet include HF hadronic transport processes. The calculations are compared with measurements from the ALICE and CMS Collaborations at the LHC and the STAR Collaboration at RHIC. 
In Sec.~\ref{ssec:model-effects}, we investigate the impact of several model components on the HF observables, specifically the transition from charm quarks to hadrons (Sec.~\ref{sssec:c-h}), the nuclear PDFs (\ref{sssec:npdf}), the interplay of diffusion and radiation (\ref{sssec:diff-rad}), and hadronic rescattering (\ref{sssec:urqmd}).
The experimental observables relevant for the present study are summarized in Tables~\ref{tab:observables-pp} and \ref{tab:observables-AA}.

\begin{table*}[h]
\centering
\scriptsize
\setlength{\tabcolsep}{4pt}
\renewcommand{\arraystretch}{1.25}

\begin{ruledtabular}
\begin{tabular}{l l c c c c c}

Experiment
&
System
&
\makecell{$D^0$ \\$p_{\rm T}$-differential\\cross section}
&
\makecell{$D_s^+$\\$p_{\rm T}$-differential\\cross section}
&
\makecell{$\Lambda_c^+$\\$p_{\rm T}$-differential\\cross section}
&
$D_s^+/D^0$
&
$\Lambda_c^+/D^0$
\\

\hline

ALICE
&
\makecell[l]{$pp$, $\sqrt{s}=5.02\,{\rm TeV}$}
&
\cite{ALICE:2019nxm}
&
\cite{ALICE:2019nxm}
&
\cite{ALICE:2020wla}
&
\cite{ALICE:2019nxm}
&
\cite{ALICE:2020wfu,ALICE:2020wla}
\\

CMS
&
\makecell[l]{$pp$, $\sqrt{s}=5.02\,{\rm TeV}$}
&
\cite{CMS:2017qjw}
&
\cite{CMS:2019cxm}
&
\cite{CMS:2023frs}
&
\cite{CMS:2019cxm}
&
\cite{CMS:2023frs}
\\

STAR
&
\makecell[l]{$pp$, $\sqrt{s}=200\,{\rm GeV}$}
&
\cite{STAR:2012dpa}
&
--
&
--
&
--
&
--
\\

\end{tabular}
\end{ruledtabular}

\caption{
Open-charm observables in $pp$ collisions used in this work. Each entry lists the corresponding experimental reference. The CMS $D_s^+$ measurement is based on a public preliminary note. For STAR, a published $pp$ baseline exists for $D^0$ and $D^{*}$ production at $\sqrt{s}=200\,{\rm GeV}$, while no corresponding published STAR measurements are available for $D_s^+$, $\Lambda_c^+$, $D_s^+/D^0$, or $\Lambda_c^+/D^0$.
}
\label{tab:observables-pp}
\end{table*}

\begin{table*}[h]
\centering
\scriptsize
\setlength{\tabcolsep}{2.5pt}
\renewcommand{\arraystretch}{1.25}

\begin{tabular*}{\textwidth}{@{\extracolsep{\fill}}llcccccccc}
\toprule
Experiment
& System
& \makecell{$D^0$\\$R_{\rm AA}$}
& \makecell{$D^0$\\$v_2$}
& \makecell{$D_s^+$\\$R_{\rm AA}$}
& \makecell{$D_s^+$\\$v_2$}
& $D_s^+/D^0$
& \makecell{$\Lambda_c^+$\\$R_{\rm AA}$}
& \makecell{$\Lambda_c^+$\\$v_2$}
& $\Lambda_c^+/D^0$
\\
\midrule

ALICE
& PbPb, $\sqrt{s_{\rm NN}}=5.02\,{\rm TeV}$
& \cite{ALICE:2021rxa}
& \cite{ALICE:2020iug}
& \cite{ALICE:2021kfc}
& \cite{ALICE:2021kfc}
& \cite{ALICE:2021kfc}
& \cite{ALICE:2021bib}
& --
& \cite{ALICE:2021bib}
\\

ALICE
& PbPb, $\sqrt{s_{\rm NN}}=5.36\,{\rm TeV}$
& --
& \cite{ALICE:2026zcz}
& --
& \cite{ALICE:2026zcz}
& --
& --
& \cite{ALICE:2026zcz}
& --
\\

CMS
& PbPb, $\sqrt{s_{\rm NN}}=5.02\,{\rm TeV}$
& \cite{CMS:2017qjw}
& \cite{CMS:2020bnz,CMS:2017dju}
& \cite{CMS:2019cxm}
& \cite{CMS:2026how}
& \cite{CMS:2019cxm}
& --
& --
& \cite{CMS:2019uws,CMS:2023frs}
\\

CMS
& PbPb, $\sqrt{s_{\rm NN}}=5.36\,{\rm TeV}$
& --
& --
& --
& --
& --
& --
& \cite{CMS:HIN-25-022}
& --
\\

STAR
& AuAu, $\sqrt{s_{\rm NN}}=200\,{\rm GeV}$
& \cite{STAR:2018zdy}
& \cite{STAR:2017kkh}
& --
& --
& \cite{STAR:2021tte}
& --
& --
& \cite{STAR:2019ank}
\\

\bottomrule
\end{tabular*}

\caption{Open-heavy-flavor observables in heavy-ion collisions at RHIC and the LHC. Each cell lists the corresponding experimental reference. For CMS $D_s^+$ production, $R_{\rm AA}$, $D_s^+/D^0$ at $\sqrt{s_{\rm NN}}=5.02\,{\rm TeV}$  and $\Lambda_c^+$ $v_2$ at $\sqrt{s_{\rm NN}}=5.36\,{\rm TeV}$, Ref.~\cite{CMS:2019cxm} and Ref.~\cite{CMS:HIN-25-022} are public  preliminary notes.
}
\label{tab:observables-AA}
\end{table*}

\subsection{HF Spectra in $pp$ Collisions and Initial Conditions}
\label{ssec:initial}

The time evolution of the HQ phase-space distribution requires initial conditions in both coordinate and momentum space. We assume that charm quarks are produced in initial hard nucleon--nucleon collisions and distribute their production points according to the binary-collision density. The initial momentum distribution of the HQs is calculated using the Fixed Order Next-to-Leading Logarithm (FONLL) approach~\cite{Cacciari:1998it}. In $pp$ collisions, where no rescattering effects are accounted for, the subsequent conversion of charm quarks into the experimentally measured hadron species requires a fragmentation function and the relative abundances of the charm hadrons. These ingredients are constrained by the available $pp$ data, as follows.

Figure~\ref{fig:pp5020_ALICE} compares the calculated prompt charm-hadron spectra at $\sqrt{s}=5.02\,\mathrm{TeV}$ with the ALICE measurements. The left panel employs the statistical hadronization fractions calculated at $T=170$\,MeV together with the Peterson fragmentation function used in the iSCM setup, while the right panel employs the HQET-based fragmentation function used in the RRM setup.  The fragmentation parameters are chosen such that each prescription provides a reasonable description of the measured $D^0$, $D^+$, $D_s^+$, and $\Lambda_c^+$ spectra. This procedure is important for the subsequent AA calculations, since differences between the two hadronization prescriptions in nuclear collisions should not originate from an inconsistent $pp$ reference. At RHIC energy, Fig.~\ref{fig:pp200_STAR} shows the corresponding comparison with the STAR $D^0$ spectrum at $\sqrt{s}=200\,\mathrm{GeV}$. The same $pp$ baseline is used in the calculation of the corresponding Pb--Pb and Au--Au observables below.

In addition to the $pp$ baseline of the charm-quark spectrum, initial conditions for AA collisions require accounting for initial-state nuclear effects. These enter into the calculation through the nuclear PDFs, which mainly amounts to nuclear shadowing in the relevant kinematic regions. In Pb--Pb collisions, we consider an initial charm-quark spectrum with an integrated total shadowing of $25\%$ as the input for HQ in-medium transport, while an integrated total shadowing of $30\%$ is used for the Au--Au system. A more detailed description of initial-state nuclear effects is provided in Appendix~\ref{app:IS-effect}.

\begin{widetext}

\begin{figure}[t]
    \centering
    \includegraphics[width=0.85\textwidth]{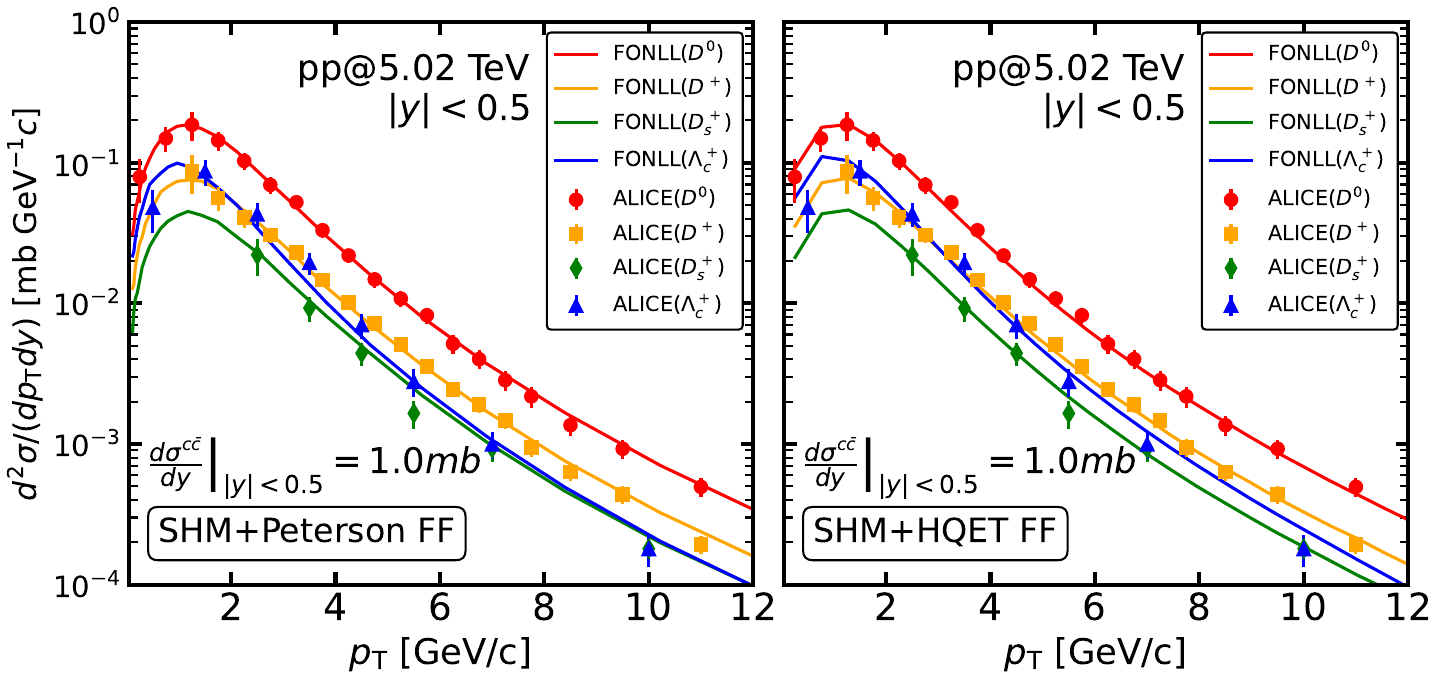}
    \caption{Transverse-momentum differential cross sections of prompt $D^0$, $D^+$, $D_s^+$, and $\Lambda_c^+$ in $pp$ collisions at $\sqrt{s}=5.02$\,TeV in the rapidity interval $-0.5<y<0.5$. Results obtained from the FONLL charm-quark spectrum followed by statistical hadronization and fragmentation are compared with ALICE data. The left panel employs the Peterson fragmentation function used together with iSCM, while the right panel employs the HQET-based fragmentation function used together with RRM.
    }
    \label{fig:pp5020_ALICE}
\end{figure}
\begin{figure}[t]
    \centering
    \includegraphics[width=0.85\textwidth]{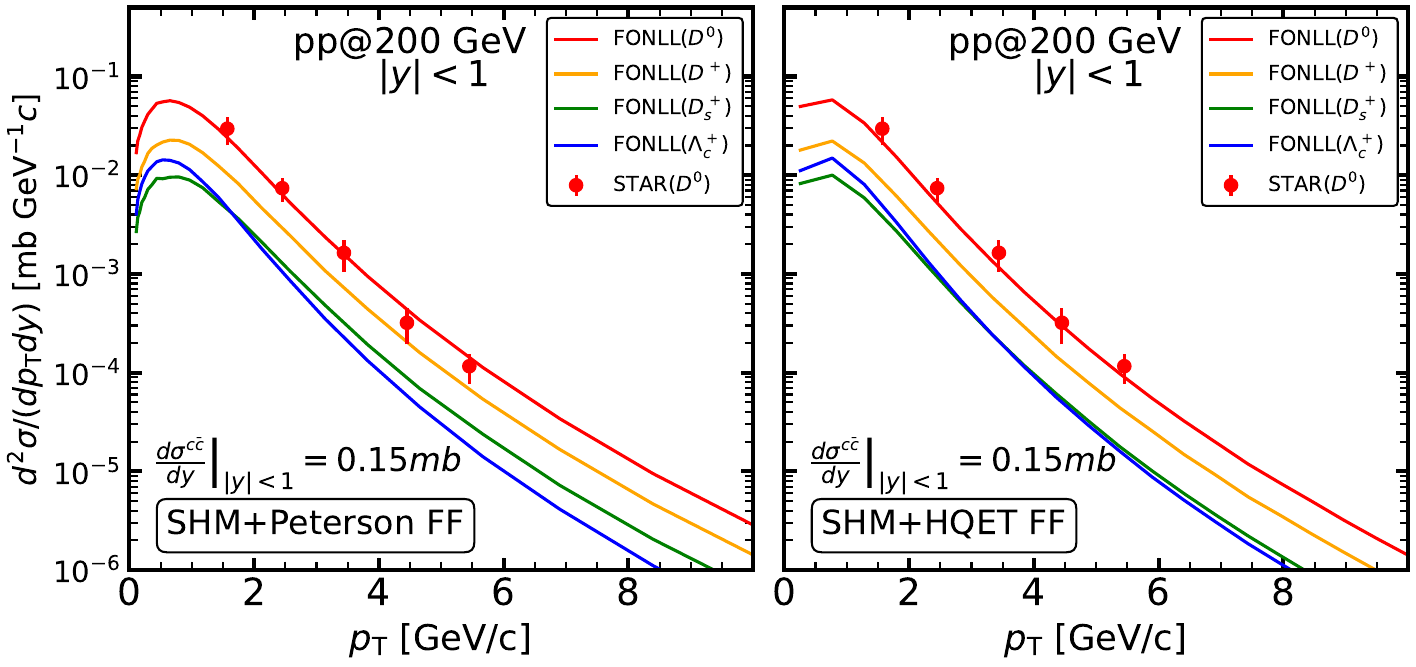}
    \caption{
    Transverse-momentum differential cross sections of prompt $D^0$, $D^+$, $D_s^+$, and $\Lambda_c^+$ in $pp$ collisions at $\sqrt{s}=200$\,GeV in the rapidity interval $-1<y<1$. Results obtained from the FONLL charm-quark spectrum followed by statistical hadronization and fragmentation are compared with STAR data \cite{STAR:2012dpa}. The left panel employs the Peterson fragmentation function used together with iSCM, while the right panel employs the HQET-based fragmentation function used together with RRM.
    }
    \label{fig:pp200_STAR}
\end{figure}

\end{widetext}

\begin{figure}[h]
\centering
\includegraphics[width=0.7\textwidth]{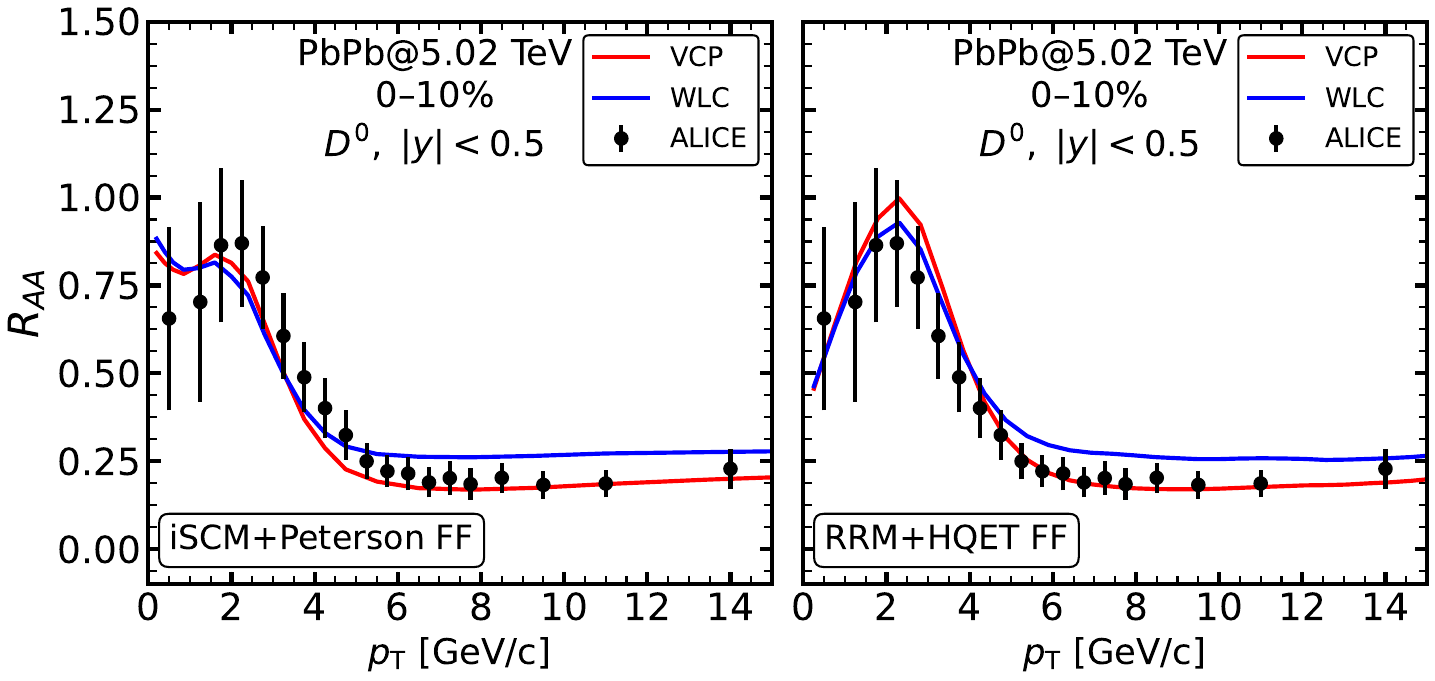}
\includegraphics[width=0.7\textwidth]{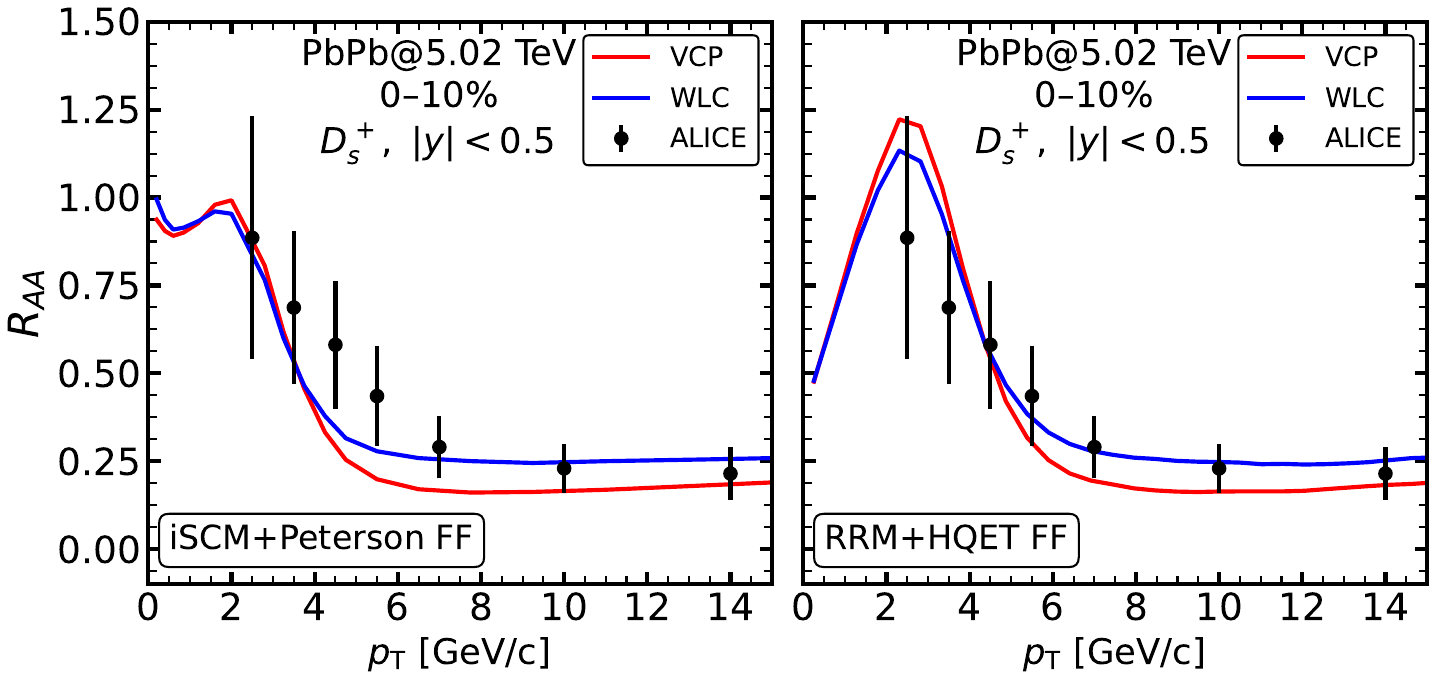}
\includegraphics[width=0.7\textwidth]{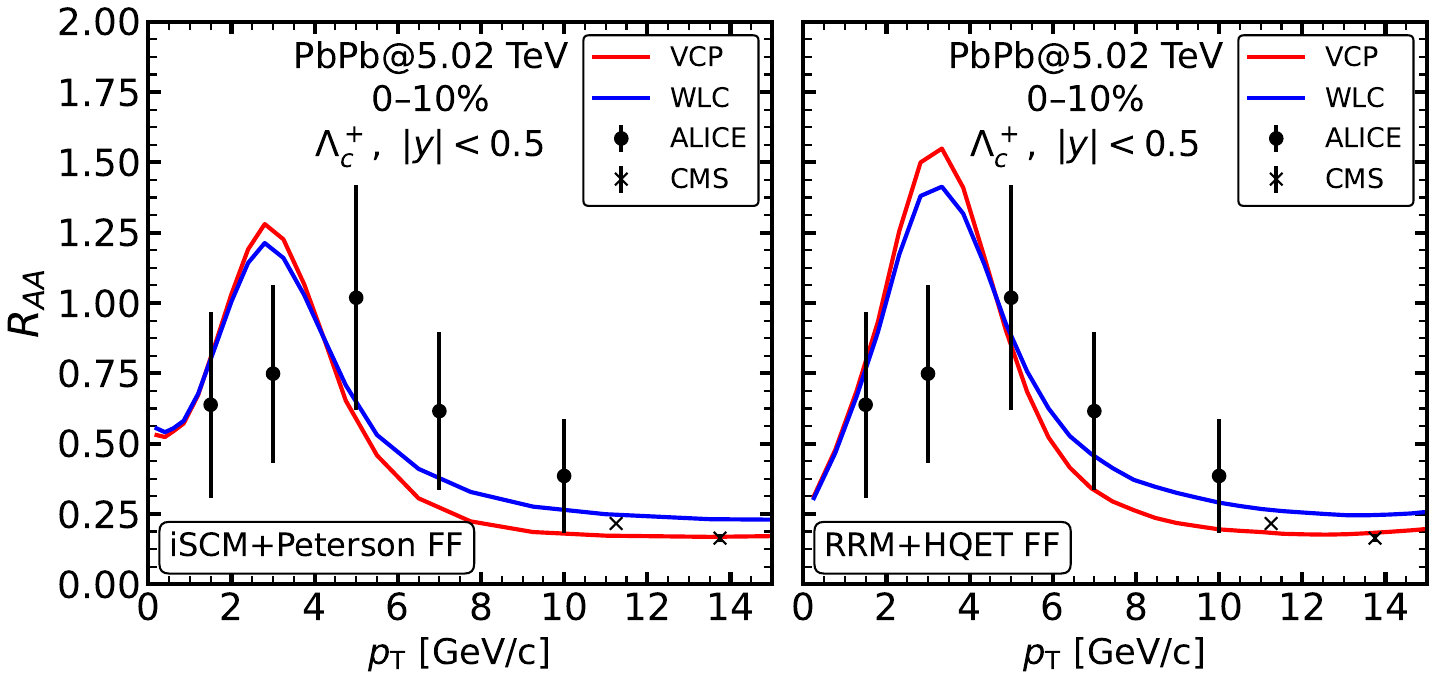}
\caption{Nuclear modification factor, $\raa$, of $D^0$, $D_s^+$, and $\Lambda_c^+$ at central rapidity in 0--10\% Pb--Pb collisions at $\sqrt{s_{\NN}}=5.02\,\mathrm{TeV}$. The experimental data are from ALICE~\cite{ALICE:2021rxa,ALICE:2021kfc,ALICE:2021bib}. The red and blue curves correspond to the VCP and WLC transport coefficients, respectively. The left panels employ iSCM+Peterson FF, while the right panels employ RRM+HQET FF.}
\label{fig:RAA-PbPb-0to10}
\end{figure}
%

\subsection{Nuclear Modification Factor}
\label{ssec:raa}

\begin{figure}[h]
\centering
\includegraphics[width=0.7\textwidth]{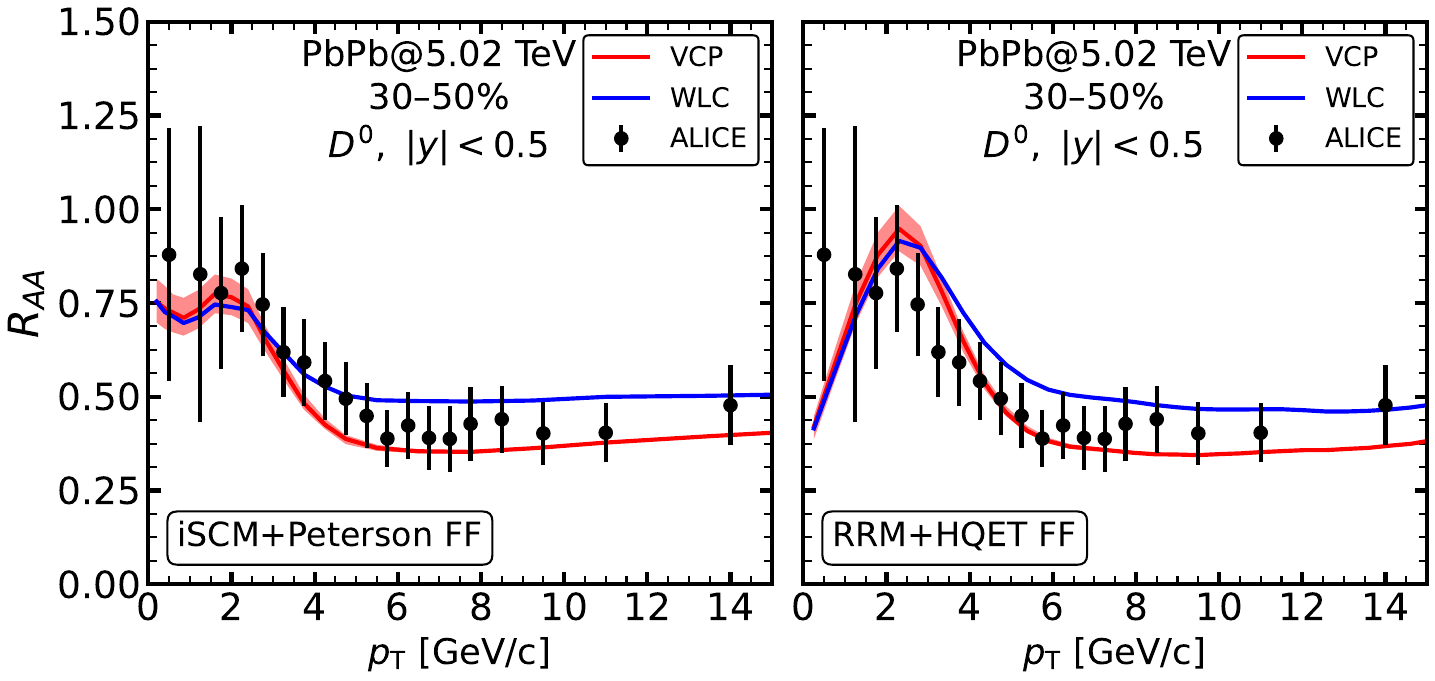}
\includegraphics[width=0.7\textwidth]{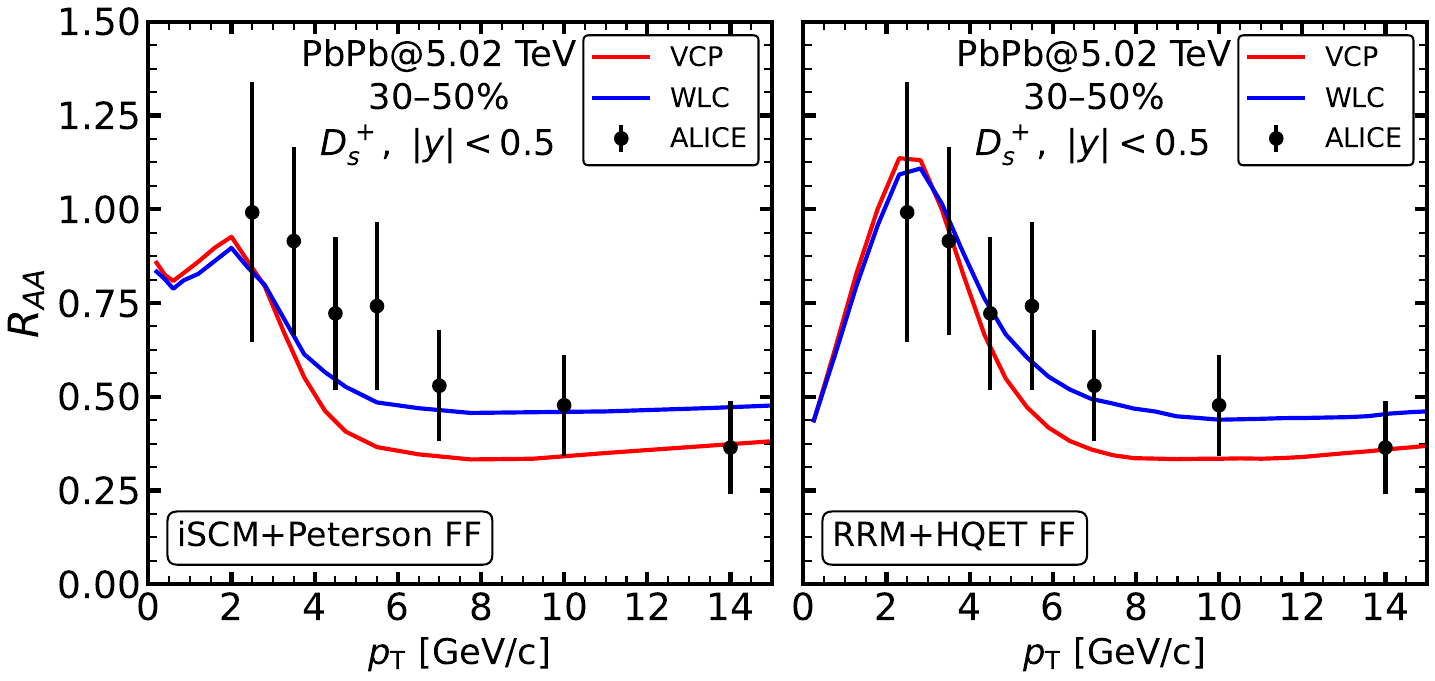}
\includegraphics[width=0.7\textwidth]{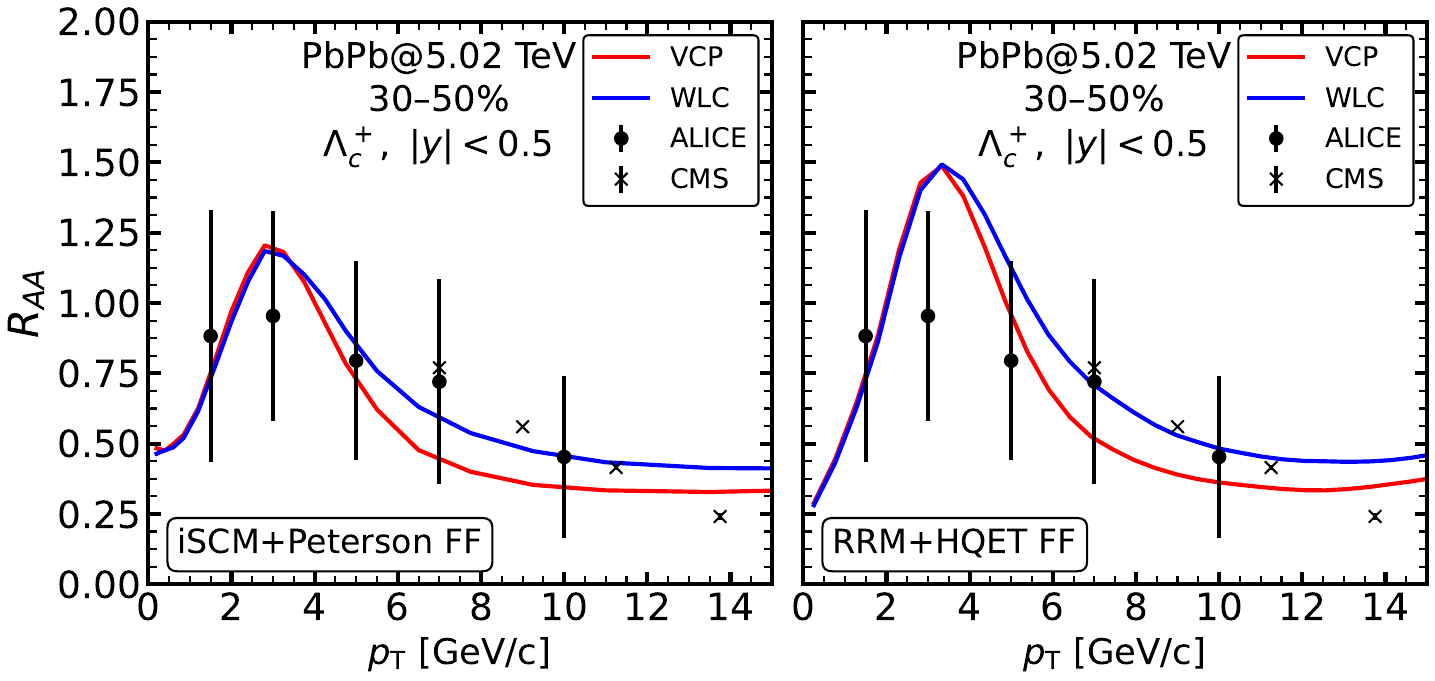}
\caption{Nuclear modification factor, $\raa$, of $D^0$, $D_s^+$, and $\Lambda_c^+$ at central rapidity in 30--50\% Pb--Pb collisions at $\sqrt{s_{\NN}}=5.02$\,TeV. The experimental data are from ALICE~\cite{ALICE:2021rxa,ALICE:2021kfc,ALICE:2021bib}. The red and blue curves correspond to the VCP and WLC transport coefficients, respectively. The left panels employ iSCM+Peterson FF, while the right panels employ RRM+HQET FF. In the $D^0$ panels, the shaded bands illustrate the sensitivity of the VCP results to the assumed initial-state nuclear shadowing strength, varied from 20\% to 30\% around the 25\% baseline.}
\label{fig:RAA-PbPb-30to50}
\end{figure}

The nuclear modification factor is defined as the ratio of the momentum spectrum of HF hadrons in AA collisions to that in $pp$ collisions, normalized by the number of binary collisions,
\begin{equation}
\raa=\frac{1}{\langle N_{\text{coll}}\rangle}
\frac{\rmd N_{\text{AA}}/\rmd p_T}
     {\rmd N_{\text{pp}}/\rmd p_T},
\end{equation}
where $\langle N_{\text{coll}}\rangle$ is the average number of binary nucleon--nucleon collisions in the pertinent centrality class.
In Fig.~\ref{fig:RAA-PbPb-30to50}
(Fig.~\ref{fig:RAA-PbPb-0to10}) we show our results for the $\raa$ of 
$D^0$, $D_s^+$, and $\Lambda_c^+$ in 30--50\% (0--10\%) Pb--Pb collisions at $\sqrt{s_{\NN}}=5.02\,\mathrm{TeV}$, in comparison to ALICE measurements \cite{ALICE:2021rxa,ALICE:2021kfc,ALICE:2021bib}. The left panels show the results obtained with the improved sudden coalescence model and Peterson fragmentation (iSCM+Peterson FF), while the right panels show those obtained with the resonance recombination model and HQET-based fragmentation (RRM+HQET FF). The red and blue curves correspond to the VCP and WLC transport coefficients, respectively.
The same conventions are adopted in the following figures unless stated otherwise.

Overall, the model calculations give a good description of the experimental data for all species of charmed hadrons at different centrality classes. The $D^0$ results show the expected centrality dependence: the suppression at intermediate and high $\pT$ is stronger in 0--10\% than in 30--50\% central collisions, due to the larger space-time extent and higher temperatures of the medium in central collisions, which increase the interaction of charm quarks with the QGP. The two hadronization prescriptions produce somewhat different shapes at low and intermediate $\pT$. In the iSCM + Peterson calculation, the $\raa$ essentially decreases from its low-$\pT$ value toward the suppressed high-$\pT$ region, whereas the RRM + HQET calculation develops a more pronounced ``flow bump" around $\pT \simeq 2\,\mathrm{GeV}/c$. Thus, although both calculations reproduce the overall suppression pattern, their agreement with the measured low- and intermediate-$\pT$ spectra is not equivalent. The position and magnitude of the maximum depend on the conversion of the (same underlying) transported charm-quark distributions into the final $D^0$ spectrum, including the fragmentation prescription and the momentum acquired through recombination. 

The VCP and WLC calculations are relatively close at low $\pT$, while their separation becomes more visible toward larger momenta. In central and semi-central collisions, the WLC interactions generally give a larger $D^0$ $\raa$ than the VCP at intermediate and high $\pT$, with similar dependence on the hadronization model. The comparison therefore does not establish a unique preference for either interaction from $\raa$ alone. In addition, the shadowing variation shown in Fig.~\ref{fig:RAA-PbPb-30to50} provides an additional initial-state sensitivity that must be considered when interpreting differences between the curves at low-$\pT$. The separate contributions of the initial-state modification and hadronization are further examined in Sec.~\ref{ssec:model-effects}. A quantitative discrimination of the two interactions requires the combined information from other observables, in particular the elliptic flow as discussed in Sec.~\ref{ssec:v2}.

\begin{figure}[h]
    \centering
    \includegraphics[width=0.78\linewidth]{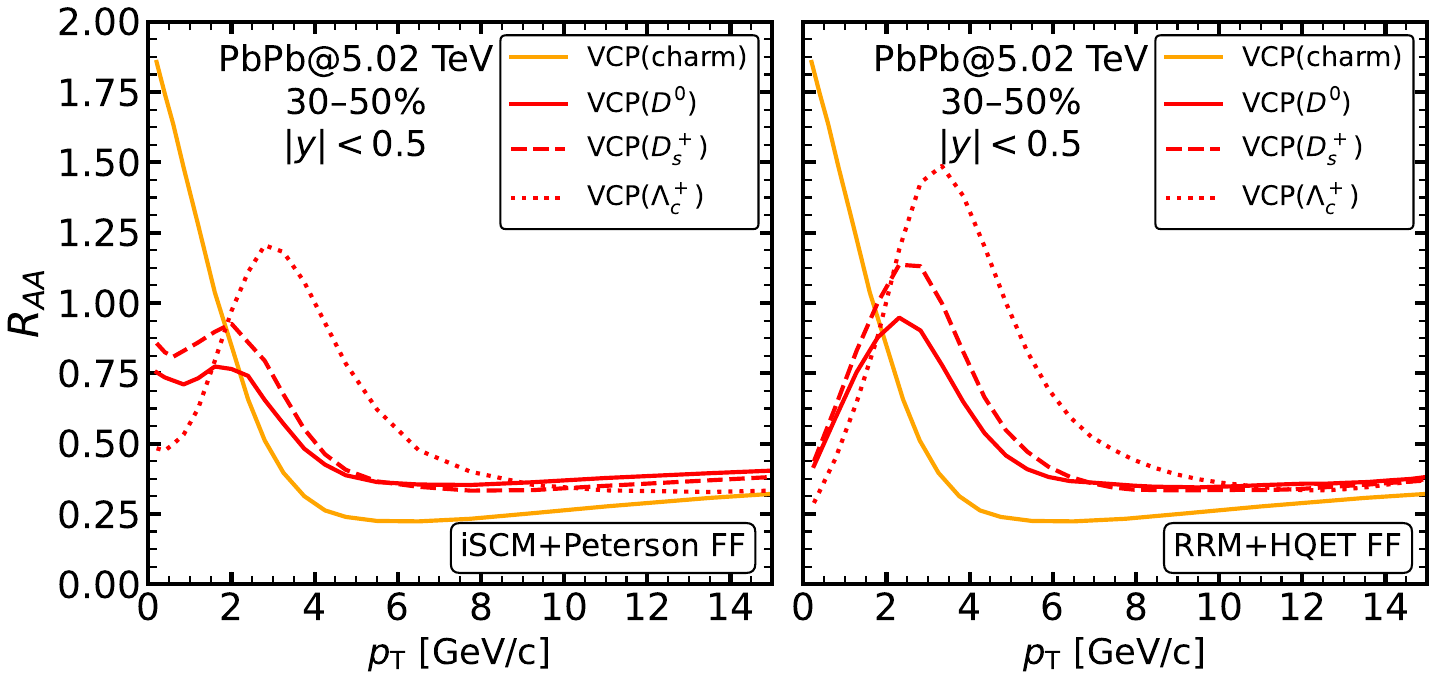}
    \caption{A comparison of the nuclear modification factor of all species of particles of interest, with the left panel corresponding to ``iSCM+Peterson FF" and the right to ``RRM+HQET FF". }
    \label{fig:Raa_VCP_all_species}
\end{figure}

A comparison with identified $D_s^+$ and $\Lambda_c^+$ observables can tell us more about the species dependence, which is a genuine hadronization effect; for convenience, we also provide a direct comparison of $\raa$ of different hadron species in Fig.~\ref{fig:Raa_VCP_all_species} for semi-central Pb--Pb collisions with the VCP interaction. The $D_s^+$ results exhibit a similar structure to the $D^0$ spectra. In the central and semi-central collisions, the calculations rise toward a maximum (\ie, a flow bump) at low or intermediate momentum before descending into the suppressed high-$\pT$ region. Both hadronization models give a very similar overall structure, especially at higher $ p_T$, but the magnitude of the maximum at low $\pT$ depends on the hadronization model. Recombination with strange quarks from the thermal medium increases the probability for a transported charm quark to emerge in the $D_s^+$ channel and transfers additional momentum from the light constituent to the final meson. The ALICE
measurements are broadly compatible with the calculated trends within their uncertainties.  

The $\Lambda_c^+$ results show a considerably stronger enhancement in the few-GeV region. Both hadronization models produce a maximum that can exceed unity, followed by a decrease toward higher $\pT$. The width and magnitude of this structure differ between iSCM + Peterson and RRM + HQET. The enhancement should not be interpreted as evidence for a smaller in-medium energy loss of charm quarks that eventually form $\Lambda_c^+$ baryons. For a given transport scenario, the preceding charm-quark evolution is the same for the different hadron species. The observed species dependence, as clearly exemplified in Fig.~\ref{fig:Raa_VCP_all_species}, is generated in the subsequent hadronization stage through the relative hadronization probabilities, the momentum carried by recombining light quarks, and the decay of excited charm-hadron states. The relative importance of these contributions is more directly assessed through the charm-hadron yield ratios discussed in Sec.~\ref{ssec:chem}.

\begin{figure}[H]
\centering
\includegraphics[width=0.78\textwidth]{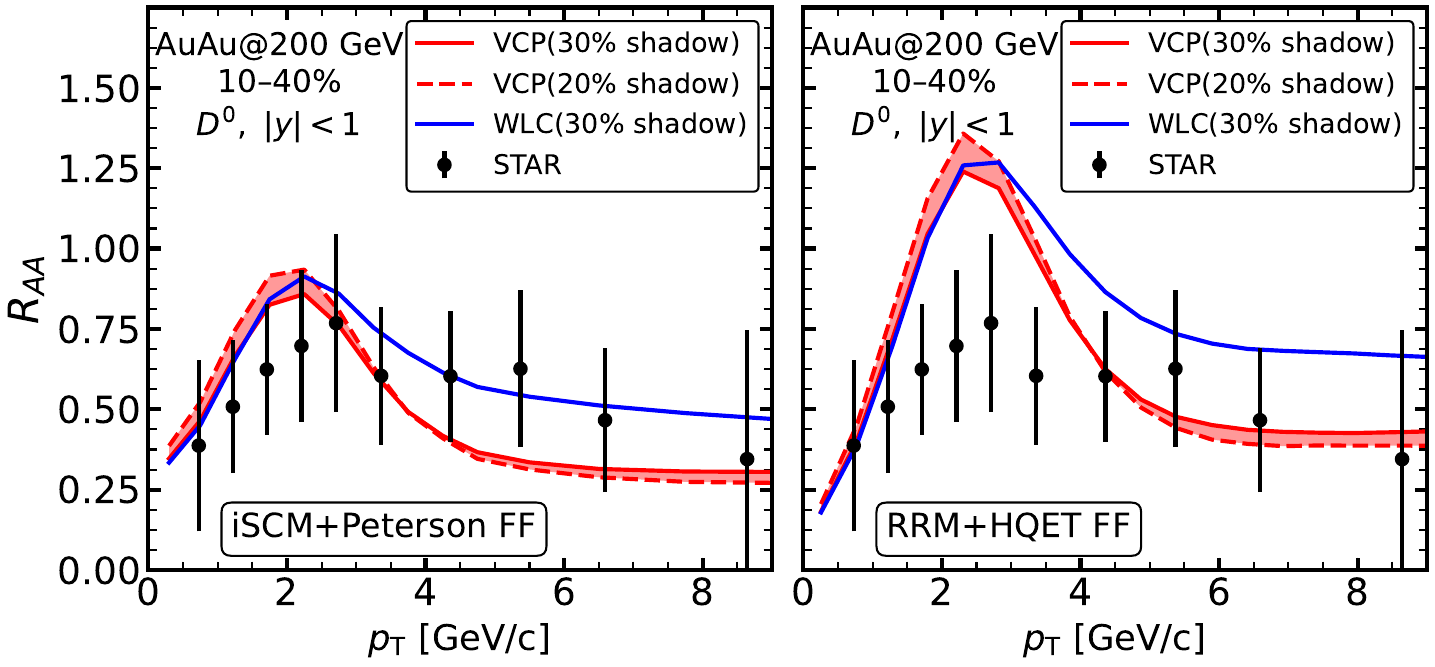}
\includegraphics[width=0.78\textwidth]{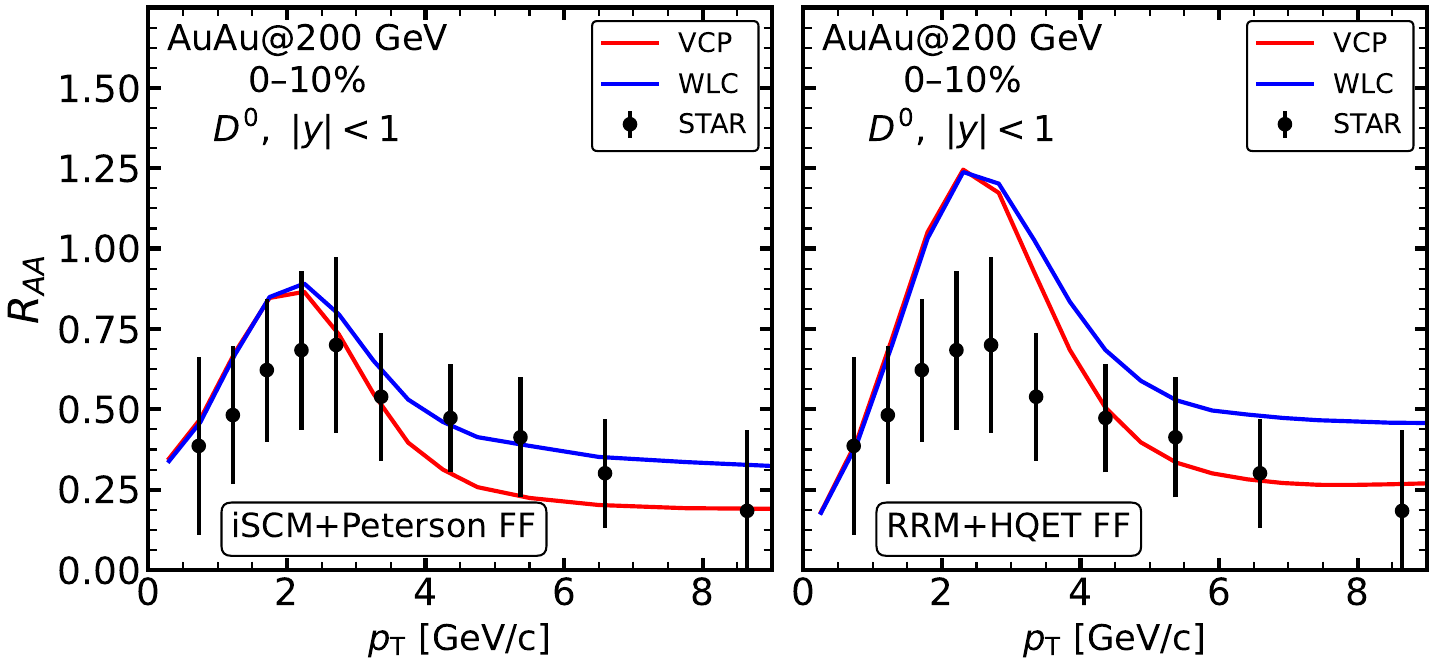}
\caption{
Nuclear modification factor, $\raa$, of $D^0$ at central rapidity in Au--Au collisions at $\sqrt{s_{\NN}}=200\,\mathrm{GeV}$ for the 10--40\% (upper) and 0--10\% (lower) centrality classes. The experimental data are from STAR~\cite{STAR:2018zdy}. The red and blue curves correspond to the VCP and WLC transport coefficients, respectively. The left panels employ iSCM+Peterson FF, while the right panels employ RRM+HQET FF. In the 10-40\% panels, the shaded bands illustrate the sensitivity of the VCP results to the assumed initial-state nuclear shadowing strength, varied from 20\% to 30\%.
}
\label{fig:RAA-AuAu}
\end{figure}

Turning to RHIC energies, the corresponding $D^0$ results for Au--Au collisions at $\sqrt{s_{\NN}}=200\,\mathrm{GeV}$ are shown in Fig.~\ref{fig:RAA-AuAu} for the 10--40\% and 0--10\% centrality classes. In the former, both hadronization prescriptions produce a maximum around $\pT\simeq2$--$3~\mathrm{GeV}/c$, but the stronger flow bump
generated by RRM+HQET FF overshoots the central values of the STAR measurements, whereas iSCM+Peterson FF gives a more moderate enhancement ({part of the overshoot in the RRM+HQET scenario is caused by an underestimate of the pertinent $pp$ spectra in the $\pT\simeq$2-3\,GeV region which figure in the denominator of the $\raa$, recall the right panel of Fig.~\ref{fig:pp200_STAR})}. This is consistent with the stronger flow bump generated by RRM in Pb--Pb collisions. In the 0--10\% class, however, the two hadronization models produce similar peaks, which are both above the measured central values. At higher $\pT$, the WLC interaction generally gives a larger $\raa$ than VCP; the former is closer to the semi-central data, while the latter is closer to the highest-$\pT$ central-collision data points. The shadowing variation in the upper panels mainly affects the low- and intermediate-$\pT$ region. Therefore, the RHIC comparison provides a useful test of the framework at a lower collision energy, where both the initial charm spectrum and the temperature and lifetime of the medium differ from their LHC counterparts, but does not by itself determine the preferred $T$-matrix interaction. A more restrictive assessment requires the simultaneous comparison of the $\raa$, $v_2$, and charm-hadron yield ratios. 

\begin{figure}[h]
\centering
\includegraphics[width=0.71\textwidth]{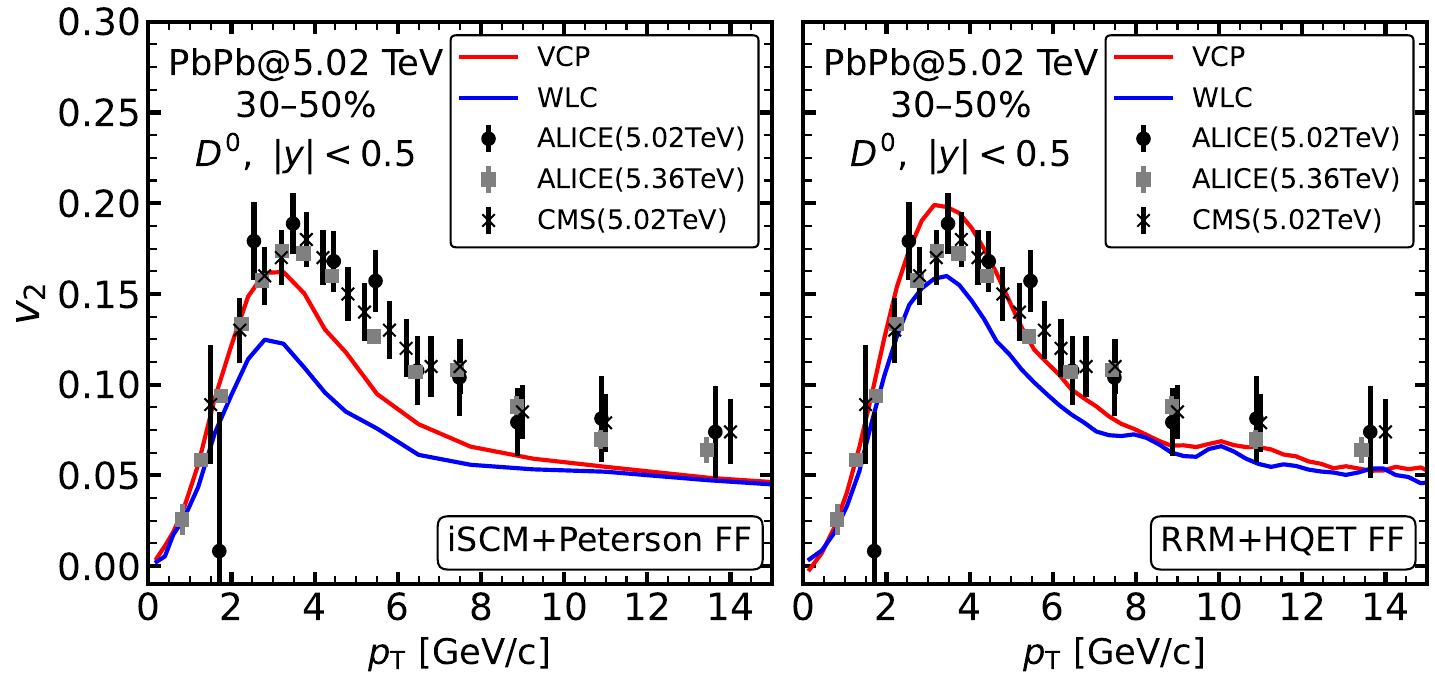}
\includegraphics[width=0.71\textwidth]{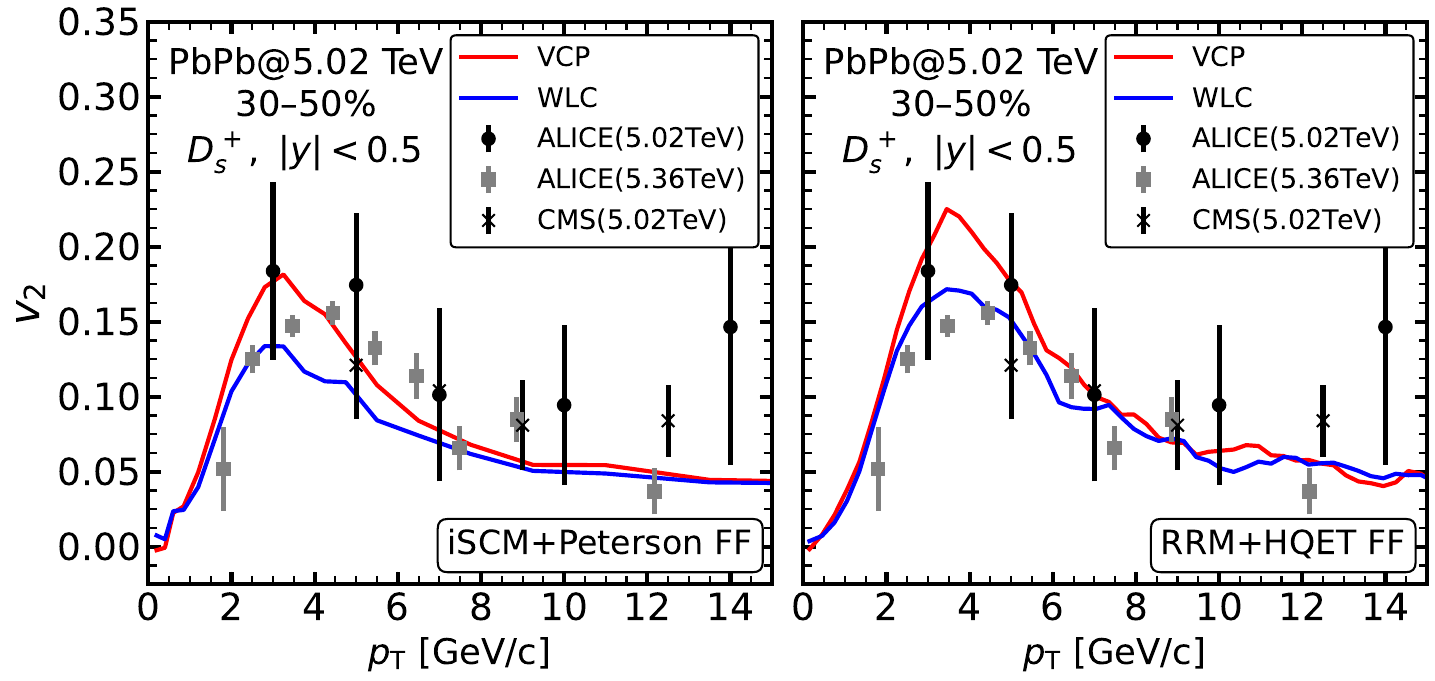}
\includegraphics[width=0.71\textwidth]{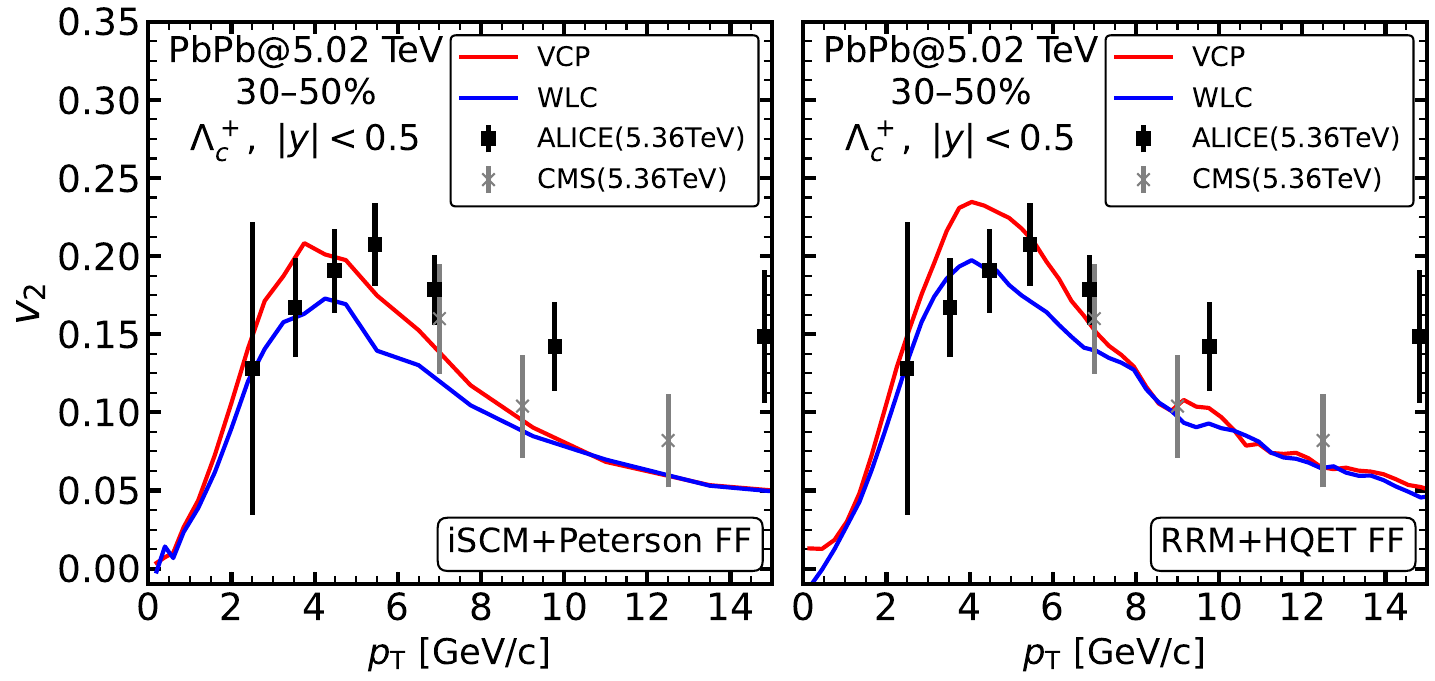}
\caption{
Elliptic flow, $v_2$, of $D^0$, $D_s^+$, and $\Lambda_c^+$ at central rapidity in 30--50\% Pb--Pb collisions. The calculations are performed at $\sqrt{s_{\NN}}=5.02\,\mathrm{TeV}$. The experimental data are from ALICE at
$\sqrt{s_{\NN}}=5.02\,\mathrm{TeV}$~\cite{ALICE:2020iug}, ALICE at $\sqrt{s_{\NN}}=5.36\,\mathrm{TeV}$~\cite{ALICE:2026zcz},
and CMS at $\sqrt{s_{\NN}}=5.02\,\mathrm{TeV}$~\cite{CMS:2020bnz}, as applicable to the individual hadron species. The red and blue curves correspond to the VCP and WLC transport coefficients, respectively. The left panels employ iSCM+Peterson FF, while the right panels employ RRM+HQET FF.
}
\label{fig:v2-PbPb5020}
\end{figure}
%
\subsection{Elliptic Flow}
\label{ssec:v2}
The spatial anisotropy of the initial overlap region in an AA collision is converted into a momentum-space anisotropy of the final-state particle distribution by the subsequent fireball evolution.  The pertinent anisotropic flow coefficients are defined through the Fourier expansion of the azimuthal dependence of the invariant yield,
\begin{equation}
E\frac{\rmd^3N}{\rmd^3p}
=\frac{1}{2\pi}
\frac{\rmd^2N}{p_T\rmd p_T\rmd y}
\left[
1+
2\sum_{n=1}^{\infty}
v_n\cos\left(n(\phi-\Psi_n)\right)
\right],
\end{equation}
where $E$ is the particle energy, $\phi$ its azimuthal angle relative to the symmetry angle, and $y$ its rapidity. The elliptic flow, $v_2$, is the second-order harmonic coefficient. For a smooth event-averaged medium as considered in this work, the principal axes of the initial geometry are aligned with the transverse $x$- and $y$-axes, such that the second-order symmetry plane can be chosen as $\Psi_2=0$. The elliptic flow coefficient can then be evaluated directly from the final-state particle momenta as
\begin{align}
v_2
=
\left\langle \cos 2\phi \right\rangle
=
\left\langle
\frac{p_x^2-p_y^2}{p_x^2+p_y^2}
\right\rangle.
\end{align}
Here the average is taken over particles within the kinematic range of interest. For the differential elliptic flow, $v_2(p_T)$, the same average is performed separately for particles in each transverse-momentum bin. 

Figure~\ref{fig:v2-PbPb5020} shows the calculated $v_2$ of $D^0$, $D_s^+$, and $\Lambda_c^+$ in 30--50\% Pb--Pb collisions. The $D^0$ results are compared with measurements from ALICE and CMS at $\sqrt{s_{\NN}}=5.02\,\mathrm{TeV}$, while the more recent $\sqrt{s_{\NN}}=5.36\,\mathrm{TeV}$ ALICE measurements are also displayed as an additional reference (theoretically, only marginal variation relative to 5.02\,TeV is expected).
The semi-central $D^0$ elliptic flow provides a particularly sensitive constraint on the interaction strength. In the intermediate-$\pT$ region, the VCP interaction produces a larger $v_2$ than the WLC interaction. This ordering can be related to the temperature dependence of the underlying $T$-matrix friction coefficients shown in Fig.~\ref{fig:Ap_and_Ds2piT}. 
Since the elliptic flow at low and intermediate $\pT$ is largely drawn from the collectively expanding medium, its buildup receives an important contribution from the later stages of the QGP evolution, when the bulk momentum anisotropy is well developed.  The stronger interaction of the VCP interaction in the lower-temperature region therefore provides a natural explanation for its larger HF $v_2$ relative to the WLC scenario. 

For $D^0$ in the 30--50\% centrality class, this leads to an important complementarity between $\raa$ and $v_2$. Within both the iSCM and RRM setups, the calculated $\raa$ can broadly describe the data, as the difference between VCP and WLC results is small compared to experimental uncertainties, whereas the larger anisotropy generated with VCP lies closer to the measured $v_2$. The tension is reduced in the RRM setup, for which the VCP result provides a more consistent simultaneous description of the suppression and elliptic flow over the phenomenologically most relevant intermediate-$\pT$ region. This also illustrates that the RRM scheme can generate stronger flow for charm hadrons via the recombination process, which is consistent with the fact that RRM has a larger coalescence probability compared to iSCM (see Fig.~\ref{fig:Pcoal}).

\begin{figure}[H]
    \centering
    \includegraphics[width=0.85\linewidth]{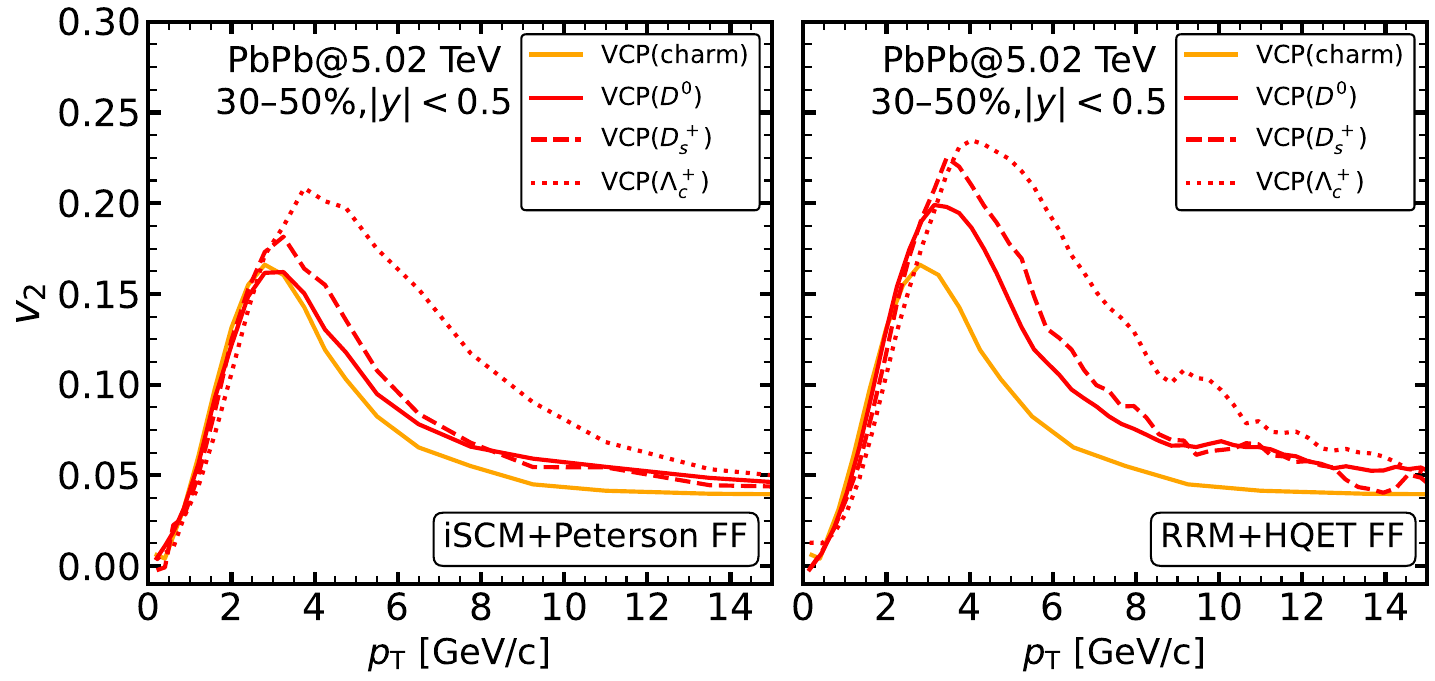}
    \caption{A comparison of the elliptic flow of all species of particles of interest, with the left panel corresponding to ``iSCM+Peterson FF" and the right to ``RRM+HQET FF".}
    \label{fig:v2_VCP_all_species}
\end{figure}
The $D_s^+$ and $\Lambda_c^+$ elliptic flows provide a further probe of hadronization. Recombination with collectively flowing light quarks transfers additional momentum anisotropy to the final-state charm hadrons.  For $D_s^+$, both in-medium potential scenarios generate a maximum at intermediate $\pT$, with the magnitude retaining a visible dependence on the transport coefficients and a slight dependence on the hadronization model. The present measurements are broadly compatible with such a sizable anisotropy, although their uncertainties are larger than those for $D^0$. The $\Lambda_c^+$ exhibits an even stronger sensitivity to recombination because two light constituents participate in the formation of charm baryons. The available data indicate substantial $\Lambda_c^+$ elliptic flow, but their current precision does not yet permit a quantitative discrimination between the different model components.
We also provide a direct comparison of the $v_2$ for different hadron species to that of the $c$-quark just prior to hadronization in semi-central Pb--Pb collisions in Fig.~\ref{fig:v2_VCP_all_species}. It is clear that color-neutralization of the charm quarks imparts significant additional elliptic flow on the hadrons via recombination, which also generates a hadron-specific hierarchy.  
\begin{figure}[t]
\centering
\includegraphics[width=0.7\textwidth]{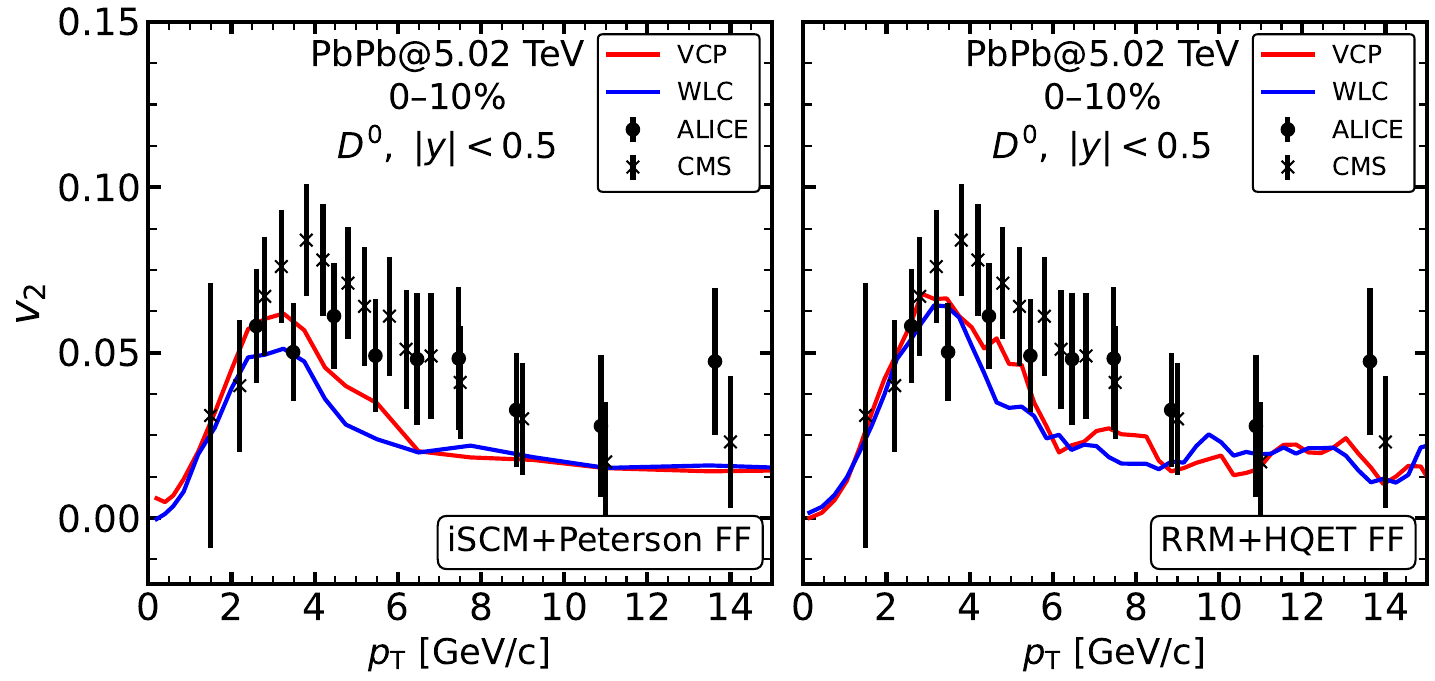}
\includegraphics[width=0.7\textwidth]{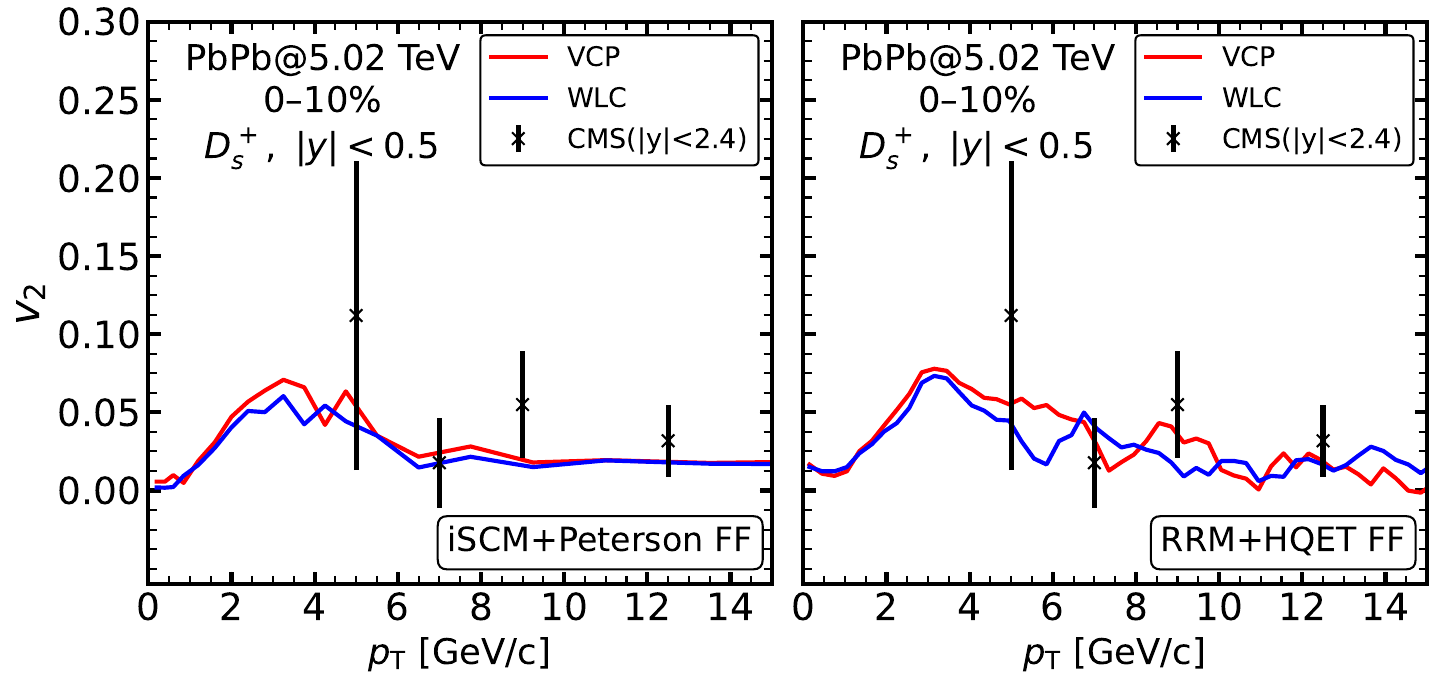}
\caption{
Elliptic flow, $v_2$, of $D^0$ and $D_s^+$ at central rapidity in 0--10\% Pb--Pb collisions at $\sqrt{s_{\NN}}=5.02\,\mathrm{TeV}$. The experimental data are from ALICE~\cite{ALICE:2020iug,ALICE:2021kfc} and CMS~\cite{CMS:2020bnz}. The red and blue curves correspond to the VCP and WLC transport coefficients, respectively. The left panels employ iSCM+Peterson FF, while the right panels employ RRM+HQET FF.}
\label{fig:v2-PbPb-0to10}
\end{figure}
\begin{figure}[t]
\centering
\includegraphics[width=0.7\textwidth]{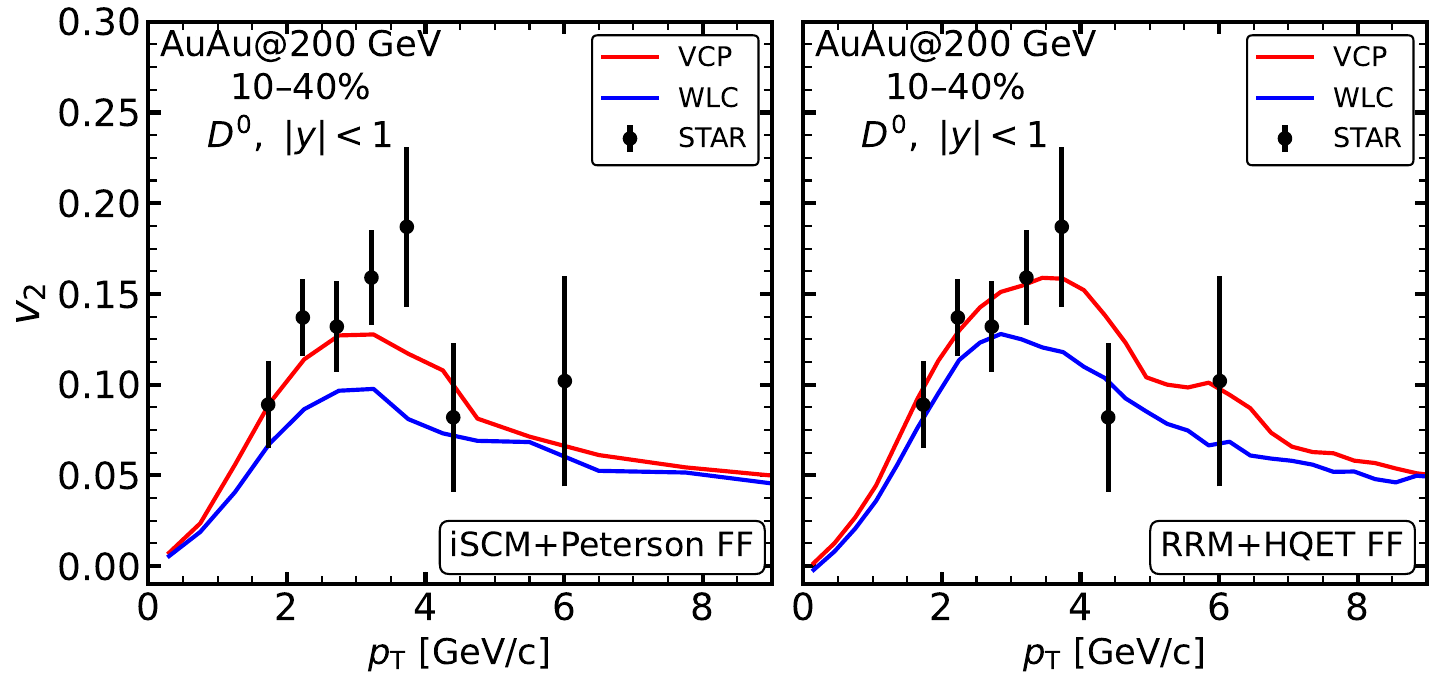}
\caption{
Elliptic flow, $v_2$, of $D^0$ at central rapidity in 10--40\% Au--Au collisions at $\sqrt{s_{\NN}}=200\,\mathrm{GeV}$. The experimental data are from STAR~\cite{STAR:2017kkh}. The red and blue curves correspond to the VCP and WLC transport coefficients, respectively. The left panel employs iSCM+Peterson FF, while the right panel employs RRM+HQET FF.
}
\label{fig:v2-AuAu-10to40}
\end{figure}

Figure~\ref{fig:v2-PbPb-0to10} shows the corresponding results in central Pb--Pb collisions. As expected from the smaller initial geometric eccentricity, the elliptic flow in the 0--10\% centrality class is substantially reduced relative to the 30--50\% results. The calculated $D^0$ $v_2$ is comparable to the ALICE and CMS measurements, while the difference between the VCP and WLC scenarios is less pronounced than in semi-central collisions. The available $D_s^+$ data have considerably larger uncertainties even for a broader rapidity interval and therefore provide a weaker constraint on the transport and hadronization scenarios in this centrality class for now. 

The corresponding $D^0$ elliptic flow in 10--40\% Au--Au collisions is shown in Fig.~\ref{fig:v2-AuAu-10to40}. At RHIC energies, the same ordering of the two interaction scenarios is observed, with VCP producing the larger elliptic flow. The VCP result is closer to the STAR data around the maximum of the measured $v_2$, although the calculated anisotropy tends to remain below the experimental central values in the few-GeV region for the iSCM hadronization. Together with the reasonable description of the corresponding $\raa$, this again demonstrates that the simultaneous reproduction of suppression and elliptic flow provides a more restrictive test than either observable alone.

\subsection{Hadro-Chemistry}
\label{ssec:chem}

The relative abundances of different charm-hadron species provide a complementary probe of the HQ hadronization process. Compared with absolute yields, part of the uncertainty associated with the overall charm production yields and with the common preceding charm-quark transport evolution cancels in ratios such as $D_s^+/D^0$ and $\Lambda_c^+/D^0$. These observables are therefore particularly sensitive to how transported charm quarks are redistributed among the different meson and baryon species during hadronization.

\begin{figure}
    \centering
    \includegraphics[width=0.7\linewidth]{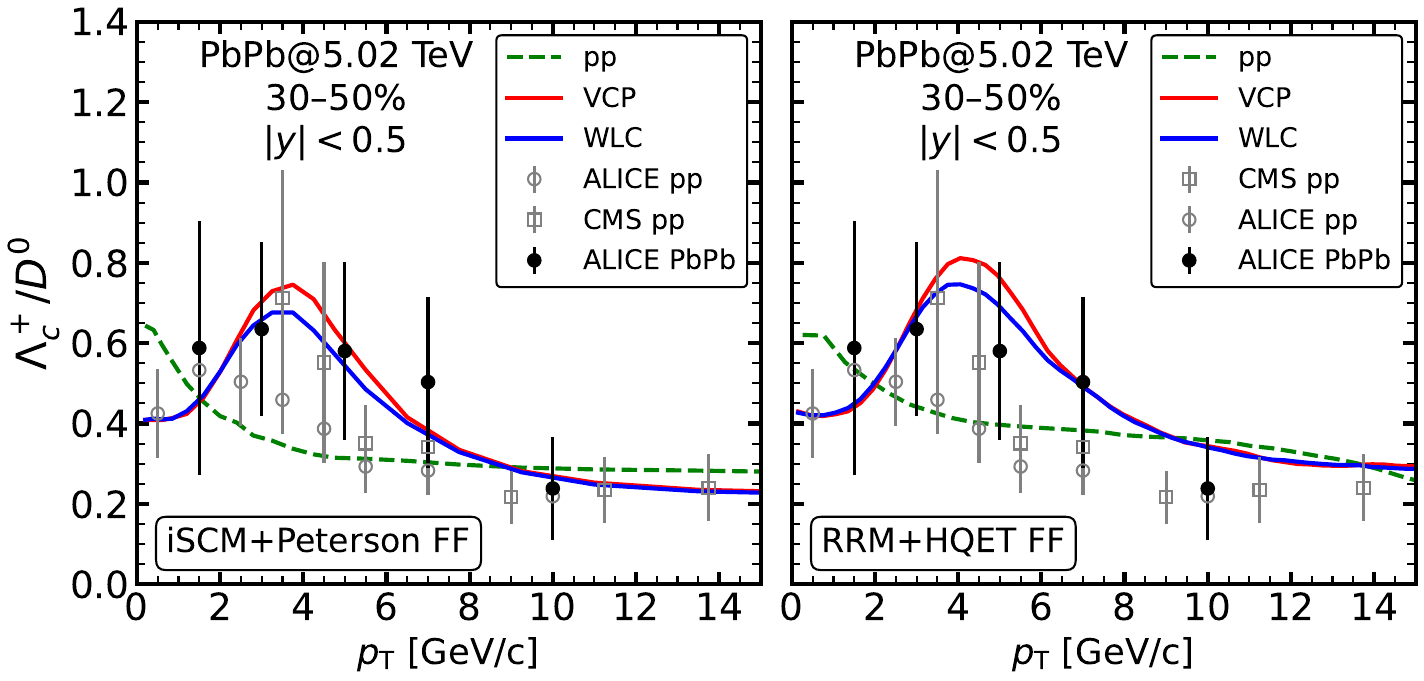}
    \includegraphics[width=0.7\linewidth]{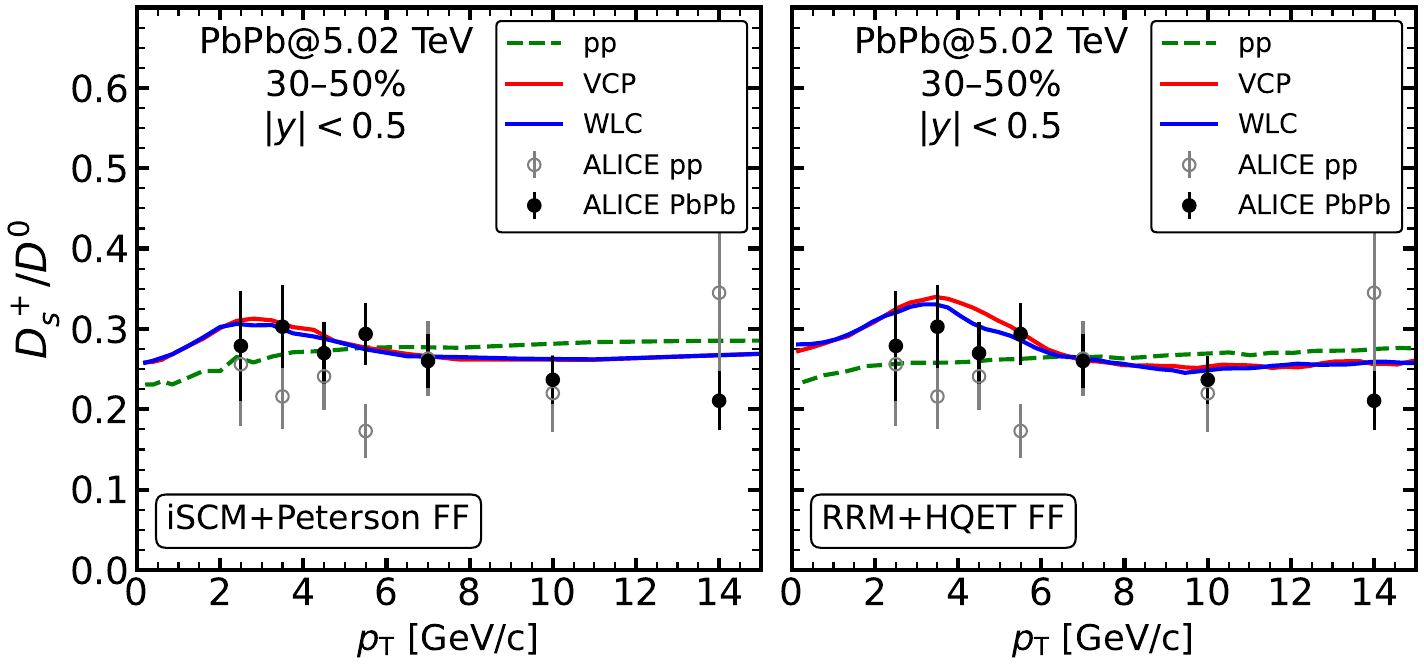}
    \includegraphics[width=0.7\linewidth]{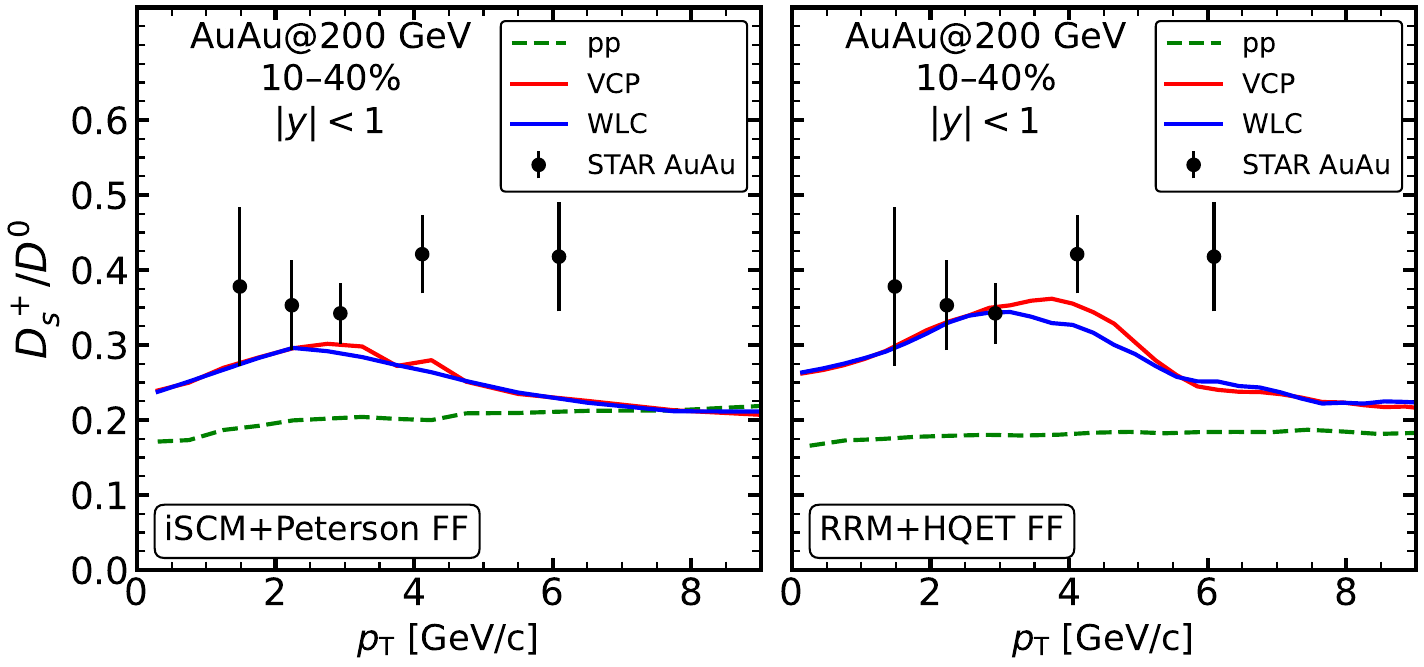}
    \caption{
    Charm-hadron yield ratios at central rapidity.
    Top: $\Lambda_c^+/D^0$ in 30--50\% Pb--Pb collisions at $\sqrt{s_{\NN}}=5.02\,\mathrm{TeV}$, as well as the pp baselines.
    Middle: $D_s^+/D^0$ in 30--50\% Pb--Pb collisions at $\sqrt{s_{\NN}}=5.02\,\mathrm{TeV}$.
    Bottom: $D_s^+/D^0$ in 10--40\% Au--Au collisions at $\sqrt{s_{\NN}}=200\,\mathrm{GeV}$.
    The left and right panels correspond to iSCM+Peterson FF and RRM+HQET FF, respectively, while the red and blue curves denote the VCP and WLC transport scenarios. The corresponding experimental measurements are taken from
    Refs.~\cite{ALICE:2021bib,ALICE:2021kfc,STAR:2021tte}.
    }
    \label{fig:hadrochemistry}
\end{figure}

The upper panels of Fig.~\ref{fig:hadrochemistry} show the $\Lambda_c^+/D^0$ ratio in 30--50\% Pb--Pb collisions. Both hadronization models generate a pronounced enhancement at intermediate $\pT$ relative to the corresponding $pp$ baseline. The ratio rises when recombination becomes important, reaching a maximum around 4\,GeV and subsequently decreasing as fragmentation becomes increasingly dominant toward higher $\pT$. This qualitative structure is common to both hadronization models, but the magnitude and width of the enhancement differ slightly. The $\Lambda_c^+/D^0$ ratio therefore exposes a model dependence that is much less apparent in the $D^0$ spectrum alone. For comparison, we also plot measured data in $pp$ collisions at ALICE and CMS with the results obtained from the statistical hadronization + Peterson/HQET FFs, which have provided enhancements in the low-$\pT$ region, as has been studied in Ref.\cite{He:2019tik}. These enhancements in $pp$ collisions are caused by considering the decay of excited states in the relativistic quark model into the $\Lambda_c^+$ ground state. 

The middle panels of Fig.~\ref{fig:hadrochemistry} show the $D_s^+/D^0$ ratio in 30--50\% Pb--Pb collisions. Compared with the corresponding $pp$ baselines, both hadronization schemes predict an enhancement in the intermediate $\pT$ region. This enhancement is a characteristic feature of recombination of HQs with strange quarks from the thermal medium.  As the transverse momentum increases, the recombination contribution gradually decreases and fragmentation becomes dominant, causing the ratio to approach its corresponding $pp$ value. The magnitude of the enhancement slightly depends on the hadronization model, and results from both iSCM+ Peterson FF and RRM+HQET FF are close to the central values of the Pb--Pb measurements over most of the measured intermediate-$\pT$ region. The difference between the VCP and WLC transport scenarios is rather small, indicating that this observable is more sensitive to the hadronization prescription than to the variation of the HQ transport coefficients considered here. 

The lower panels of Fig.~\ref{fig:hadrochemistry} present the $D_s^+/D^0$ ratio in 10--40\% Au--Au collisions at $\sqrt{s_{\NN}}=200$\,GeV, and a similar qualitative behavior is observed from the model calculations. Neither hadronization scheme can fully describe the STAR measurements. 
Taken together with the $\Lambda_c^+/D^0$ results, these comparisons demonstrate that charm-hadron yield ratios provide direct sensitivity to the implementation of HQ hadronization, whereas the dependence on the underlying transport scenarios is rather modest.
 
\subsection{Impact of Model Components on HF Observables}
\label{ssec:model-effects}

The comparisons shown in the previous sections reflect the cumulative impact of multiple stages in the evolution of the system and respective assumptions regarding the underlying physics. To better pinpoint the contributions of the various stages and physics choices seen in the final $D^0$ spectra and elliptic flow, we now vary individual elements of the framework while keeping others fixed. We first examine the modification introduced in converting transported charm quarks into $D^0$ mesons (Sec.~\ref{sssec:c-h}), followed by the effects of the initial nuclear PDFs (Sec.~\ref{sssec:npdf}), the relative roles of diffusion and radiation in the QGP (Sec.~\ref{sssec:diff-rad}), and finally the rescattering of charm hadrons in the hadronic phase (Sec.~\ref{sssec:urqmd}). These controlled variations are carried out within both hadronization frameworks, allowing us to assess whether a given effect is robust against the details of the hadronization prescription.

\subsubsection{From Charm Quarks to Hadrons}
\label{sssec:c-h}
\begin{figure}[H]
    \centering
    \includegraphics[width=0.74\linewidth]
    {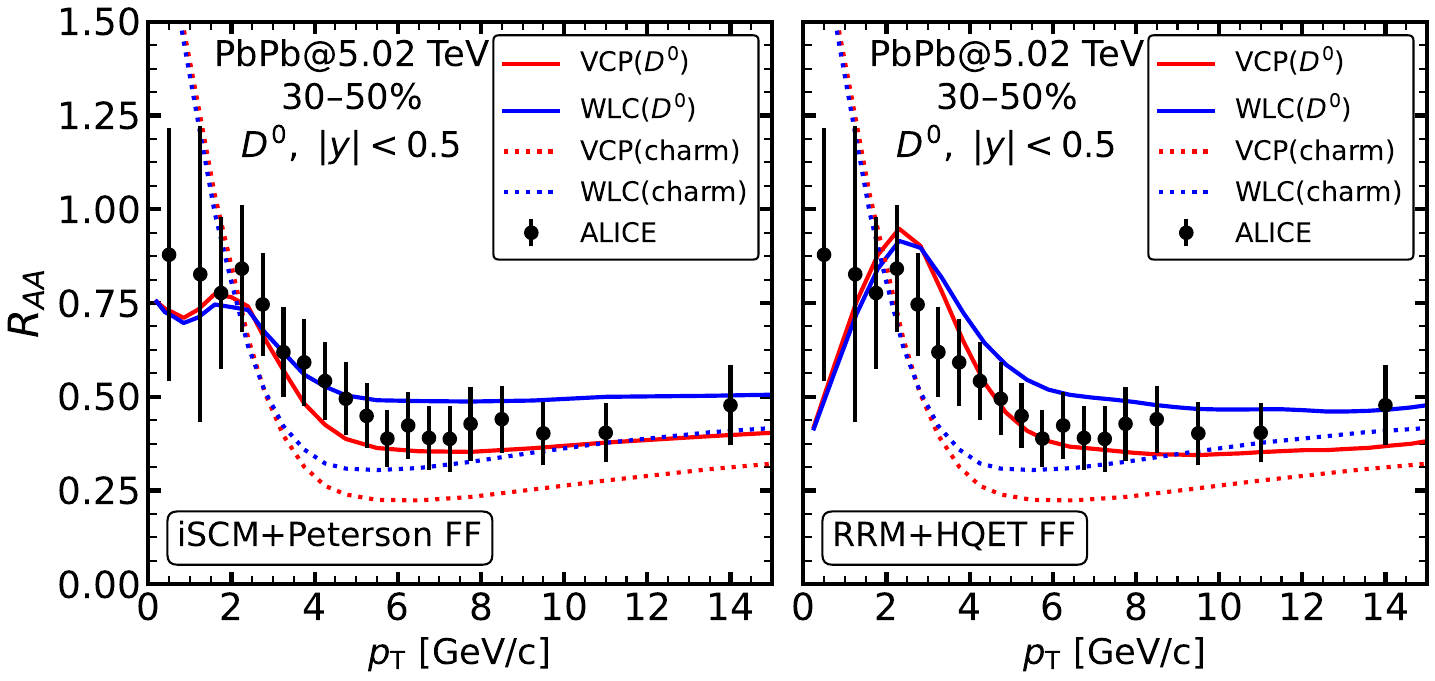}
    \includegraphics[width=0.74\linewidth]
    {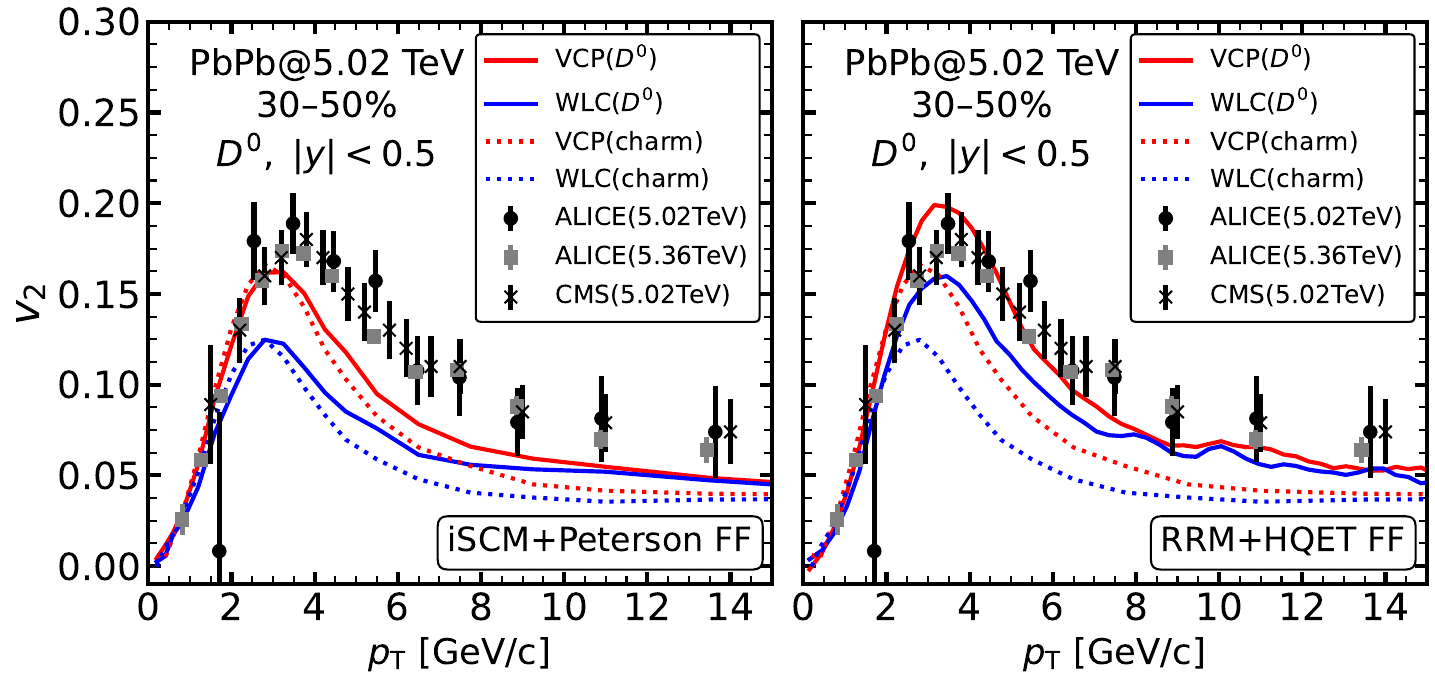}
    \caption{Comparison of the charm-quark and final $D^0$ nuclear modification factor (left) and elliptic flow (right) in 30--50\% Pb--Pb collisions at $\sqrt{s_{\NN}}=5.02\,\mathrm{TeV}$. The comparison illustrates the modification generated by hadronization after the charm-quark transport through the QGP.
    }
    \label{fig:hadronization-effect-PbPb}
\end{figure}

The first comparison isolates the modification generated by the conversion of the transported charm-quark distribution into the final $D^0$ distribution. The results for 30--50\% Pb--Pb collisions are shown in Fig.~\ref{fig:hadronization-effect-PbPb}.

\begin{figure}[H]
    \centering
    \includegraphics[width=0.49\linewidth]
    {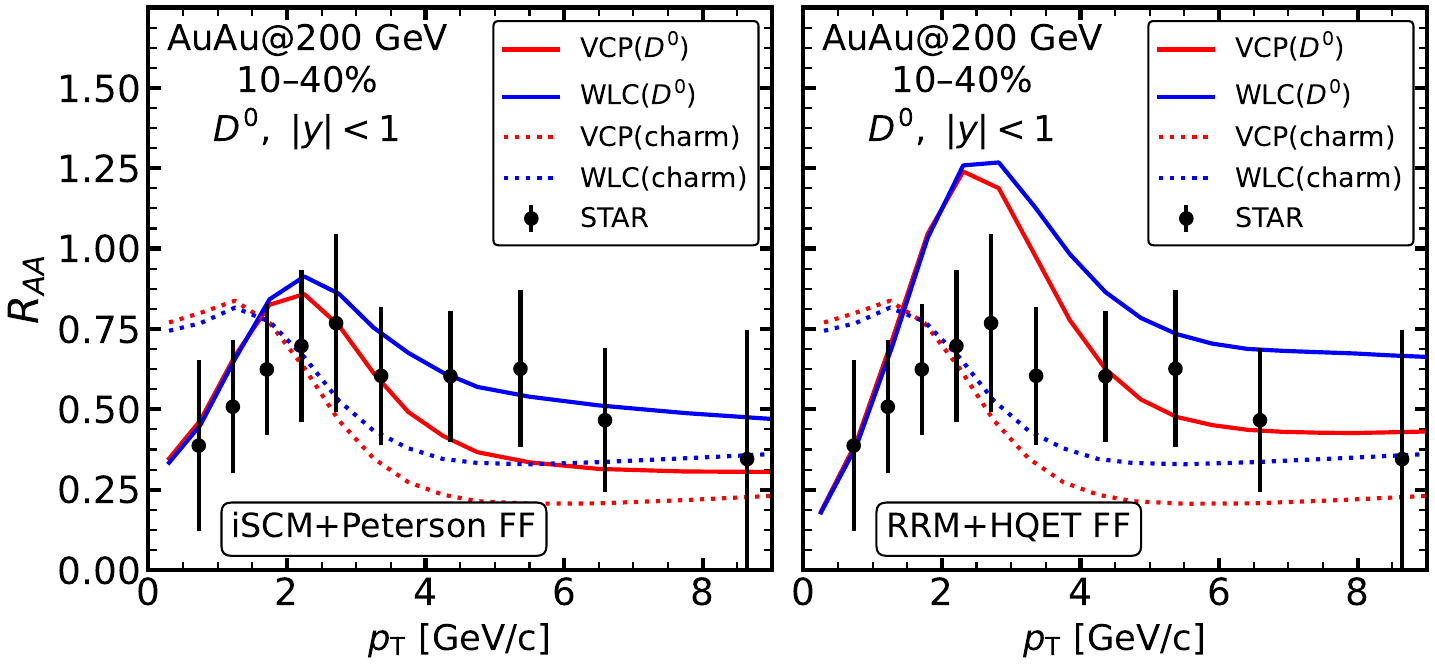}
    \includegraphics[width=0.49\linewidth]
    {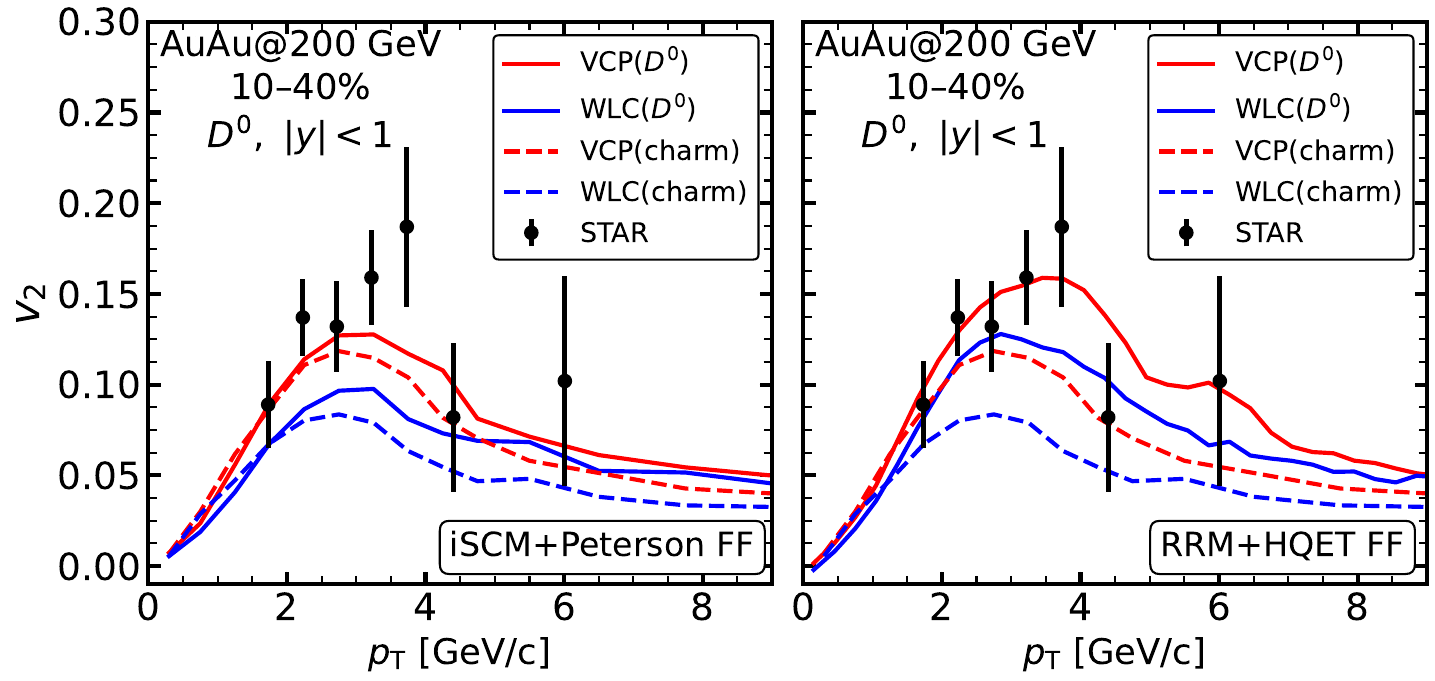}
    \caption{Comparison of the charm-quark and final $D^0$ nuclear modification factor (left) and elliptic flow (right) in 10--40\% Au--Au collisions at $\sqrt{s_{\NN}}=200\,\mathrm{GeV}$.
    }
    \label{fig:hadronization-effect-AuAu}
\end{figure}

In the momentum region where recombination is important, the light constituent generally supplies additional momentum to the final $D^0$ meson and changes the mapping between the charm-quark and hadron $\pT$. This reshapes the $D^0$ $\raa$ at low and intermediate momenta. The effect is more apparent in the elliptic flow, especially in the resonance recombination model. The light quarks participating in recombination are part of the collectively expanding bulk medium and therefore carry sizable momentum anisotropy. Consequently, the produced $D^0$ inherits additional $v_2$ during hadronization, resulting in a larger anisotropy than that of the charm quark prior to hadronization over the momentum region where recombination is important. Toward high $\pT$, the recombination probability decreases, and fragmentation becomes dominant, reducing the difference between the charm-quark and $D^0$ observables. The same comparison at RHIC energy is shown in Fig.~\ref{fig:hadronization-effect-AuAu}, where the qualitative role of hadronization is found to persist.

\subsubsection{Initial-State Nuclear-PDF Effects}
\label{sssec:npdf}

\begin{figure}[H]
    \centering
    \includegraphics[width=0.74\linewidth]{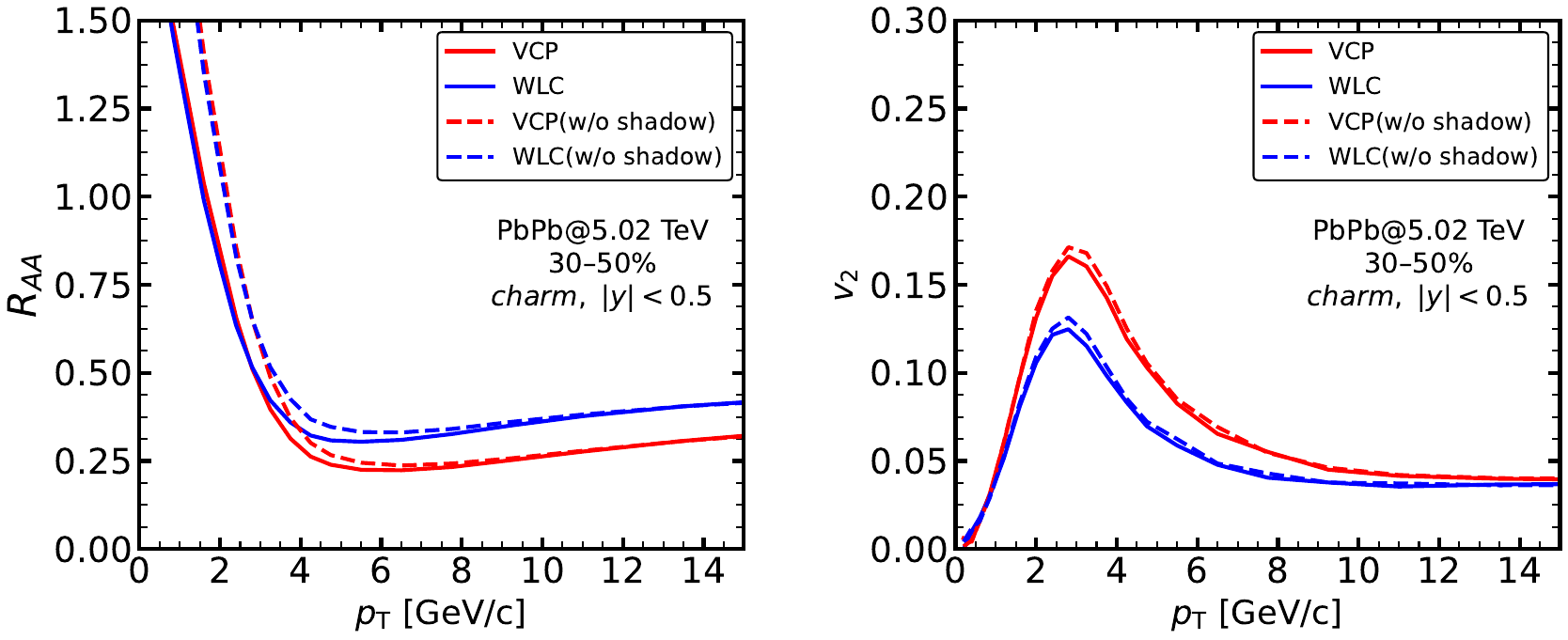}
    \caption{Impact of the initial-state nuclear-PDF modification on the charm-quark $\raa$ (left) and $v_2$ (right) in 30--50\% Pb--Pb collisions at
    $\sqrt{s_{\NN}}=5.02\,\mathrm{TeV}$.}
    \label{fig::npdf-effect-PbPb-charm}
\end{figure}

Next, we examine the impact of the initial-state nuclear modification of the parton distributions. Before hadronization, we compare the charm-quark $\raa$ and $v_2$ obtained with and without shadowing after the QGP transport evolution. The results are presented in Fig.~\ref{fig::npdf-effect-PbPb-charm}, in which the dashed curves are associated with an un-shadowed initial-charm spectrum, as those used in $pp$ collisions at the same collision energy and rapidity. The reduction of the charm yield due to the initial-state nuclear shadowing is most significant in the $\raa$ being at low and intermediate transverse momenta and gradually diminishing toward high $\pT$. Its influence on $v_2$ is more subtle. For both VCP and WLC interactions, the calculation without shadowing exhibits a slightly larger $v_2$ in the intermediate $\pT$ region than the result including shadowing. This behavior originates from the interplay between the momentum dependence of the shadowing correction and the subsequent transport evolution of charm quarks in the medium. Since shadowing is stronger at lower $\pT$ and becomes weaker toward higher momenta, it modifies not only the overall charm yield but also the shape of the initial momentum distribution. The final charm population at a given $\pT$ receives contributions from different initial momenta, with different weights for propagation along the short and long directions of the medium. Consequently, the momentum-dependent shadowing correction can modify the relative in-plane and out-of-plane populations and thereby affect the resulting elliptic flow. 
 
\begin{figure}[H]
    \centering
    \includegraphics[width=0.74\linewidth]
    {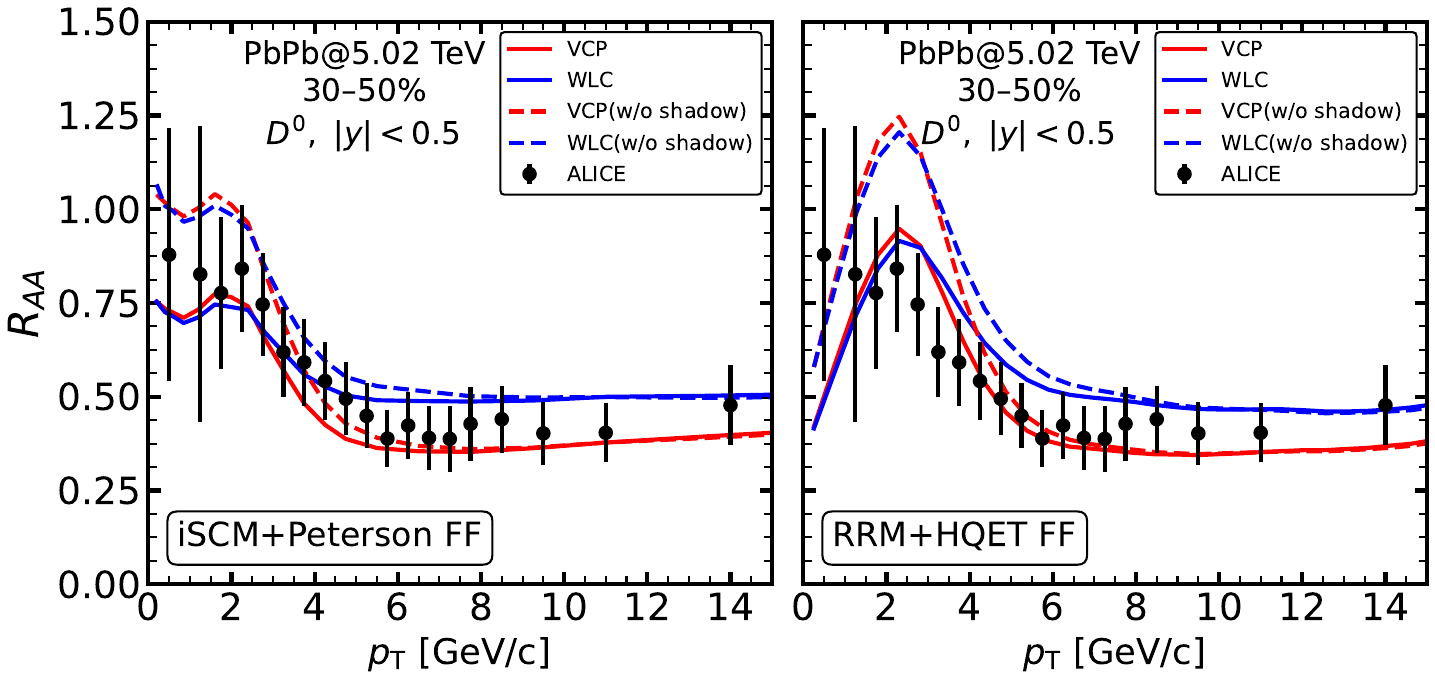}
    \includegraphics[width=0.74\linewidth]
    {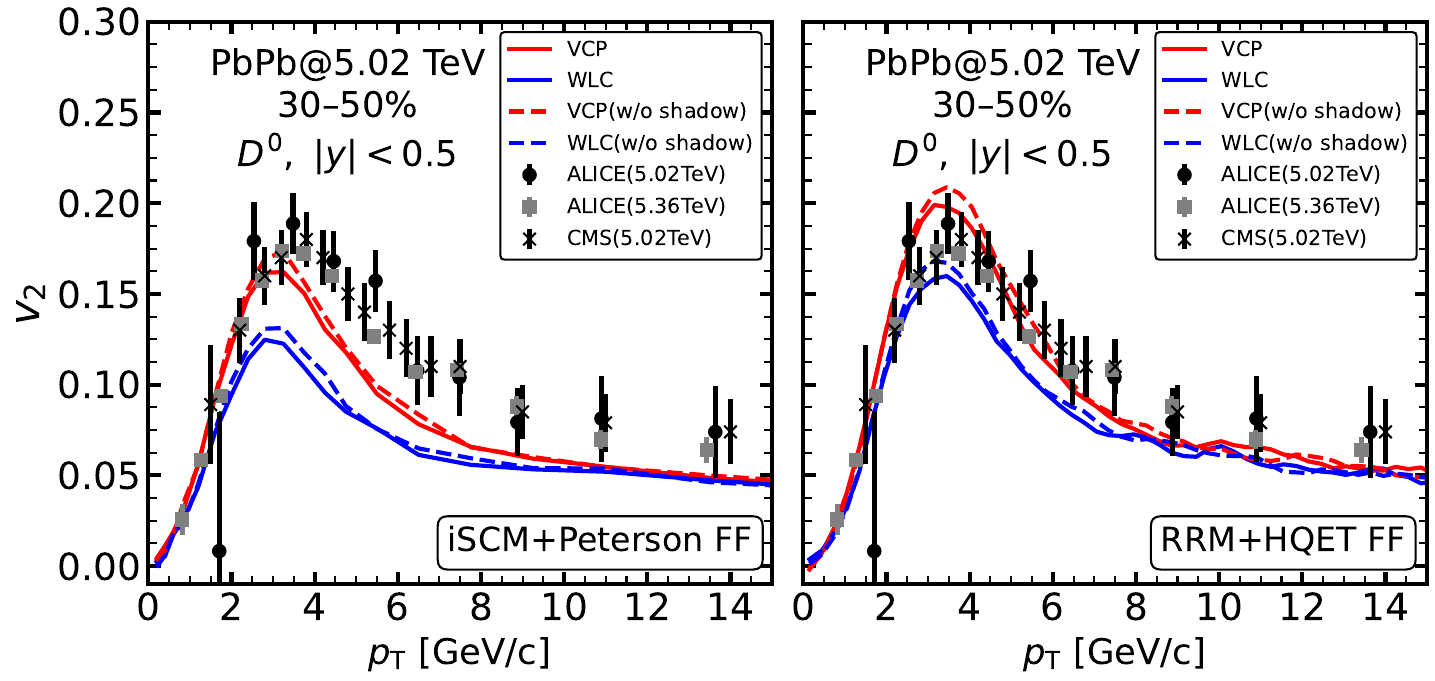}
    \caption{
    Impact of the initial-state nuclear-PDF modification on the $D^0$ nuclear modification factor (upper) and elliptic flow (lower) in 30--50\% Pb--Pb collisions at $\sqrt{s_{\NN}}=5.02\,\mathrm{TeV}$.
    }
    \label{fig:npdf-effect-PbPb}
\end{figure}
Proceeding with hadronization, the $D^0$ results obtained with and without the nuclear-PDF modification are compared in Fig.~\ref{fig:npdf-effect-PbPb}. Nuclear shadowing suppresses $\raa$ in the low-$\pT$ region, bringing the $D^0$ results closer to the central value of the data, which is a more pronounced effect for the RRM model and essential to describe the data at low or intermediate $\pT$. For $v_2$, the difference between the shadowed and unshadowed calculations follows a similar pattern with relatively small impact as at the charm-quark level.

\subsubsection{Diffusion and Radiative Energy Loss}
\label{sssec:diff-rad}
\begin{figure}[H]
    \centering
    \includegraphics[width=0.72\linewidth]
    {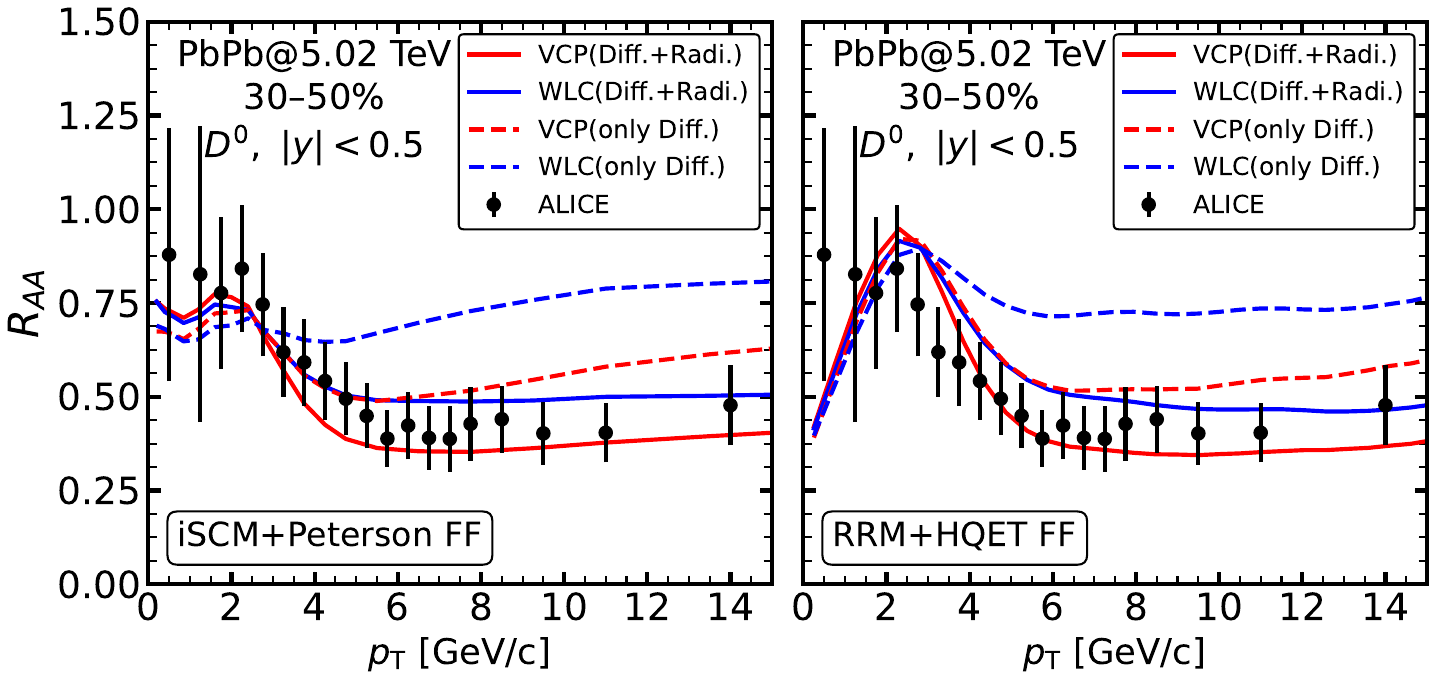}
    \includegraphics[width=0.72\linewidth]
    {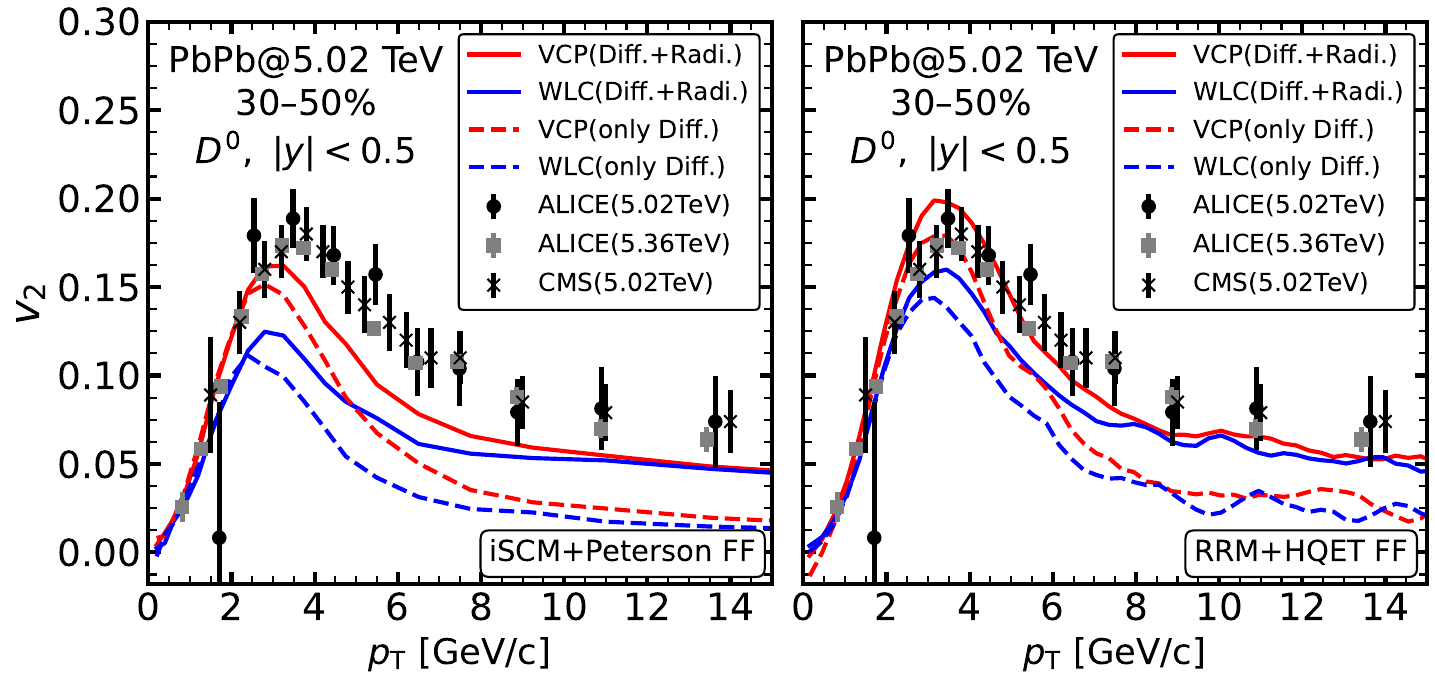}
    \caption{Comparison of the full diffusion + radiation calculation with the diffusion-only calculation for the $D^0$ nuclear modification factor (upper) and elliptic flow (lower) in 30--50\% Pb--Pb collisions at  $\sqrt{s_{\NN}}=5.02\,\mathrm{TeV}$.
    }
    \label{fig:radiation-effect-PbPb}
\end{figure}

We now isolate the role of medium-induced gluon radiation by comparing the full calculation, which contains both diffusion and radiation, with a calculation retaining elastic diffusion processes only. The results are shown in Fig.~\ref{fig:radiation-effect-PbPb}.

Diffusion provides the dominant modification of the charm spectrum at low and intermediate momentum, where the Brownian motion description of the HQ is most effective. Toward larger $\pT$, the diffusion-only calculation produces insufficient
suppression relative to the full result. The inclusion of medium-induced radiation lowers the $\raa$ increasingly at intermediate and high $\pT$, demonstrating the growing importance of the inelastic energy loss as the HQ momentum increases.

Radiation also affects the elliptic flow. In a non-central collision, the radiative energy loss depends on the HQ trajectory through the expanding medium. Heavy quarks propagating out of plane generally have a larger path length than those propagating in plane, resulting in stronger radiative suppression in the out-of-plane direction. This azimuthal-angle dependent suppression provides an additional positive contribution to $v_2$, complementing the anisotropy acquired through diffusive interactions with the collectively expanding medium. Consequently, the inclusion of radiation increases the $v_2$, particularly at intermediate and high $\pT$ where radiative processes become increasingly important. The simultaneous inclusion of diffusion and radiation is therefore necessary for describing both the magnitude of the suppression and its azimuthal dependence over the full momentum range considered here.

\begin{figure}[H]
    \centering
    \includegraphics[width=0.72\linewidth]{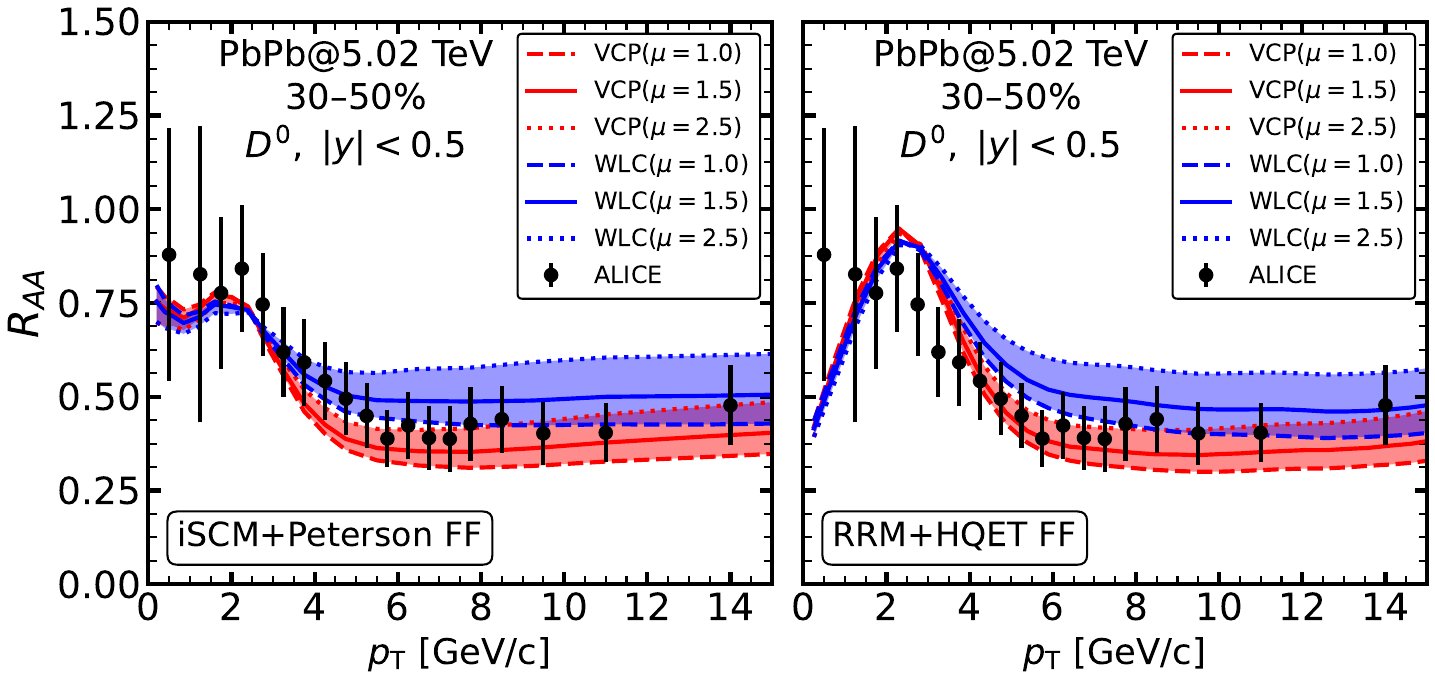}
    \includegraphics[width=0.72\linewidth]{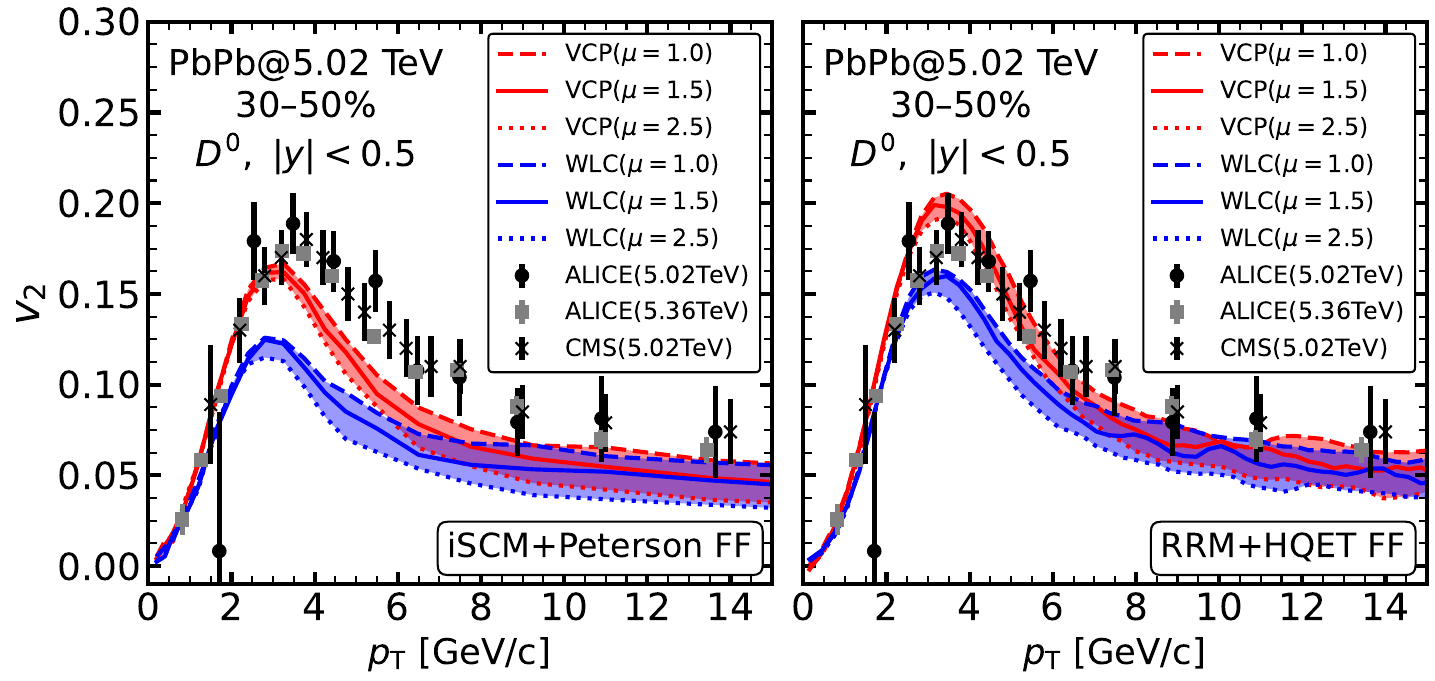}
    \caption{The $D^0$-meson $\raa$ (left) and $v_2$ (right) in 30--50\% Pb--Pb collisions at  $\sqrt{s_{\NN}}=5.02\,\mathrm{TeV}$, with different medium scales, $\mu$, figuring in Eq.~\eqref{eq:alphas}. }
    \label{fig:mu-effect-PbPb}
\end{figure}
To further assess the uncertainty associated with the radiative sector, we vary the medium-scale parameter $\mu$ entering the running coupling in Eq.~\eqref{eq:alphas} and show the results in Fig.~\ref{fig:mu-effect-PbPb}. The central value is taken to be $\mu=1.5$, while the upper and lower edges of each uncertainty band correspond to $\mu=1.0$ and $\mu=2.5$, respectively.  Since increasing $\mu$ raises the minimum scale of the running coupling and thereby weakens the perturbative parton--medium interaction, its effect is most relevant in the momentum region where medium-induced radiation is important. A smaller $\mu$ strengthens the radiative interaction, resulting in a stronger suppression of $\raa$ and a larger $v_2$ through the enhanced path-length dependence of the energy loss. At low $\pT$, the bands become much narrower because the evolution is dominated by the nonperturbative diffusion sector.

\subsubsection{Hadronic Rescattering}
\label{sssec:urqmd}

\begin{figure}[H]
    \centering
    \includegraphics[width=0.72\linewidth]
    {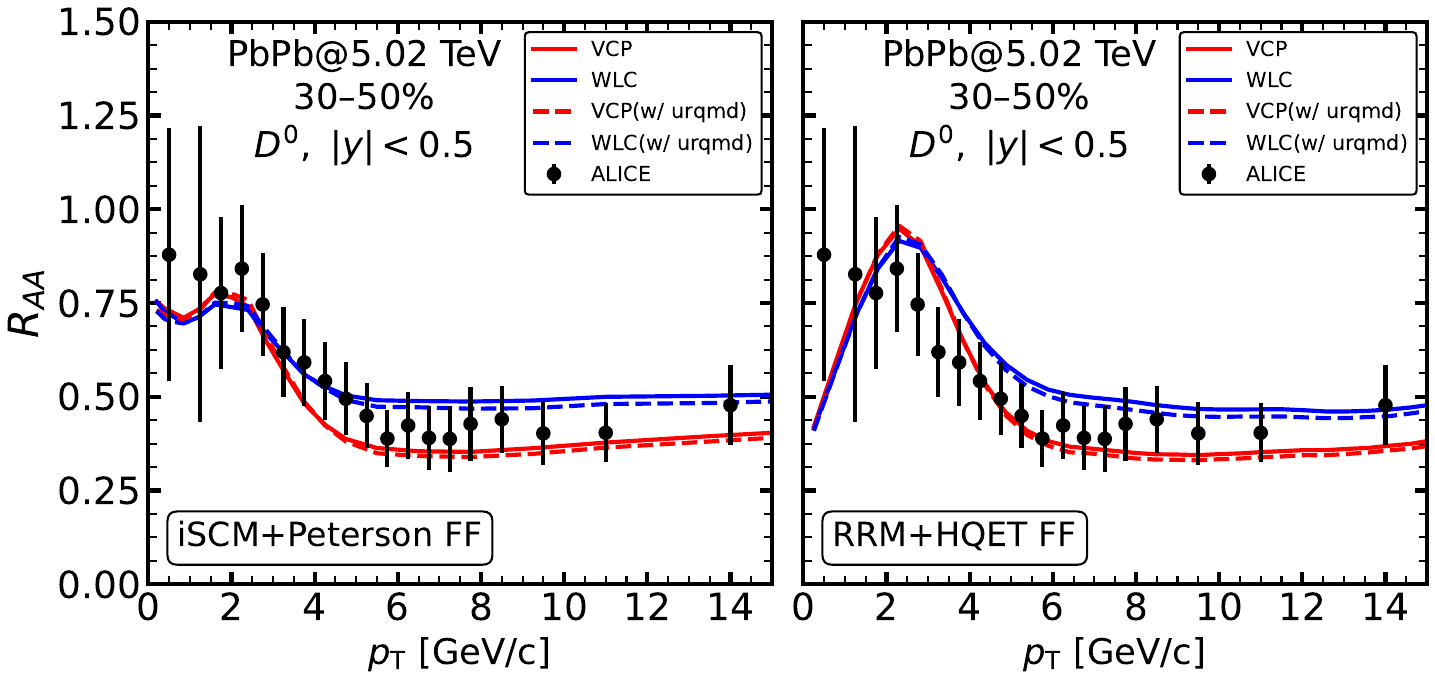}
    \includegraphics[width=0.72\linewidth]
    {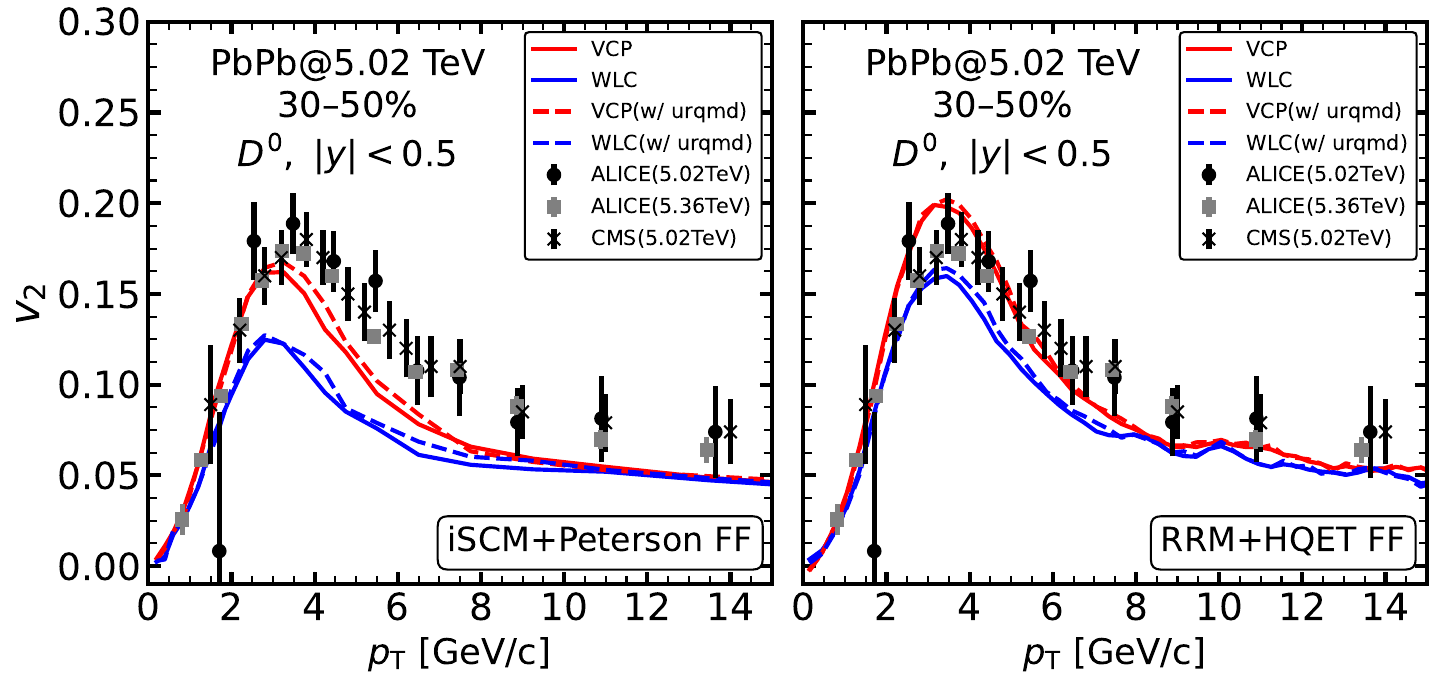}
    \caption{Impact of hadronic rescattering on the $D^0$ nuclear modification factor (upper) and elliptic flow (lower) in 30--50\% Pb--Pb collisions at   $\sqrt{s_{\NN}}=5.02\,\mathrm{TeV}$. Results with and without the subsequent UrQMD hadronic evolution are compared.}
    \label{fig:urqmd-effect-PbPb}
\end{figure}

The final contribution that we want to assess in the present framework is the rescattering of charm hadrons in the hadronic medium  (focusing on the $D^0$ meson in this work). We note that this part has not been included in the calculations presented in previous sections, while its implementation has been discussed in Sec.~\ref{ssec:hadron}. The effects of the UrQMD evolution on $D^0$ observables in Pb--Pb collisions are displayed in Fig.~\ref{fig:urqmd-effect-PbPb}.

For the elastic cross sections of {5/7.5~mb} employed in this work, in connection with a rather low cutoff on the CM energies, hadronic rescattering produces quite a small modification of the $D^0$ $\raa$ compared to the changes accumulated during the QGP evolution and hadronization. {The effect on the elliptic flow, $v_2$, is a bit more significant but mostly restricted to the $\pT\simeq3-6$\,GeV region}.  By the time the system enters the hadronic phase, a sizable collective anisotropy has already developed in the bulk medium. Continued rescattering therefore transfers additional anisotropic flow to the charm hadrons. The result illustrates that $v_2$ retains some sensitivity to the late stages of the collision evolution even when the modification of the azimuthally averaged spectrum is comparatively modest.

\begin{figure}[H]
    \centering
    \includegraphics[width=0.49\linewidth]
    {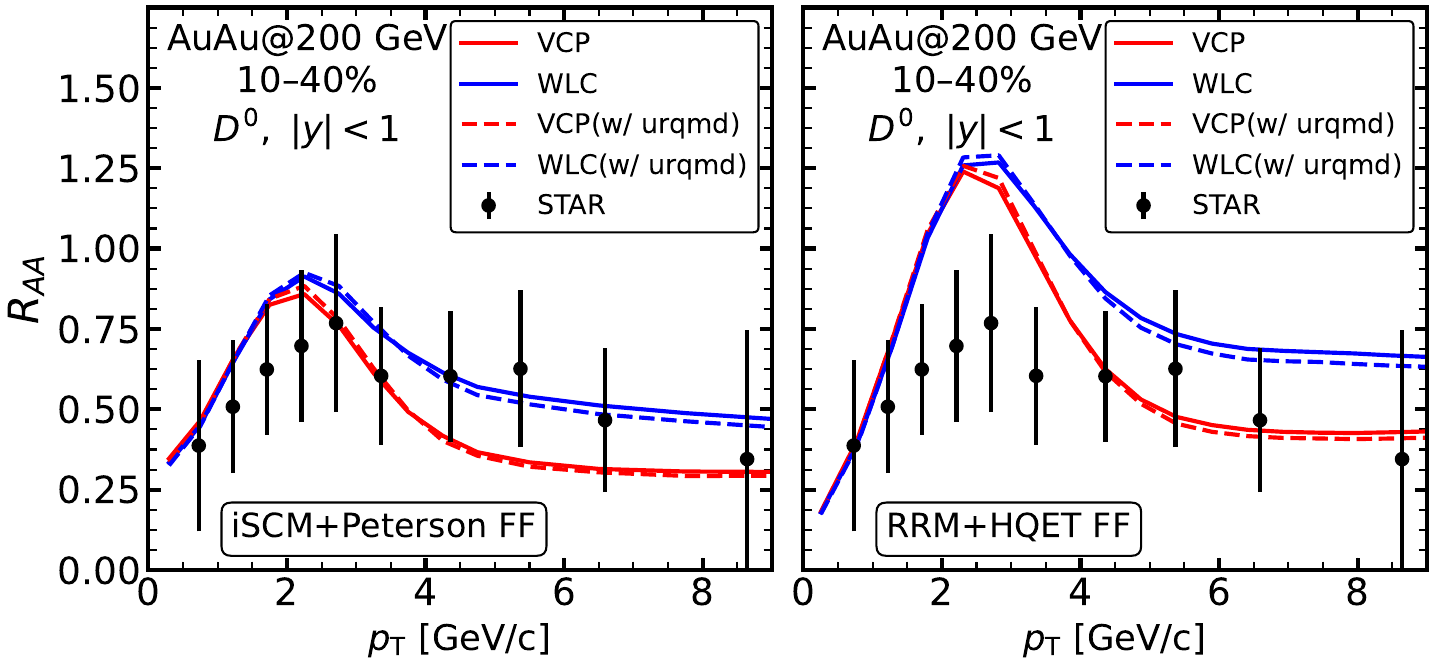}
    \includegraphics[width=0.49\linewidth]
    {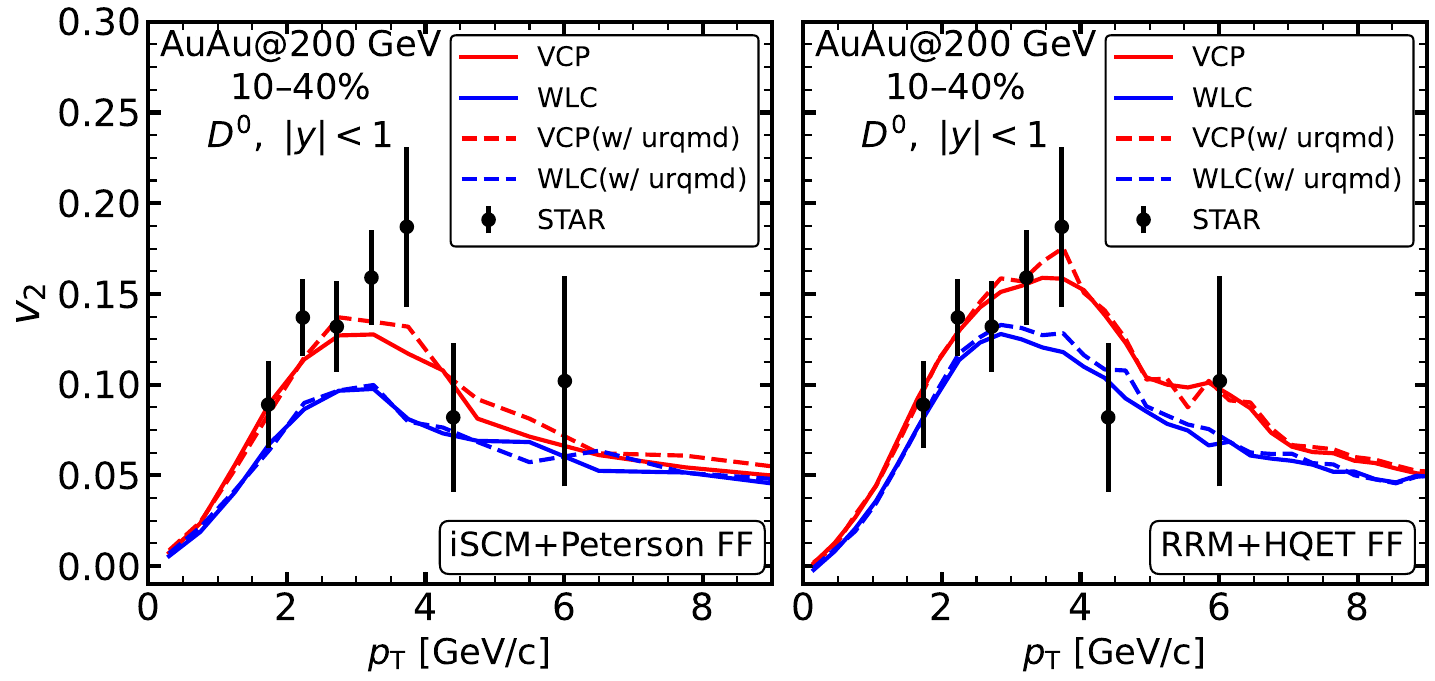}
    \caption{
    Impact of hadronic rescattering on the $D^0$ nuclear modification factor
    (left) and elliptic flow (right) in 10--40\% Au--Au collisions at
    $\sqrt{s_{\NN}}=200\,\mathrm{GeV}$.
    Results with and without the subsequent UrQMD hadronic evolution are
    compared.
    }
    \label{fig:urqmd-effect-AuAu}
\end{figure}

The corresponding comparison at RHIC energy is shown in
Fig.~\ref{fig:urqmd-effect-AuAu}. The Au--Au calculation provides an additional test of the relative importance of the hadronic stage at lower collision energy. Although the hadronic rescattering remains a correction to the much larger modifications accumulated during the QGP transport and hadronization stages, its contribution is not negligible, in particular for the final elliptic flow.

\vspace{0.5cm}

Let us conclude this Sec.~\ref{ssec:model-effects} on trying to highlight the role of individual components of our transport framework in their impact on the HF observables; let us reiterate the main points. The low-$\pT$ $\raa$ receives contributions from both nuclear shadowing and hadronization, while the increasing suppression toward higher momentum requires the radiative component of the QGP transport. Recombination plays a central role in converting the transported charm-quark distributions into the measured hadron spectra and provides an important contribution to $v_2$. Hadronic rescattering induces smaller but non-negligible 
late-stage modifications, especially on the elliptic flow. Consequently, a simultaneous interpretation of $\raa$, $v_2$, and charm-hadron chemistry is essential for quantitatively disentangling the microscopic HQ interactions from the other stages of the evolution in nuclear collisions.

\section{Summary}
\label{sec:sum}
We have presented an HF transport framework for ultrarelativistic heavy-ion collisions which combines up-to-date theoretical and phenomenological descriptions of the different stages relevant for the evolution of charm quarks and the hadrons containing them. The propagation of HQs in the QGP is described by relativistic Langevin dynamics, where the transport coefficients are determined from elastic interactions calculated in the thermodynamic $T$-matrix approach. We have considered two $T$-matrix interactions, denoted as VCP and WLC, reflecting current theoretical uncertainties in implementations of constraints from lattice QCD. 
The elastic interactions are supplemented with medium-induced gluon radiation, including coherence effects. The transport simulations have been embedded into a $2+1$D viscous hydrodynamic evolution of the bulk medium. Charm-quark hadronization following the QGP evolution has been treated in two variants, using either an improved sudden-coalescence model or the resonance recombination model, both supplemented with fragmentation. The fragmentation parameters and the charm-hadron fractions are constrained by the corresponding $pp$ spectra. Finally, the rescattering of $D^0$'s in the hadronic phase has been estimated using the UrQMD transport model.

We have applied this setup to a broad range of open-charm observables in Pb--Pb collisions at $\sqrt{s_{\NN}}=5$\,TeV and Au--Au collisions at $\sqrt{s_{\NN}}=200$\,GeV. The calculated $D^0$ nuclear modification factors reproduce the main features of the measured spectra in central and semi-central collisions, with stronger suppression in more central collisions, as expected. The calculations at RHIC energy also describe the broad structure of the STAR $D^0$ data, with appropriately amended initial-charm spectra and medium evolution. The comparison between the VCP and WLC results suggests that the $\raa$ alone does not provide a unique discrimination between the two interactions. In particular, the relative agreement with the data also depends on the hadronization model, as recombination modifies the mapping between the transported charm-quark spectrum and the final $D$-meson spectrum.

The elliptic flow of HQs provides an additional and possibly stronger constraint on the charm-medium interaction. For semi-central Pb--Pb collisions, the VCP interaction generally gives a larger $D^0$ $v_2$ than the WLC interaction in the intermediate-$\pT$ region. This difference can be understood from the temperature dependence of the corresponding friction coefficients: while the two interactions are rather similar at a temperature around 300\,MeV, the VCP is stronger toward lower temperatures and thus is more effective at building up the elliptic flow, which is largest in the later evolution of the QGP fireball. 
Hadronization generates additional modifications, where the RRM produces a larger enhancement of the elliptic flow than the iSCM, consistent with the larger recombination probability in RRM. A similar ordering of the VCP and WLC results is also found for the $D^0$ elliptic flow at RHIC.

The $D_s^+$ and $\Lambda_c^+$ results further illuminate the role of hadronization. Recombination with thermal light quarks changes both the momentum distribution and the anisotropy of the final charm hadrons. For $D_s^+$, recombination with strange quarks enhances the yield at low and intermediate $\pT$ relative to the strongly suppressed high-$\pT$ region and also increases the elliptic flow. The effect is more pronounced for the $\Lambda_c^+$, where two light quarks participate in the recombination process. Both hadronization models produce an enhancement of the $\Lambda_c^+$ $\raa$, and of the $\Lambda_c^+/D^0$ ratio for $\pT$'s around a few GeV. Since the preceding charm-quark transport is the same for the different charm-hadron species, this enhancement mainly reflects the distribution of charm among the different hadronic channels, together with the momentum carried by the recombining light quarks.

We have further conducted a study of selected individual components in the calculations. Nuclear modifications of the initial parton distributions mainly affect the low-$\pT$ part via a reduction of the $D^0$ $\raa$ and have a much smaller effect on $v_2$. 
Diffusion gives a sizable modification of the charm spectrum at low and intermediate momentum, while at larger $\pT$, medium-induced radiation becomes increasingly important to produce sufficient suppression and increase the $v_2$, through the HQ path length dependence in the spatially anisotropic fireball.
Concerning hadronization, our results for $D^0$ $v_2$ indicate that the RRM scheme has greater power than iSCM in increasing elliptic flow via recombination. Hadronic rescattering has been found to produce a small but noticeable change in the $\raa$ plots, but its effect on $v_2$ is more visible since the hadronic medium carries substantial collective flow.  The $D_s^+$ and $\Lambda_c^+$ observables, in particular their yield ratios, 
provide complementary information on the hadronization process. 
A simultaneous description of these observables is instrumental for quantitative determinations of the HQ interaction in the QCD medium.

The framework presented here achieves a fair overall description of available HF data at RHIC and the LHC using lattice-QCD-based interactions that do not require any scaling factors. This opens up the possibility to delineate remaining uncertainties in the microscopic description of transport parameters. Further improvements are expected from upcoming developments aimed at completing our approach by including: (i) an assessment of the transport in the pre-equilibrium stage, (ii) accounting for event-by-event fluctuations in the bulk evolution, and (iii) a refined treatment of HF hadron transport by bridging the current gap between hadronization (at $\Th=160$\,MeV) and the transition from hydrodynamics to UrQMD (at $T_{\rm sw}=152$\,MeV), using a smoothly matched diffusion set-up based on microscopic interactions with a possibly extended hydrodynamic medium evolution.

Several future applications are of considerable interest, especially in view of upcoming measurements. In a  systematic deployment to the bottom sector, the mass dependence promises a higher sensitivity to the transport coefficients, while in smaller collision systems, such as O-O collisions, a shorter-lived QGP system may increase the sensitivity in the charm sector. Beyond single-particle observables, the medium modifications of HF pair correlations from their initial production can inform on the momentum broadening, energy loss, and de-correlation of HQs in the medium. Work in some of these directions is underway.

\section{Acknowledgment}
We thank Cameron Dean, Yen-Jie Lee, Ivan Vitev, and Ramona Vogt for valuable discussions, and Zhanduo Tang for providing the $T$-matrix transport coefficients. This work was supported by the U.S. Department of Energy through the Topical Collaboration in Nuclear Theory on Heavy-Flavor Theory (HEFTY) for QCD Matter under award no. DE-SC0023547, DOE Office of Nuclear Physics, grant No. DE-FG02-05ER41367 (SAB), and by the U.S. National Science Foundation under grant no. PHY-2514775. This research used resources of the National Energy Research Scientific Computing Center (NERSC), a U.S. Department of Energy Office of Science User Facility under Contract No. DE-AC02-05CH11231 using NERSC Award No.
NP-ERCAP0026721. WK is supported by NSFC under Grant No. 12575140.

\newpage
\appendix 

\section{Model Uncertainties in Radiative Energy Loss}
\label{app:rad}

In this appendix, we discuss two sources of uncertainty associated with the implementation of the dead-cone effect in radiative energy loss, as well as the cutoff dependence of the incoherent radiation rates.

\paragraph{The dead-cone effect in the presence of LPM interference}
Although Ref.~\cite{Ke:2018jem} has shown that the implementation of radiative energy loss of light quarks and gluons in LIDO agrees with weak-coupling theoretical calculations within 10\%, incorporating finite-mass effects into the dynamical treatment of LPM interference is more complicated. In particular, the mass dependence cannot generally be accounted for by simply multiplying the entire radiation spectrum by the vacuum dead-cone suppression factor. To assess the impact of this model uncertainty on our results, we consider two choices for the dead-cone exponent, $n=1$ and $n=2$, in Eq.~\eqref{eq:Pdc}. As discussed previously, $n=2$ is expected for incoherent diffusion-induced radiation  (Bethe-Heitler regime), whereas $n=1$ provides better agreement with the coherent radiation spectrum in a finite medium obtained from SCET$_{\rm G}$ calculations.

\begin{figure}[H]
    \centering
    \includegraphics[width=0.7\linewidth]{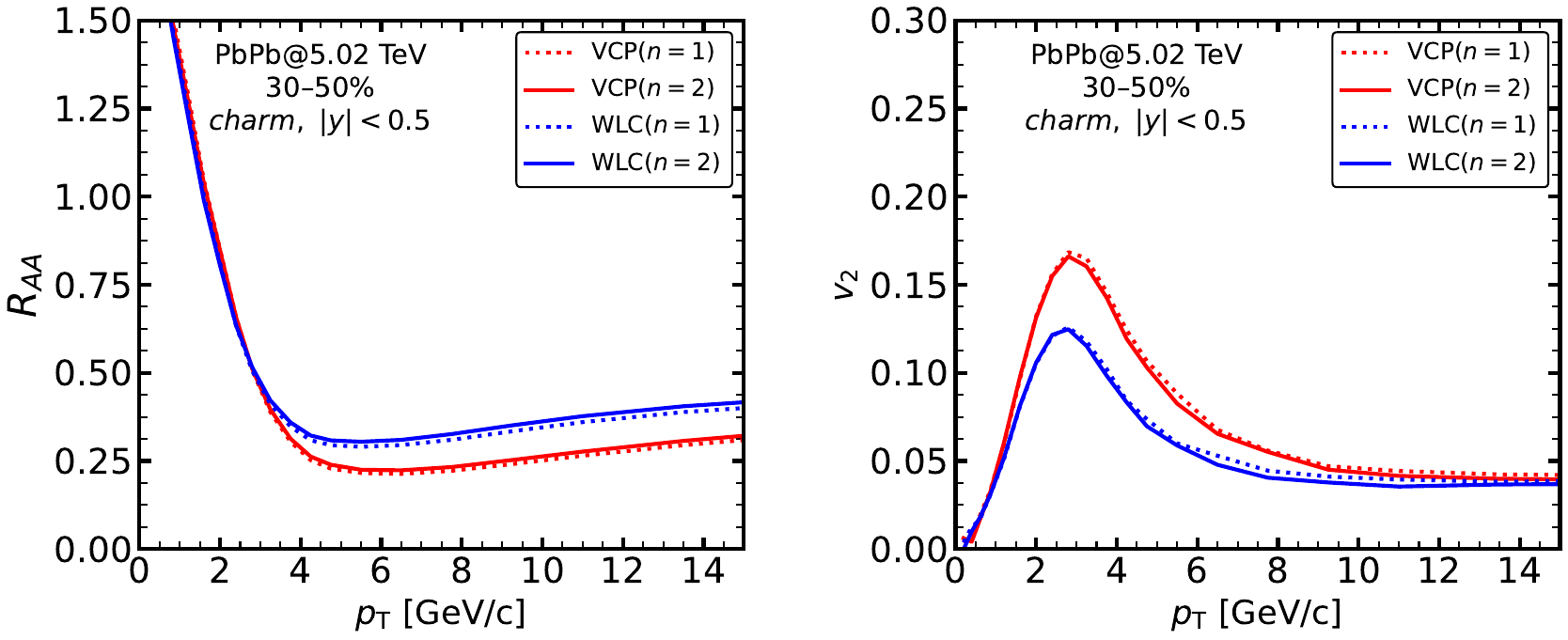}
  \caption{Sensitivity of the charm-quark nuclear modification factor $\raa$ (left) and elliptic flow $v_2$ (right) to the dead-cone prescription for 30--50\% Pb--Pb collisions at $\sqrt{s_{\NN}}=5.02~\mathrm{TeV}$ and $|y|<0.5$. Red and blue curves denote the VCP and WLC transport scenarios, respectively, with dotted and solid curves corresponding to $n=1$ and the default choice of $n=2$ for the exponent in the dead-cone suppression factor of Eq.~(\ref{eq:Pdc}).}
    \label{fig:DC-power}
\end{figure}

Figure~\ref{fig:DC-power} compares the resulting $\raa$ and $v_2$ of charm quarks after transport through the QGP fireball in 30--50\% central Pb--Pb collisions at the LHC. The small differences between the two prescriptions indicate that the charm-quark spectra are only weakly sensitive to the choice of dead-cone exponent over the momentum range considered. This weak sensitivity can be understood from two considerations. First, within the $\pT$ range studied, radiative energy loss is already subleading compared to collisional effects. Second, both prescriptions introduce the same characteristic dead-cone angular scale, $\theta E \sim M$, which suppresses gluon radiation at small angles. Varying the exponent primarily changes the sharpness of this suppression without altering its characteristic scale. Consequently, the total energy loss, and hence the resulting charm-quark spectra, remain relatively insensitive to the choice of dead-cone exponent in the $\pT$ range considered.

\paragraph{Modeling uncertainty associated with $Q_{\rm cut}$}

Another source of uncertainty arises from the parameter $Q_{\rm cut}$ used in the calculation of diffusion-induced incoherent radiation. This parameter controls the magnitude of the perturbative transport coefficient, $\hat{q}_S$, given by
\begin{align}
\hat{q}_S = \alpha_s(Q_{\rm cut}^2) C_R m_D^2 T \ln\left(1+\frac{Q_{\rm cut}^2}{m_D^2}\right) \ .
\end{align}
Since the HQ diffusion coefficient is already determined by the $T$-matrix calculation, the dependence on $Q_{\rm cut}$ enters primarily through its influence on HQ radiative energy loss. In particular, $\hat{q}_S$ determines the characteristic mean free path entering the Bethe-Heitler radiation rate, $\lambda\sim m_D^2/\hat{q}_S$. In the deep LPM regime, however, the suppression factor effectively replaces the mean free path $\lambda$ with the radiation formation time, which is determined self-consistently from the transverse-momentum broadening dynamically accumulated during the simulation. Therefore, varying $Q_{\rm cut}$ is expected to affect the radiative energy loss primarily through its influence on the Bethe-Heitler regime. Furthermore, the use of a running coupling weakens the dependence of $\hat{q}_S$ on $Q_{\rm cut}$.

\begin{figure}[H]
    \centering
    \includegraphics[width=0.7\linewidth]{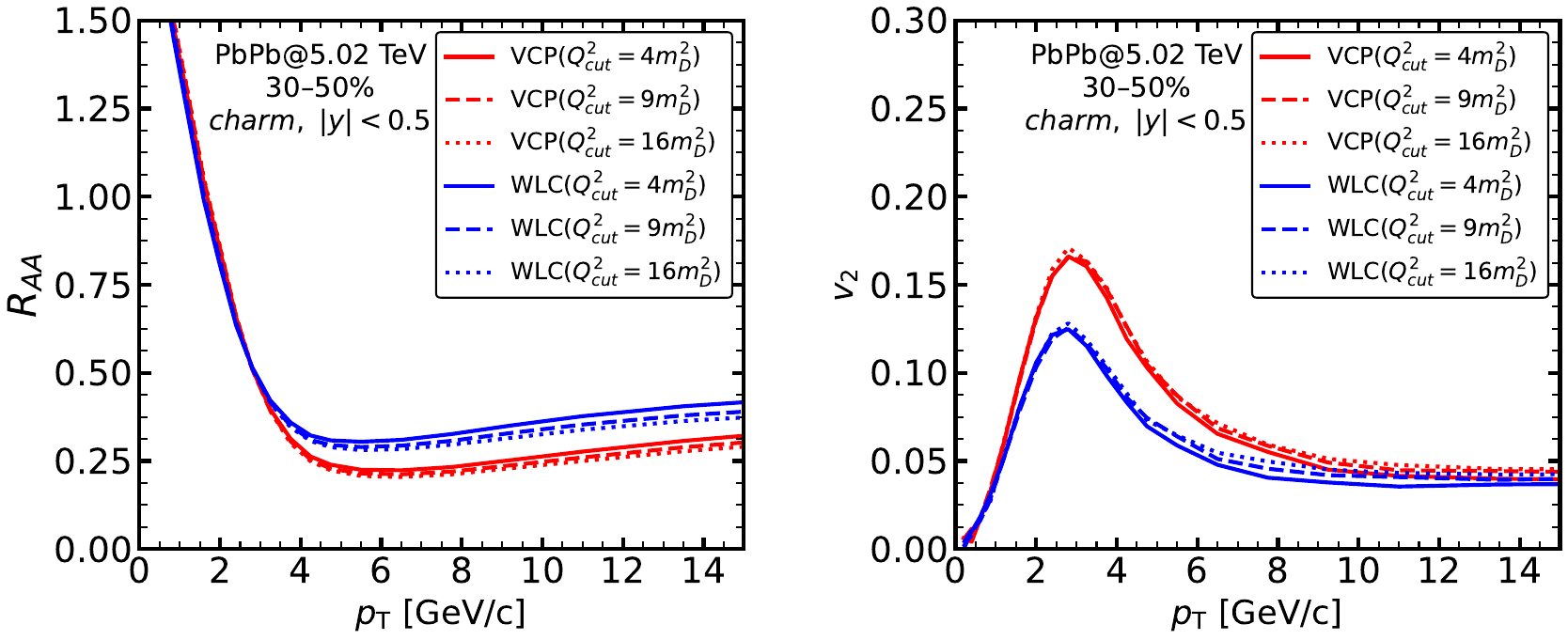}
  \caption{Sensitivity of the charm-quark nuclear modification factor $\raa$ (left) and elliptic flow $v_2$ (right) to the cutoff scale $Q_{\rm cut}$, shown for 30--50\% Pb--Pb collisions at $\sqrt{s_{\NN}}=5.02~\mathrm{TeV}$ and $|y|<0.5$. Red and blue curves denote the VCP and WLC transport scenarios, respectively. Solid, dashed, and dotted curves correspond to $Q_{\rm cut}^2=4m_D^2$, $9m_D^2$, and $16m_D^2$, respectively. Owing to the logarithmic dependence on $Q_{\rm cut}^2$, the variation of the cutoff scale produces only a small effect on both observables.}
    \label{fig:Qcut}
\end{figure}

In Fig.~\ref{fig:Qcut}, we verify that the resulting $\raa$ and $v_2$ are largely insensitive to a factor-of-two variation in $Q_{\rm cut}$, from $2m_D$ to $4m_D$. This indicates that adopting a fixed value of $Q_{\rm cut}$ in the main analysis does not significantly affect the conclusions of our study.

\section{Initial-State Nuclear Effects}
\label{app:IS-effect}
In the calculations for AA systems presented in this work, the initial charm-quark spectrum is modified relative to the corresponding $pp$ baseline to account for cold-nuclear-matter effects in the incoming nuclei. The purpose of this appendix is to specify the momentum-dependent modification used for Pb--Pb and Au--Au collisions and to explain how its overall normalization is chosen.

For HQ production at the collision energies considered here, an important initial-state effect arises from the nuclear modification of the parton distribution functions (PDFs). These modifications are conventionally characterized by
\begin{equation}
R_i^A(x,Q^2)
=
\frac{f_i^{p/A}(x,Q^2)}
     {f_i^p(x,Q^2)},
\end{equation}
where $f_i^{p/A}(x,Q^2)$ denotes the nuclear PDF (nPDF) of parton species $i$ in a
proton bound inside nucleus $A$ and $f_i$ the corresponding PDF
in a free proton; $x$ is the longitudinal momentum fraction carried by the incoming parton and $Q$ the factorization scale.

Existing nPDF parameterizations come with sizable uncertainties, and a systematic assessment of those is beyond the scope of the present work. Since the total number of charm quarks is believed to be approximately conserved throughout the subsequent medium evolution, the experimentally measured momentum-integrated $\raa$ can be used as a constraint on the overall magnitude of the total suppression from initial-state effects. 
For Pb--Pb collisions at LHC energy, where such measurements are available, we implement the nPDF effect through a $\pT$-dependent multiplicative factor. Specifically, we use the shadowing parametrization adopted from Ref.~\cite{Emelyanov:1997guf},
\begin{equation}
S_c(\pT)
=
a_1 e^{-b_1(\pT-d_1)} + c_1
+ a_2 e^{-b_2(\pT-d_2)},
\label{eq:Sc}
\end{equation}
where the parameters are given in Table~\ref{tab:shadowing_parameters} for the considered percentages of the shadowing.
\begin{table}[H]
\centering
\renewcommand{\arraystretch}{1.5} 
\setlength{\tabcolsep}{12pt}      
\begin{tabular}{|c|c|c|c|c|c|c|c|}
\hline
$\%$ shadowing &$a_1$ & $b_1$ & $d_1$ & $a_2$ & $b_2$ & $d_2$ & $c_1$ \\
\hline
$20\%$ &$-0.20$ & $0.17$ & $1.00$ & $-65.39$ & $0.52$ & $-11.14$ & $1.02$ \\
\hline
$25\%$ &$-0.27$ & $0.17$ & $1.00$ & $-65.39$ & $0.52$ & $-11.14$ & $1.02$ \\
\hline
$30\%$ &$-0.33$ & $0.17$ & $1.00$ & $-65.39$ & $0.52$ & $-11.14$ & $1.02$ \\
\hline
\end{tabular}
\caption{Parameters used in the shadowing parametrization of Eq.~\eqref{eq:Sc}.}
\label{tab:shadowing_parameters}
\end{table}

\begin{figure}[H]
    \centering
    \begin{minipage}[c][5cm][c]{0.32\linewidth}
        \centering
        \includegraphics[width=\linewidth,height=5cm,keepaspectratio]{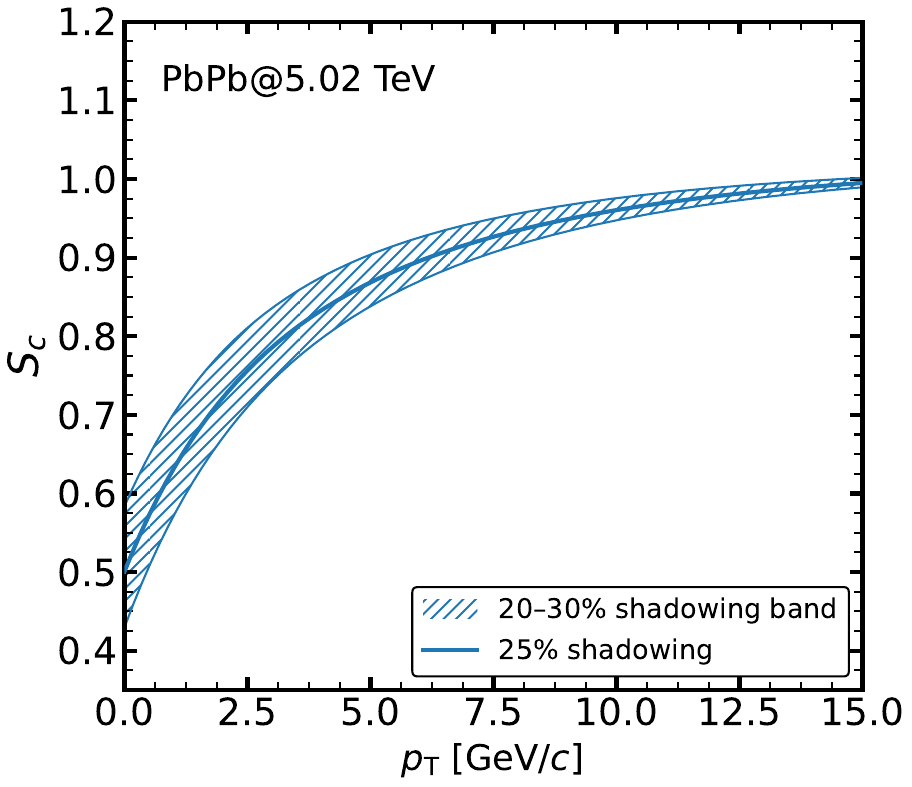}
    \end{minipage}\hfill
    \begin{minipage}[c][5cm][c]{0.32\linewidth}
        \centering
        \includegraphics[width=\linewidth,height=5cm,keepaspectratio]{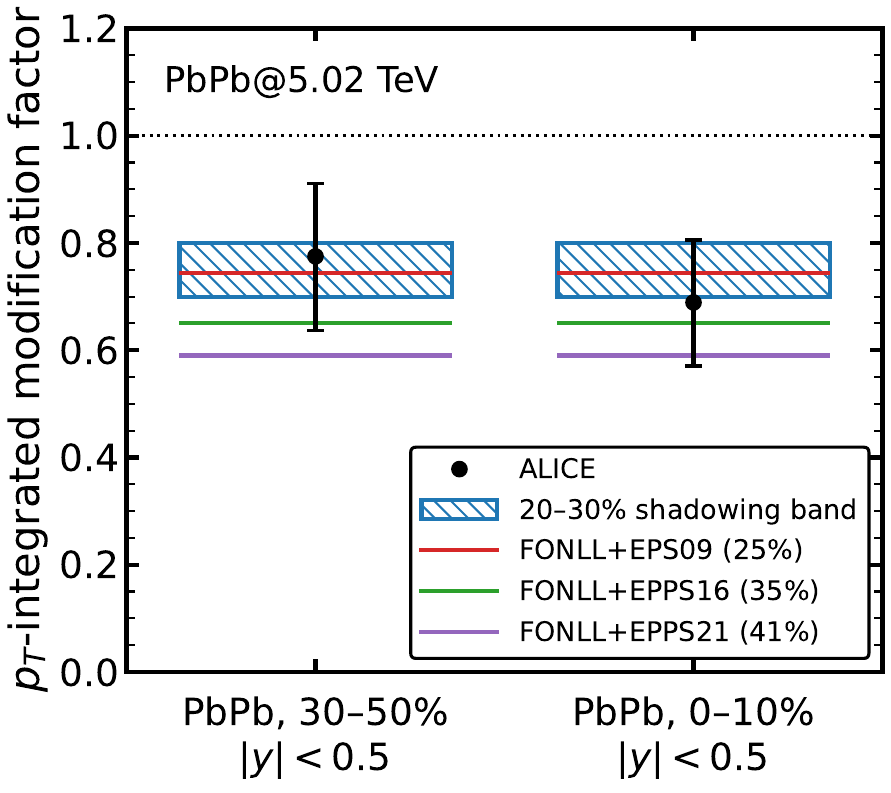}
    \end{minipage}\hfill
    \begin{minipage}[c][5cm][c]{0.32\linewidth}
        \centering
        \includegraphics[width=\linewidth,height=5cm,keepaspectratio]{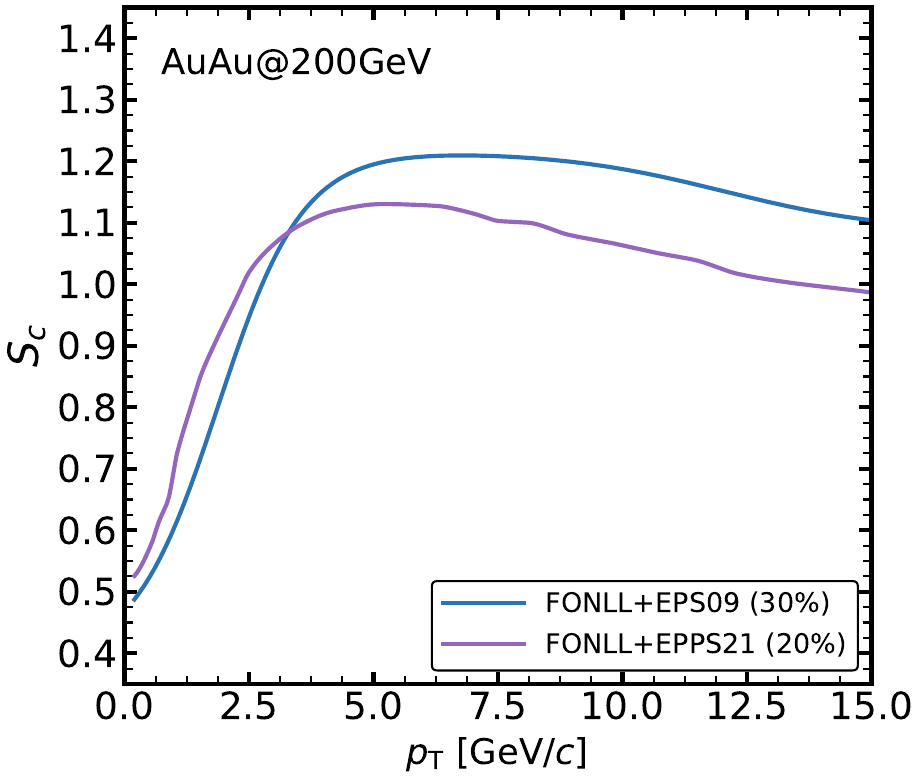}
    \end{minipage}
    \caption{Nuclear modification factors for charm quarks. Left panel: $\pT$ dependence of the  20--30\% shadowing envelope and its 25\% central value for Pb--Pb collisions at $\sqrt{s_{\mathrm{NN}}}=5.02$\,TeV as used in this work. 
 Middle panel: $\pT$-integrated shadowing scenarios compared with ALICE $\pT$-integrated nuclear modification factor measurements of $D^0$ in 2 centrality bins as measured in Pb--Pb collisions at $\sqrt{s_{\mathrm{NN}}}=5.02$\,TeV.
Right panel: $\pT$-dependent shadowing from scaled FONLL+EPS09 (used in this work) and unscaled FONLL+EPPS21 for Au-Au collisions at $\sqrt{s_{\mathrm{NN}}}=200$\,GeV.}
    \label{fig:shadowing}
\end{figure}

A 25\% inclusive shadowing is used as the baseline initial-state modification for Pb--Pb collisions at $\sqrt{s_{\NN}}=5.02$\,TeV (after folding with the initial FONLL charm spectrum); it is close to the central value obtained with the EPS09NLO nPDF~\cite{Eskola:2009uj}, and the resulting $\pT$ dependence is shown in the left panel of Fig.~\ref{fig:shadowing}. To estimate the sensitivity to the overall strength of the initial-state modification, we vary the normalization of inclusive charm suppression from 20\% to 30\%, as illustrated for $D^0$ observables in Fig.~\ref{fig:RAA-PbPb-30to50} in our Pb--Pb calculations. In the middle panel of Fig.~\ref{fig:shadowing}, we compare our band to the $\pT$-integrated central values from FONLL calculations with EPPS16~\cite{Eskola:2016oht} and EPPS21~\cite{Eskola:2021nhw} nPDFs, as well as ALICE data~\cite{ALICE:2021rxa}.
For simplicity, we neglect the centrality dependence of the shadowing correction by applying the same parametrization to 0--10\% and 30--50\% Pb--Pb collisions.

At RHIC energies, assuming $\Lambda_c^+/D^0 \approx 0.25$ (as discussed in Sec.~\ref{ssec:frag}), the total $c\bar{c}$ production cross section in $pp$ collisions amounts to approximately $150\,\mu$b. In Au--Au collisions, where our hadronization model predicts $\Lambda_c^+/D^0 \approx 0.5$, reproducing the measured $D^0$ yield would require an effective integrated suppression of nearly $40\%$ from the $pp$ baseline. However, since the nPDF parameterizations considered here do not predict such a large suppression, we adopt a somewhat smaller value of $30\%$. We then use the $\pT$-dependence from FONLL+EPS09 calculations, normalized to this assumed integrated suppression. As an illustration of the uncertainty, we compare our results to a FONLL+EPPS21 prediction (central values), corresponding to an integrated suppression of approximately $20\%$,  in the right panel of Fig.~\ref{fig:shadowing}.  
 
\section{Feeddown Uncertainty from Excited $\Lambda_c$ States in RRM}
\label{app:feeddown-uncertainty}
The RRM employed in this work includes an
extended spectrum of excited charm-baryon states~\cite{He:2019tik}. In the default implementation used for the results presented in the main text, the excited $\Lambda_c$ states are assumed to feed down to the ground-state $\Lambda_c$. 
For sufficiently massive excited $\Lambda_c$ states, however, strong decays into a ground-state $D$ meson and a nucleon become kinematically allowed once the corresponding $DN$ threshold is exceeded. While there is some experimental evidence for decays of excited $\Lambda_c$ states into $DN$ channels~\cite{LHCb:2017jym,Belle:2024cmc}, the branching fractions of these high-lying states are not well constrained. Therefore, assuming that all such states eventually feed down to the ground-state $\Lambda_c$ introduces uncertainty in the final charm-hadron composition. This uncertainty is considered only for the RRM calculation. 
The iSCM, by comparison, contains a much more limited set of excited $\Lambda_c$ states, and the corresponding sensitivity to possible $DN$ decay channels is found to be negligible for the observables considered here.

\begin{figure}[H]
    \centering
    \includegraphics[width=0.45\linewidth]{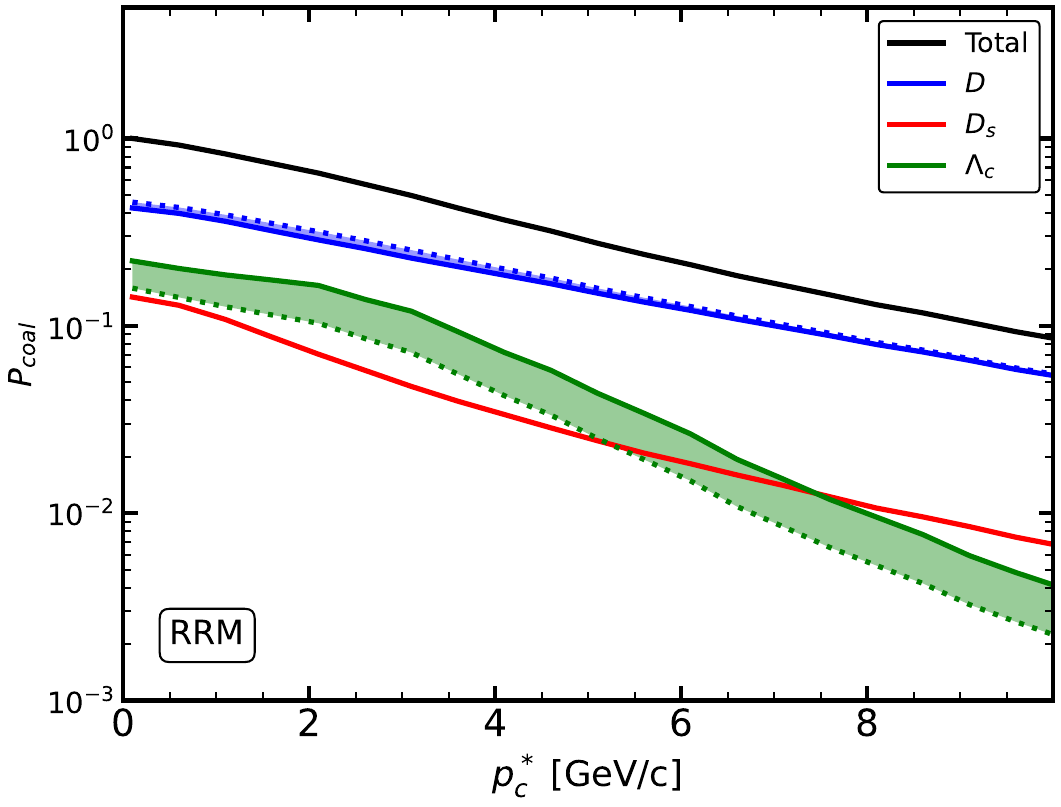}
    \caption{
    Species-resolved recombination probabilities in the RRM as functions of the charm-quark momentum in the local fluid rest frame. Solid curves correspond to the default feeddown prescription used in the main text, while dotted curves show the alternative prescription in which excited $\Lambda_c$ states above the $DN$ threshold have a 50\% probability to feed down through the $DN$ channel.
    }
    \label{fig:rrm_feeddown_pcoal}
\end{figure}

To estimate the sensitivity of the calculated observables to this feeddown uncertainty, we consider a 50\% fraction of feeddown to the ground-state $\Lambda_c$ and 50\% to the $DN$ channel, with the latter contributing to the final ground-state $D$-meson yield. All other ingredients of the hadronization calculation, including the recombination probabilities and the fragmentation prescription, are kept unchanged. Figure~\ref{fig:rrm_feeddown_pcoal} illustrates how this variation redistributes the feeddown contribution among the final charm-hadron
species. While the total RRM recombination probability is unchanged, the alternative prescription decreases the $\Lambda_c$ contribution and correspondingly increases the non-strange $D$-meson contribution.

Since this variation does not change the total charm yield, its main effect is to redistribute charm among the final hadron species, decreasing the final $\Lambda_c$ and increasing the ground-state $D$-meson yields. As a consequence, observables involving $\Lambda_c$ are expected to be particularly sensitive to this variation, while ratios with $D^0$ in the denominator can also be modified.

\begin{figure}[H]
    \centering
    \includegraphics[width=0.7\linewidth]{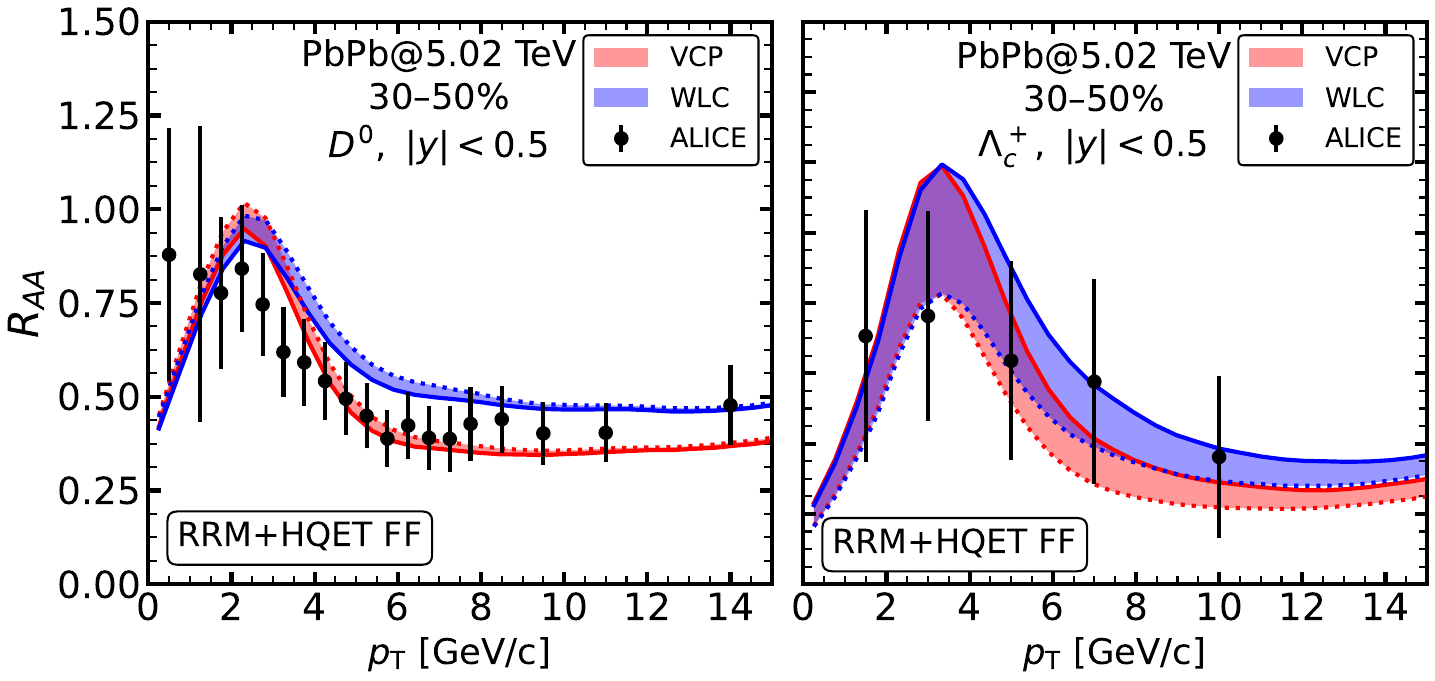}
    \caption{Nuclear modification factors of $D^0$ and $\Lambda_c^+$ in 30--50\% Pb--Pb collisions at $\sqrt{s_{\NN}}=5.02$\,TeV within the RRM+HQET-FF framework. Solid curves correspond to the default prescription in which the excited $\Lambda_c$ states feed down to the ground-state $\Lambda_c$. Dotted curves show the alternative prescription in which excited $\Lambda_c$ states above the $DN$ threshold are assigned equal probabilities to feed down to the ground-state $\Lambda_c$ and to the $DN$ channel. Red and blue curves denote the VCP and WLC transport scenarios, respectively.
    }
    \label{fig:feeddown_raa}
\end{figure}

Figure~\ref{fig:feeddown_raa} shows the resulting $D^0$ and $\Lambda_c^+$ $\raa$'s in 30--50\% Pb--Pb collisions at $\sqrt{s_{\NN}}=5.02$\,TeV within the RRM + HQET-FF framework. The modified feeddown prescription produces a visible redistribution between the $\Lambda_c$ and $D$ sectors. The $\Lambda_c^+$ $\raa$ is reduced, most noticeably in the intermediate-$\pT$ region where recombination and resonance feeddown are important. At the same time, the additional contribution from the $DN$ channel leads to a modest increase in the final $D^0$ yield and therefore raises the corresponding $D^0$ $\raa$. Toward larger $\pT$, where fragmentation becomes increasingly dominant and excited-state recombination decreases, the difference between the two feeddown prescriptions becomes smaller.

\begin{figure}[H]
    \centering
    \includegraphics[width=0.7\linewidth]{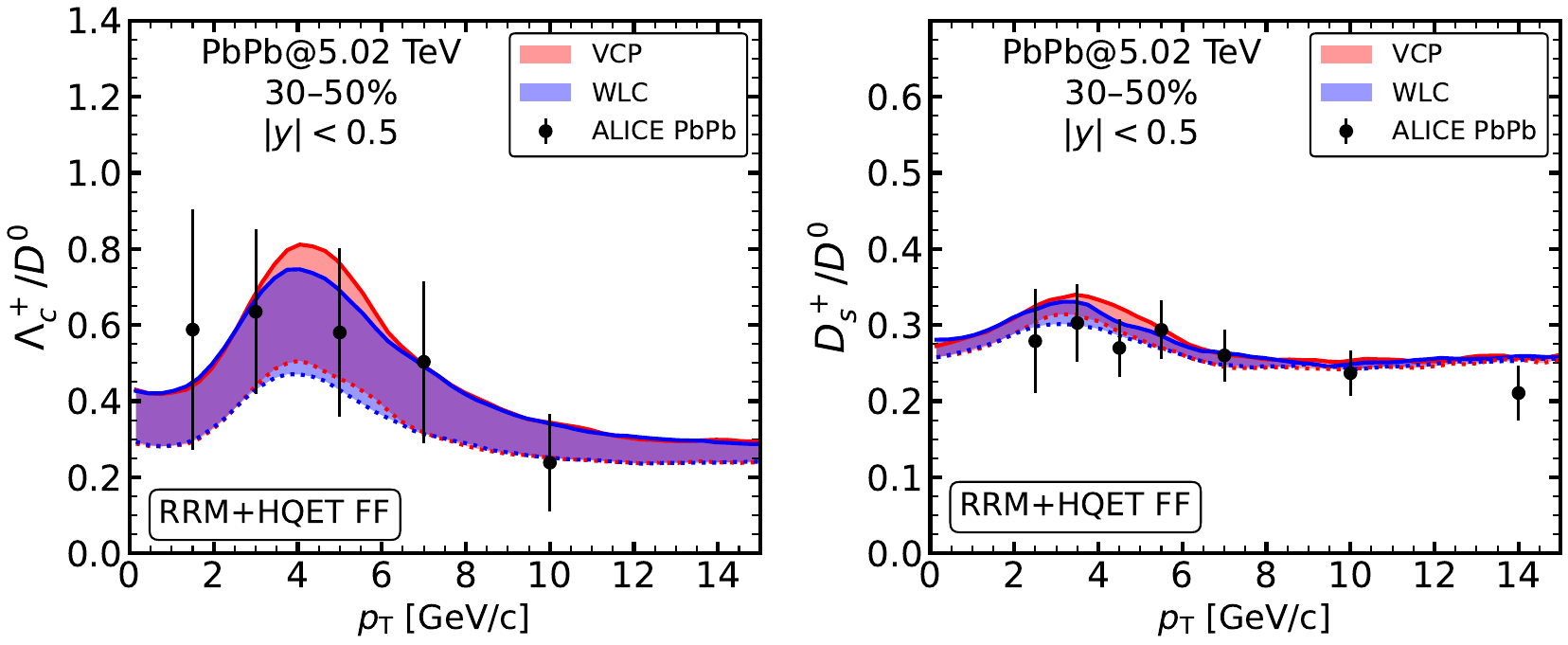}
    \caption{
    Charm-hadron yield ratios $\Lambda_c^+/D^0$ and $D_s^+/D^0$ in 30--50\% Pb--Pb collisions at $\sqrt{s_{\NN}}=5.02$\,TeV within the RRM+HQET-FF framework. Solid curves correspond to the default excited-$\Lambda_c$ feeddown prescription, while dotted curves show the alternative 50\%--50\% feeddown scenario for states above the $DN$ threshold. Red and blue curves denote the VCP and WLC transport scenarios, respectively.
    }
    \label{fig:feeddown_ratios}
\end{figure}

The effect on the charm-hadron yield ratios is shown in Fig.~\ref{fig:feeddown_ratios}. The $\Lambda_c^+/D^0$ ratio is reduced through the combined effect of a smaller $\Lambda_c^+$ numerator and a larger $D^0$ denominator. Likewise, the $D_s^+/D^0$ ratio is also reduced. The largest sensitivity appears in the intermediate-$\pT$ region, where recombination contributes substantially to charm-hadron production and a sizable fraction of the final yield originates from excited states. The effect is especially pronounced for $\Lambda_c^+/D^0$, since the alternative prescription modifies both the numerator and the denominator toward a smaller ratio. 
These results thus indicate that poorly constrained decays of excited $\Lambda_c$ states constitute a modest but noticeable hadronization uncertainty for observables involving charm baryons.

\bibliography{HEFTY-bib}

\end{document}